\documentclass[11pt]{article}
\pdfoutput=1

\usepackage{amssymb,amsfonts}
\usepackage{amsthm}
\usepackage{amsmath}
\usepackage{hhline}
\usepackage{picins}

\usepackage{times}
\usepackage{calc}
\usepackage[compact]{titlesec}

\usepackage{float} %
\usepackage{nicefrac} %
\usepackage{paralist} %
\usepackage[usenames,dvipsnames]{color} %
\usepackage{boxedminipage} %
\usepackage{fullpage} %
\usepackage[pdftex,bookmarks=true,pdfstartview=FitH,colorlinks,linkcolor=blue,filecolor=blue,citecolor=blue,urlcolor=blue,pagebackref=true]{hyperref}
\usepackage{upgreek} %
\usepackage{bm} %
\usepackage{ifthen} %
\usepackage{wrapfig} %
\usepackage{mathdots,yhmath}

\usepackage{tikz}
\usetikzlibrary{fit, matrix, positioning, calc}

\newcommand{\Q}{\cQ} %

\newcommand{\G}{\mathsf{Game}}

\newcommand{\pow}[1]{^{(#1)}}
\newcommand{\ii}{\pow{i}}
\newcommand{\iim}{\pow{i-1}}
\newcommand{\iip}{\pow{i+1}}

\renewcommand{\th}{^\text{th}}
\newcommand{\st}{^\text{st}}

\usepackage{bm, xspace}

\newcommand{\pr}{\mathsf{P}}
\usepackage[compact]{titlesec}

\newcommand{\remove}[1]{}

\newcommand{\OWF}{\ensuremath{{\mathsf{OWF}}}\xspace}
\newcommand{\OWFz}{\ensuremath{{\mathsf{OWF}_\zeta}}\xspace}
\newcommand{\SFE}[1]{\ensuremath{{{\mathsf{SFE}}_{#1}}}\xspace}

\newcommand{\gz}{\ensuremath{g^\zeta}\xspace}

\newcommand{\Adv}{\ensuremath{\mathsf{Adv}}\xspace}

\newcommand{\ol}{\ensuremath{\overline}}
\newcommand{\wt}[1]{\widetilde{#1}}
\newcommand{\wh}[1]{\widehat{#1}}
\newcommand{\nin}{\not \in}

\newcommand{\se}{\subseteq}

\newcommand{\rv}[1]{\mathbf{#1}} %

\newcommand{\ie}{i.e.,\ }

\newcommand{\eg}{e.g.,\ }

\newcommand{\set}[1]{\{ #1 \}}

\newcommand{\size}[1]{\left\vert#1\right\vert}
\newcommand{\abs}[1]{\size{#1}}

\newcommand{\adjSize}[1]{\left|#1\right|}
\newcommand{\adjAbs}[1]{\adjSize{#1}}

\newcommand{\zo}{\{0,1\}}
\newcommand{\bits}{\{0,1\}}

\renewcommand{\implies}{\Rightarrow}
\renewcommand{\iff}{\Leftrightarrow}

\newcommand{\getsr}  {\mathbin{\stackrel{\mbox{\,\tiny \$}}{\gets}}}

\newcommand{\cindn}[1]{\stackrel{}{\approx_\mathsf{c}}}

\newcommand{\from}{\gets}
\newcommand{\To}{\rightarrow}

\newcommand{\SD}{\mathsf{SD}}

\renewcommand{\S}{\ensuremath{{S}}\xspace}
\renewcommand{\L}{\ensuremath{{L}}\xspace}
\newcommand{\U}{\ensuremath{{U}}\xspace}

\newcommand{\C}{\ensuremath{{C}}\xspace}
\newcommand{\W}{\ensuremath{{W}}\xspace}

\newcommand{\N}{\ensuremath{{\mathbb N}}\xspace}

\newcommand{\cO}{\ensuremath{\mathcal O}\xspace}
\newcommand{\cP}{\ensuremath{\mathcal P}\xspace}
\newcommand{\cQ}{\ensuremath{\mathcal Q}\xspace}

\newcommand{\cU}{\ensuremath{\mathcal U}\xspace}
\newcommand{\cV}{\ensuremath{\mathcal V}\xspace}
\newcommand{\cW}{\ensuremath{\mathcal W}\xspace}
\newcommand{\cX}{\ensuremath{\mathcal X}\xspace}
\newcommand{\cY}{\ensuremath{\mathcal Y}\xspace}
\newcommand{\cZ}{\ensuremath{\mathcal Z}\xspace}

\newcommand{\bfQ}{\mathbf{Q}}

\newcommand{\bfU}{\mathbf{U}}
\newcommand{\bfV}{\mathbf{V}}

\newcommand{\bfa}{\mathbf{a}}
\newcommand{\bfb}{\mathbf{b}}

\newcommand{\bfm}{\mathbf{m}}

\newcommand{\bfu}{\mathbf{u}}
\newcommand{\bfv}{\mathbf{v}}

\newcommand{\bfx}{\mathbf{x}}
\newcommand{\bfy}{\mathbf{y}}

\newcommand{\rva}{\rv{a}}
\newcommand{\rvb}{\rv{b}}
\newcommand{\rvc}{\rv{c}}

\newcommand{\rvr}{\rv{r}}

\newcommand{\rvx}{\rv{x}}
\newcommand{\rvy}{\rv{y}}

\newcommand{\eps}{\ensuremath{\varepsilon}}

\newcommand{\poly}{\ensuremath{\operatorname{poly}}\xspace}

\newcommand{\negl}{\ensuremath{\operatorname{negl}}\xspace}
\newcommand{\Supp}{\ensuremath{\operatorname{Supp}}\xspace}

\newtheorem{theorem}{Theorem}[section]

\theoremstyle{plain}
\newtheorem{claim}[theorem]{Claim}

\newtheorem{lemma}[theorem]{Lemma}
\newtheorem{corollary}[theorem]{Corollary}

\theoremstyle{definition}
\newtheorem{definition}[theorem]{Definition}

\theoremstyle{definition}

\newenvironment{restatethm}[1]{

\medskip\noindent {\bf #1} (Restated.) \em}{\em \medskip}
\newcommand{\namedref}[2]{\hyperref[#2]{\text{#1}~\ref*{#2}}}

\newcommand{\torestate}[3]{%
\expandafter \def \csname BBRESTATE #2 \endcsname{#3}
\theoremstyle{plain}
\newtheorem{BBRESTATETHMNUM#2}[theorem]{#1}
\begin{BBRESTATETHMNUM#2}\label{#2}\csname BBRESTATE #2 \endcsname   \end{BBRESTATETHMNUM#2}
\newtheorem*{BBRESTATETHMNONNUM#2}{\namedref{#1}{#2}}
}

\newcommand{\restate}[1]{\begin{BBRESTATETHMNONNUM#1}[Restated] \csname BBRESTATE #1 \endcsname
\end{BBRESTATETHMNONNUM#1}}

\newcommand{\sectionref}[1]{\namedref{Section}{sec:#1}}
\newcommand{\appendixref}[1]{\namedref{Appendix}{app:#1}}
\newcommand{\theoremref}[1]{\namedref{Theorem}{thm:#1}}
\newcommand{\corollaryref}[1]{\namedref{Corollary}{cor:#1}}

\newcommand{\lemmaref}[1]{\namedref{Lemma}{lem:#1}}
\newcommand{\claimref}[1]{\namedref{Claim}{clm:#1}}

\newcommand{\figureref}[1]{\namedref{Figure}{fig:#1}}
\newcommand{\equationref}[1]{\namedref{Eq.}{eq:#1}}
\newcommand{\footnoteref}[1]{\namedref{Footnote}{foot:#1}}

\newcommand{\func}[1]{\ensuremath{{\mathcal F}_{\textsc{#1}}}\xspace}
\newcommand{\Fcom}{\func{com}}

\newcommand{\prot}{\ensuremath{\Pi}\xspace}

\newcommand{\numround}{\ensuremath{N}\xspace}
\newcommand{\spar}{\ensuremath{\kappa}\xspace}

\newcommand{\Eve}{\ensuremath{\mathsf{Eve}}\xspace}
\newcommand{\error}{\ensuremath{\varepsilon}\xspace}

\newcommand{\ancest}{\mathop{\preceq}}
\newcommand{\strictancest}{\mathop{\prec}}

\newcommand{\children}{\ensuremath{\mathsf{chldrn}}\xspace}
\newcommand{\Ancest}{\ensuremath{\mathsf{ancstrs}}\xspace}
\newcommand{\Aparent}{\ensuremath{\mathsf{Apred}}\xspace}
\newcommand{\Bparent}{\ensuremath{\mathsf{Bpred}}\xspace}
\newcommand{\Anode}{\ensuremath{\mathsf{Anodes}}\xspace}
\newcommand{\Bnode}{\ensuremath{\mathsf{Bnodes}}\xspace}
\newcommand{\Achild}{\ensuremath{\mathsf{Achildren}}\xspace}
\newcommand{\Bchild}{\ensuremath{\mathsf{Bchildren}}\xspace}
\newcommand{\simerr}{\ensuremath{{\nu_0}}\xspace}

\newcommand{\augtree}{\ensuremath{{\mathbb{T}}^+}\xspace}
\newcommand{\FX}{\ensuremath{F^\minwt_{X}}\xspace}
\newcommand{\FY}{\ensuremath{F^\minwt_{Y}}\xspace}
\newcommand{\FXsimple}{\ensuremath{F_{X}}\xspace}
\newcommand{\FYsimple}{\ensuremath{F_{Y}}\xspace}
\newcommand{\FXnomin}{\ensuremath{F^0_{X}}\xspace}
\newcommand{\FYnomin}{\ensuremath{F^0_{Y}}\xspace}
\newcommand{\highFXreal}{\ensuremath{\breve{F}_{X}}\xspace}
\newcommand{\highFYreal}{\ensuremath{\breve{F}_{Y}}\xspace}

\newcommand{\highFX}{\ensuremath{\widetilde{F_{X}}}\xspace}
\newcommand{\highFY}{\ensuremath{\widetilde{F_{Y}}}\xspace}
\newcommand{\highFXX}{\ensuremath{\widehat{F_{X}}}\xspace}
\newcommand{\highRXX}{\ensuremath{\widehat{R_{X}}}\xspace}
\newcommand{\highSXX}{\ensuremath{\widehat{S_{X}}}\xspace}
\newcommand{\xx}[1]{\ensuremath{\hat{x}_{#1}}\xspace}
\newcommand{\yy}[1]{\ensuremath{\hat{y}_{#1}}\xspace}
\newcommand{\VA}[2]{\ensuremath{{V_{A,{#1}}}(#2)}\xspace}
\newcommand{\VB}[2]{\ensuremath{{V_{B,{#1}}}(#2)}\xspace}
\newcommand{\VE}[1]{\ensuremath{{V_E}(#1)}\xspace}
\newcommand{\dVA}[2]{\ensuremath{{\bfV_{A,{#1}}}(#2)}\xspace}
\newcommand{\dVB}[2]{\ensuremath{{\bfV_{B,{#1}}}(#2)}\xspace}

\newcommand{\VVA}[3]{\ensuremath{{V^{#1(#2,#3)}_A}}\xspace}
\newcommand{\VVB}[3]{\ensuremath{{V^{#1(#2,#3)}_B}}\xspace}

\newcommand{\expNodes}[1]{\ensuremath{\mathcal{E}^{#1}_{\yy0,\yy1}}\xspace}
\newcommand{\game}[3]{\ensuremath{\mathsf{G}^{#3}({#1},{#2})}\xspace}
\newcommand{\exptcurious}[1]{\ensuremath{\wh{\mathsf{G}}(#1)}\xspace}
\newcommand{\pcurious}[2]{\ensuremath{\wh{\mathsf{P}}_{#2}[#1]}\xspace}
\newcommand{\exptcompare}[1]{\ensuremath{\mathsf{G}'(#1)}\xspace}
\newcommand{\pcompare}[2]{\ensuremath{\mathsf{P}'_{#2}[#1]}\xspace}
\newcommand{\safe}[1]{\ensuremath{\mathsf{safe}(#1)}\xspace}
\newcommand{\unsafe}[1]{\ensuremath{\overline{\safe{#1}}}\xspace}
\newcommand{\deltax}{\ensuremath{\delta'}\xspace}
\newcommand{\deltay}{\ensuremath{\delta''}\xspace}
\newcommand{\safeerr}{\ensuremath{\wh\eps_1}\xspace}
\newcommand{\indeperr}{\ensuremath{\wh\eps_0}\xspace}
\newcommand{\minwt}{\ensuremath{\theta}\xspace}

\newcommand{\missFX}{\ensuremath{\overline{\FX}}\xspace}
\newcommand{\missFY}{\ensuremath{\overline{\FY}}\xspace}
\newcommand{\missFYnomin}{\ensuremath{\overline{\FYnomin}}\xspace}

\newcommand{\missfxconst}{\ensuremath{c_0}\xspace}
\newcommand{\missfxpoly}{\ensuremath{\zeta_0}\xspace}
\newcommand{\orderpoly}{\ensuremath{\xi}\xspace}

\newcommand{\distadv}{\ensuremath{\rho}\xspace}

\def\tx{\tilde x}

\begin{document}

\title{Limits of Random Oracles in Secure Computation}

\author{
	Mohammad Mahmoody%
	    \thanks{Cornell,
	        \href{mailto:mohammad@cs.cornell.edu}{$\mathtt{mohammad@cs.cornell.edu}$}. Supported in part by NSF Award CCF-0746990, AFOSR Award FA9550-10-1-0093, and DARPA and AFRL under contract FA8750-11-2-0211. The views and conclusions contained in this document are those of the author and should not be interpreted as representing the official policies, either expressed or implied, of the Defense Advanced Research Projects Agency or the US government.
	    } \and
    Hemanta K.\ Maji%
        \thanks{Univ. of California, Los Angeles.
            Work partially done when at Univ. of Illinois, Urbana-Champaign.
            Partially supported by NSF grants CNS 07-47027 and CNS 07-16626.
            \href{mailto:hmaji@cs.ucla.edu}{$\mathtt{hmaji@cs.ucla.edu}$}.
        } \and
    Manoj Prabhakaran%
        \thanks{Univ. of Illinois, Urbana-Champaign.
		Supported by NSF grant CNS 07-47027.
            \href{mailto:mmp@cs.uiuc.edu}{$\mathtt{mmp@cs.uiuc.edu}$}.
        }
}

\maketitle

\begin{abstract}

The seminal result of Impagliazzo and Rudich (STOC 1989) gave a black-box separation between one-way functions and public-key encryption: informally, a public-key encryption scheme cannot be constructed using one-way functions as the sole source of computational hardness. In addition, this implied a black-box separation between one-way functions and protocols for certain Secure Function Evaluation (SFE) functionalities (in particular, Oblivious Transfer). Surprisingly, however, {\em since then there has been no further progress in separating one-way functions and SFE functionalities} (though several other black-box separation results were shown). In this work, we present the complete picture for deterministic 2-party SFE functionalities. We show that one-way functions are black-box separated from {\em all such SFE functionalities}, except the ones which have unconditionally secure protocols (and hence do not rely on any computational hardness), when secure computation against semi-honest adversaries is considered. In the case of security against active adversaries, a black-box one-way function is indeed useful for SFE, but we show that it is useful only as much as access to an ideal commitment functionality is useful.

Technically, our main result establishes the limitations of random oracles for secure computation. We show that a two-party deterministic functionality $f$ has a secure function evaluation protocol in the random oracle model that is (statistically) secure against semi-honest adversaries if and only if $f$ has a protocol {\em in the plain model} that is (perfectly) secure against semi-honest adversaries.  Further, in the setting of active adversaries, a deterministic SFE functionality $f$ has a (UC or standalone) statistically secure protocol in the random oracle model if and only if $f$ has a (UC or standalone) statistically secure protocol in the commitment-hybrid model.

Our proof is based on a ``frontier analysis'' of two-party protocols, combining it with (extensions of) the ``independence learners'' of Impagliazzo-Rudich/Barak-Mahmoody. We make essential use of a combinatorial property, originally discovered by Kushilevitz (FOCS'89), of functions that have semi-honest secure protocols in the plain model (and hence our analysis applies only to functions of polynomial-sized domains, for which such a combinatorial characterization is known).

\end{abstract}

\thispagestyle{empty}
\newpage

\newpage
\pagestyle{empty}
\tableofcontents

\newpage
\setcounter{page}{1}
\pagestyle{plain}

\pagestyle{plain}

\section{Introduction}

How useful is a random oracle in two-party secure function
evaluation (SFE)?  One obvious use of a random oracle is for implementing
commitment. We show that, remarkably, for 2-party SFE%
\footnote{We restrict our treatment to SFE functionalities with finite (or
at most polynomial-sized) domains. This is because, even without random oracles, a tight
characterization of realizable functionalities is known only with this restriction.}
{\em a random oracle by itself is only as useful as a commitment functionality}.

This result has important implications in understanding the ``complexity''
of secure function evaluation functionalities vis a vis computational
primitives like one-way functions. An important
goal in cryptography is to understand the qualitative complexity of various
cryptographic primitives.  In the seminal work of Impagliazzo and Rudich
\cite{ImpagliazzoRu89} a formal framework was established to qualitatively
separate cryptographic primitives like symmetric-key encryption and
public-key encryption from each other.  Understanding that such a separation
exists has been hugely influential in theoretical and practical
cryptographic research in the subsequent decades: to optimize on both
security and efficiency dimensions, a cryptographic construction would be
based on symmetric-key primitives when possible, and otherwise is shown to
``require'' public-key primitives.

Beyond encryption, the result in~\cite{ImpagliazzoRu89} already implies the
separation of certain SFE functionalities (in particular, Oblivious
Transfer) from one-way functions.
Surprisingly, however, {\em since then
there has been no further progress on separating SFE functionalities and
one-way functions} (though several other black-box separation results have
emerged \cite{Simon98,GertnerKaMaReVi00,GertnerReMa01,BonehPaRaVaWa08,KatzScYe11,MatsudaMa11}). In this work, we present the complete picture for deterministic
2-party SFE functionalities: we show that in the case of security against
semi-honest adversaries, {\em all of them} are black-box separated from
one-way functions, except the ones which are trivial (which have
unconditionally perfectly secure protocols).  In the case of active adversaries, a
black-box one-way function is indeed useful for SFE, but we show that it is
useful only as much as access to a commitment functionality is useful (and
explicitly characterize the functions for evaluating which it is useful).

Our work could be viewed as a confluence of two largely disjoint lines of
work --- one on black-box one-way functions, and one on the structure of
secure function evaluation functionalities. The former line essentially
started with~\cite{ImpagliazzoRu89}. The latter can be traced back to
concurrent work~\cite{ChorKu89,Beaver89,Kushilevitz89} which combinatorially
characterized which finite (2-party) functionalities have (perfectly)
semi-honest secure protocols. This property, called decomposability
\cite{Kushilevitz89} will be important for us. Several later works obtained
such combinatorial characterizations of SFE functionalities in different
contexts (e.g.,~\cite{Kilian91,BeimelMaMi99,KilianKuMiOs00,MajiPrRo10uc,KraschewskiMu11,Kreitz11}).

An important ingredient of our proof is the ``frontier analysis'' approach
from~\cite{MajiPrRo09,MajiOuPrRo11}. As we shall see, frontier analysis
provides a powerful means to explicitly work with otherwise-subtle
conditional probabilities, especially as arising in 2-party protocols.  In
essence, it is simply a means to explicitly keep track of the order in which
various events occur in a protocol (or more generally, in a sequence of
random variables).  But as we shall see, having an explicit mental picture
lets us define frontiers and reason about their properties that are {\it a
priori} not obvious (see \figureref{fronts} in \sectionref{detailed-frontierorder},
for instance).  The proof in~\cite{CleveIm93} could in fact be viewed as an
instance of frontier analysis (and is one of the earliest ones that the
authors are aware of).  An instance of such an approach in a
non-cryptographic setting is present in the recent work of Barak et
al.~\cite{BarakBrChRa10}, who consider frontiers in a protocol where
significant amounts of ``new and relevant'' information is revealed, and use
this to reduce the total amount of communication.

\subsection{Our Results}

We summarize our main results below. Our main result is the following.
\begin{theorem}
\label{thm:main}
A deterministic two-party function $f$, with a polynomially large domain,
has a semi-honest secure protocol against computationally unbounded
adversaries in the random oracle model if and only if $f$ has a perfectly
semi-honest secure protocol in the plain model.
\end{theorem}
We remark that such $f$ can be explicitly characterized as decomposable
functions  as defined in \cite{Kushilevitz89} (if $f$ is symmetric),
or more generally, as those for which 
the symmetric function $f'$ obtained as the ``common information'' part
of $f$%
\footnote{\label{foot:cifunction}For a deterministic two-party function
$f:\cX\times\cY\rightarrow \cZ_A\times\cZ_B$, the common information
function $f'$ is defined as follows (see for e.g., \cite{MajiOuPrRo11}):
consider the bipartite graph consisting of nodes of the form
$(x,a)\in\cX\times\cZ_A$ and $(y,b)\in\cY\times\cZ_B$, with an edge between
$(x,a)$ and $(y,b)$ iff $f(x,y)=(a,b)$. Then $f'$ maps $(x,y)$ to the
connected component containing $(x,a)$ and $(y,b)$ where $f(x,y)=(a,b)$.
Intuitively, $f'(x,y)$ reveals only that part of the information about
$(x,y)$ that $f$ reveals to ``commonly'' to both Alice and Bob (and so they
know that it is known to the other party as well).}
is decomposable and $f$ and $f'$ are ``isomorphic.''%
\footnote{\label{foot:isomorphism} $f_0$ and $f_1$ are isomorphic
if there is a UC and semi-honest secure protocol for evaluating either function
which uses a single instance of the other function with no other
communication. In particular, if either function has a semi-honest secure
protocol in the random oracle model (respectively, plain model), then the
other one has such a protocol too.}

In this theorem, as is conventional in much of the work on the combinatorial
structure of SFE functionalities, we restrict ourselves to functions whose
domain size is polynomial in the security parameter.  A full combinatorial
characterization of semi-honest securely realizable functions (even in the
plain model) is known only with this restriction.  In particular, there are
{undecomposable functions}, with super polynomial domain size, which
are semi-honest securely realizable.  Henceforth, unless mentioned
otherwise, whenever we consider a function we shall assume that its domain
size is polynomial in the security parameter.

The above result --- that random oracles are useless for 2-party SFE ---
does {\em not} hold in the case of security against active adversaries.  In
particular, note that the commitment functionality \Fcom, can be constructed
UC-securely in a black-box manner from random oracles, and so, all the
functions which can be UC-securely computed in the \Fcom hybrid can also be
UC-securely computed in the random oracle model. But we shall show that this
is all that a random oracle is useful for in 2-party SFE. This follows from
\theoremref{main} and a compiler from \cite{MajiPrRo09} that turns
semi-honest secure protocols to UC-secure protocols in the \Fcom-hybrid
model (see proof in \sectionref{detailed-RO-Fcom}).
\begin{theorem}
\label{thm:RO-Fcom}
A deterministic two-party function $f$, with a polynomially large domain, has a
statistically UC-secure (and equivalently, a statistically
standalone-secure) protocol in the random oracle model if and only if $f$
has a statistically UC-secure (and equivalently, a statistically
standalone-secure) protocol in the \Fcom hybrid.
\end{theorem}
We remark that such $f$ can be characterized as those for which, on
removing all ``redundant inputs''%
\footnote{\label{foot:redundant}Alice's input $x$ to $f$ is said to be
redundant (for security against active adversaries) if there is an input $x'
\neq x$ that dominates $x$: i.e., Alice can substitute $x'$ for $x$ without
Bob noticing while still being able to calculate her correct output.}
one at a time, we obtain a function of the kind in \theoremref{main}.

\paragraph{Blackbox Separations.} \emph{Black-box constructions} form a
general framework of obtaining a (more complex) cryptographic primitive
$\cQ$ (\eg pseudorandom generators) from another (perhaps simpler)
cryptographic primitive $\cP$ (\eg one-way functions) while  $\cP$ is used
in the implementation of $\cQ$ only as a black-box and the security of $\cQ$
is proved based on the security of $\cP$ also through a black-box argument.
Apart from being the most common kind of reductions used in cryptographic
constructions (with ``provable security''), black-box
reductions provides us with a framework to understand ``complexity'' of cryptographic
primitives.
This line of research was initiated in the seminal work of Impagliazzo and Rudich
\cite{ImpagliazzoRu89} who showed that public-key cryptography is strictly
more complex than symmetric-key cryptography (say, one-way functions)
under this framework.

\theoremref{main} is proven in the computationally unbounded
setting, and the honest-but-curious adversaries implicit in our proofs use
super-polynomial computational power (even if the honest parties were
polynomial time).  However, similar to the results in \cite{ImpagliazzoRu89}, 
this can be translated to a statement about black-box separation of
semi-honest SFE protocols (for functions without perfectly secure protocols) from
one-way functions, in a probabilistic polynomial time (PPT) setting.
Intuitively, this is so because a random oracle is a strong one-way
function (but for the drawback that it does not have a small code to implement it);
so, if one-way function is the sole computational primitive needed
for a construction, and it is used in a black-box manner, then it should be 
possible to base the construction on a random oracle instead. 
Hence, ruling out secure protocols in the random oracle model in the
computationally unbounded setting would rule out protocols in the PPT
setting that base
their security on one-way functions in a black-box manner.
The technicalities depend on the formal definition of black-box reduction.
We follow the definitions in \cite{ReingoldTrVa04}, with slight technical
modifications, to state our results. 
A formal statement appears in \theoremref{fully-BBsep-OWF}. We summarize
this result informally below.

\begin{theorem} (Informal.)
\label{thm:BBsep}
For a deterministic two-party function $f$, with a polynomially large domain,
there is a fully black-box reduction of semi-honest secure function
evaluation of $f$ to one-way functions
if and only if $f$ has a perfectly semi-honest secure protocol in the plain model.
\end{theorem}

Though we state the result for one-way functions, in fact, any collection of
primitives that can be constructed from a random oracle (or ideal cipher) 
or a random permutation oracle%
\footnote{We point out that \theoremref{main}
extends to a random permutation oracle, as argued in
\cite{ImpagliazzoRu89}: otherwise, we can construct an
efficient distinguisher between a length preserving random oracle and a
length preserving random permutation oracle for ``long'' inputs,
and this can be shown to be impossible (as it is improbable
to find collisions in a random oracle).} 
in a black-box manner -- one-way functions, one-way permutations,
collision resistant hash functions, block-ciphers (including exponentially
hard versions of  these primitives) -- is useless for 2-party SFE, if the
primitives are used in a fully black-box manner.

As in the case of \theoremref{main}, the above statement can be extended to
the case of security against active adversaries.
\begin{theorem} (Informal.)
\label{thm:BBsep-Fcom}
For a deterministic two-party function $f$, with a polynomially large domain,
there is a fully black-box reduction of UC (or stand-alone) secure function
evaluation of $f$ to one-way functions
if and only if $f$ has a statistically
UC (or stand-alone) secure protocol in the $\Fcom$-hybrid model.
\end{theorem}
Note that, though commitment is already known to be black-box equivalent to
one-way functions, {\em statistical} (standalone) security in the
\Fcom-hybrid is, on the face of it, more restrictive than standalone
security in the PPT setting using fully black-box commitments. Further, the
theorem holds for not only one-way functions, but also the other computational primitives
mentioned above.

\subsection{Related Work}
Impagliazzo and Rudich~\cite{ImpagliazzoRu89} showed that
random oracles are not useful against a computationally unbounded adversary
for the task of secure key agreement. This analysis was recently simplified
and sharpened in~\cite{BarakMa09}. These results and techniques are one
starting point for our result.

Following~\cite{ImpagliazzoRu89} many other black-box separation results
followed
(\eg~\cite{Simon98,GertnerReMa01,BonehPaRaVaWa08,KatzScYe11,MatsudaMa11}).  In
particular, Gertner et al.~\cite{GertnerKaMaReVi00} insightfully asked the
question of comparing oblivious-transfer (OT) and key agreement (KA) and
showed that OT is strictly more complex (in the sense of
\cite{ImpagliazzoRu89}).   Another trend of results has been to prove
lower-bounds on the efficiency of the implementation reduction in black-box
constructions
(\eg~\cite{KimSiTe99,GennaroGeKaTr05,LinTrWe05,HaitnerHoReSe07,
BarakMa07,BarakMa09,HaitnerHoReSe07}). A complementary approach has been to
find black-box reductions when they exist
(\eg~\cite{ImpagliazzoLu89,Ostrovsky91,OstrovskyWi93,
Haitner08,HaitnerNgOnReVa09}).  Also, results in the black-box separation
framework of~\cite{ImpagliazzoRu89,ReingoldTrVa04} have immediate
consequences for computational complexity theory. Indeed, separations in
this framework can be interpreted as new worlds in Impagliazzo's
universe~\cite{Impagliazzo95}.

Frontier analysis is possibly implicit in previous works on proving
impossibility or lower bounds for protocols. For instance, the analysis in
\cite{CleveIm93} very well fits our notion of what frontier analysis is.
The analysis of protocols in~\cite{ChorKu89,Beaver89,Kushilevitz89} also
have some elements of a frontier analysis, but of a rudimentary form which
was sufficient for analysis of perfect security. In~\cite{MajiPrRo09}
frontier analysis was explicitly introduced and used to prove several
protocol impossibility results and characterizations.~\cite{KunzlerMuRa09}
also presented similar results and used somewhat similar techniques (but
relied on analyzing the protocol by rounds, instead of frontiers, and
suffered limitations on the round complexity of the protocols for which the
impossibility could be shown).
We also rely on
results from~\cite{MajiOuPrRo11} to extend the result to general SFE
functionalities as opposed to symmetric SFE functionalities.

\subsection{Technical Overview}
\label{sec:overview}

We rely on a careful combination of the techniques in the black-box
separation literature (in
particular~\cite{ImpagliazzoRu89,BarakMa09,DachmanLiMaMa11}) and new
frontier analysis techniques.  Below we briefly explain the overall approach
and point out some of the highlights.

A clear starting point of our investigation is the ``independence learner''
of~\cite{ImpagliazzoRu89,BarakMa09} which shows, in a protocol between Alice
and Bob involving private queries to a random oracle, how to make several
(but polynomially many) additional queries to the random oracle and make
Alice's and Bob's views (conditioned on their inputs) independent of each
other.  However,from this independence property it is not immediate  to
conclude that random oracles are useless in SFE protocols. One conjecture
(which we are not able to prove) would be that the effect of the random
oracle can be ``securely simulated'' in the plain model, and then any
protocol in the random oracle model can be compiled into a plain-model
protocol that is as secure as the original one. This would avoid the need to
rely on combinatorial characterizations of SFE functionalities, and indeed
show that random oracles are useless for virtually any protocol (up to
small, but non-negligible errors inherent in the independence learner).
However, in this work we do not obtain such a compiler. In particular, {\em
we do not rule out the possibility that in fact random oracles could have
unsimulatable effects, and may aid in secure computation of randomized
functionalities, or functionalities with super-polynomial input domains}.%
\footnote{An earlier version of this work (presented
in~\cite{Maji11Thesis}), pursued this approach, and appeared to succeed.
However, on closer scrutiny a major gap was found in the case when both
Alice and Bob can have private inputs, which we have not been able to
repair. Indeed, based on our current understanding, {\em we do not
conjecture} that the random oracle can be compiled away from all protocols
involving private inputs to both the parties.}

This leads us to the techniques used in showing that a symmetric SFE
functionality $f$ is semi-honest securely realizable if and only if it is
decomposable. The strongest version of this result was proven using
frontier analysis in~\cite{MajiPrRo09}. However, as we shall see, we need
a significantly more sophisticated argument here.

\subsubsection{Frontier Analysis Meets Random Oracles}
\label{sec:overview-frontiers}

First we describe why
na\"{\i}ve attempts at generalizing the argument used to characterize
functions with SFE protocols in the plain model
\cite{MajiPrRo09} fail in the random oracle setting.

The plain model result crucially relies on the following ``locality''
property.  When Alice sends the next message in a plain model protocol, she
can reveal (i.e., add to the transcript) new information only about her own
input but {\em not} about Bob's inputs.  So, during the execution of the
protocol, Alice and Bob would alternately reveal information about their
inputs $x$ and $y$ respectively.  Suppose we define two frontiers:
\FXsimple, where (significant, additional) information about $x$ is first
revealed, and \FYsimple where (significant, additional) information about
$y$ is first revealed in the transcript.  By the locality property,
\FXsimple consists of nodes where Alice has just sent out a message, and
\FYsimple consists of nodes where Bob has just sent out a message.
Firstly, for the sake of correctness, information about $x$ and $y$ need to
be revealed by the end of the protocol, and hence, \FXsimple and \FYsimple
are almost ``full'' frontiers (i.e., there is only a small probability that
an execution finishes without passing through both frontiers).%
\footnote{As we shall see, {\em for undecomposable functions}, this must hold even if there are inputs for one
party (say Bob) for which the function becomes constant. That is, \FYsimple needs to be crossed even for executions
in which Bob's input is a value $y$ for which the function $f(\cdot,y)$ is
constant. This is because, by undecomposability, for certain values of Alice's input
$x$, and another input $y'$ for Bob, $f(x,y)=f(x,y')$ where $f(\cdot,y')$ is not constant, and then by
security, the execution with input $(x,y)$ has to be close to the execution
with input $(x,y')$. In the latter, information about $y$ needs to be revealed.}
To draw a contradiction we rely on the property that, {\em for an undecomposable
function, it will be insecure for either party to reveal information about
their input first}. In terms of the frontiers, this says that it will be insecure if,
 a (significantly probable) portion of \FXsimple appears
above \FYsimple, or if a (significantly probable) portion of
\FYsimple appears above \FXsimple. Combined with the fact that both frontiers
are almost full, this rules out secure protocols for undecomposable
functions.

\paragraph{Handling the Random Oracle.}
In the presence of a random oracle, we lose the locality property (that
Alice's message is independent of Bob's input, conditioned on the
transcript).  It becomes possible that a correlation is established between
Alice's and Bob's views via the common random oracle, even conditioned on
the transcript.  Indeed, given a random oracle, a secure protocol for even
OT is possible unless the curious parties query the oracle on points other
than what is prescribed by the protocol.
Hence, to be meaningful in the presence of an oracle, we must define the
information revealed by a transcript as what a curious eavesdropper making
additional (polynomially bounded) queries to the oracle, can learn.  This is
where the independence learner ``Eve'' of~\cite{ImpagliazzoRu89,BarakMa09} is
relevant. Intuitively, Eve attempts to learn as much as possible
(staying within a budget of polynomially many oracle queries), by making all
``important'' queries to the oracle after each message in the protocol.
By including the information obtained by Eve into the transcript itself, we
can ensure that the frontiers do correspond to points where certain
information is revealed, conditioned on the information obtained by Eve.
Being a
semi-honest setting, it is not relevant when these queries are performed;
but for our frontier analysis, it will be important to consider the curious
eavesdropper as running concurrently with the protocol, querying the oracle
as many times as it wants, after each message in the protocol.

\paragraph{Main Challenge.} Once the transcript is augmented with Eve's view, one could hope that
the previous analysis from~\cite{MajiPrRo09} can be applied. Indeed, in this
augmented protocol, the locality property  is restored. However, now we have
introduced new messages in the transcript (namely Eve's interaction with the
random oracle), and these messages could be correlated with {\em both} Alice's and
Bob's inputs! This is the core issue that we need to tackle.%
\footnote{This is the issue that was not correctly handled in a previous
attempt by the authors (in~\cite{Maji11Thesis}), in trying to compile
away the random oracle. The current frontier analysis based approach avoids
subtle probabilistic reasoning which is invariably fraught with dangers of
false intuition.}

\paragraph{Our Solution.}
Now we give an intuitive (but imprecise) description of our proof.
As above, we shall define
the frontiers
\FXsimple and
\FYsimple
where information about $x$ and (respectively) about $y$
is first revealed in the (augmented) transcript.
Now, information about $x$ or $y$ could be revealed when Alice sends out a
message, Bob sends out a message, or Eve obtains its answers from the
oracle.  We will be able to rule out information about $x$ being revealed by
a message from Bob, or information about $y$ being revealed by a message
from Alice (this corresponds to \claimref{independence}), but this leaves open the
possibility that an answer for an Eve query to the oracle reveals information about $x$ and
$y$ simultaneously.

To address this, we pursue the following intuition: suppose no information
about $y$ has been revealed so far, and Alice sends out a message; suppose
some information about $x$ is revealed not immediately by this message, but
after Bob (and Eve) carry out oracle queries and respond to Alice's message
(but before Alice responds again).  (Our concern is that this information
could depend on $x$ and $y$ simultaneously.) Then we demonstrate a {\em
curious Bob} strategy that can learn the same information about $x$,
irrespective of his actual input $y$.  The intuition behind this strategy is
the following: consider the point immediately after Alice sent out her
message.  Bob samples for himself a view conditioned on an alternate input
$y'$ such that an actual execution with input $(x,y')$ reveals information
about $x$ that should not be revealed when Bob's input is $y$. Bob can
simulate the execution with input $y'$ for himself, starting from this point
until the next message from Alice, without interacting with Alice; however,
the oracle Bob has access to is conditioned on the actual pair of inputs
$(x,y)$, and not $(x,y')$.  Clearly, it will be pointless to use this oracle
directly to simulate the execution with input $(x,y')$. A crucial
observation at this point (this corresponds to \claimref{safe}) is that, it is highly unlikely for an oracle query
that is not in Eve's view to be present in both Alice's view and Bob's view
(or the sampled view for Bob).  This lets Bob simulate an oracle conditioned
on $(x,y')$ as follows: if an oracle query is already answered in the
sampled view for Bob (with input $y'$), use it (it is likely not to have
been asked by Alice); else, if an oracle query is present in the original
view for Bob (but not present in the sampled view, and neither in Eve's
view), then ``undo'' the effect of the query in Bob's view by sampling a new
answer for it (again, it is unlikely to have been asked by Alice); if not,
use the actual oracle (thus ensuring that any queries already present in
Alice's view are consistently answered).  This allows curious Bob to
seamlessly replace the actual oracle with an oracle consistent with inputs
$(x,y')$, even though he does not know $x$ or Alice's view of the oracle.
What facilitates this, in addition
to the fact that Eve captures all intersection queries, is the special
``modular'' nature of the random oracle.

This essentially means that when information about $x$ is revealed,
information about $y$ must have been revealed already by the time the last
message was sent by Alice (even if the information about $x$ is revealed
only during subsequent queries to the oracle by Bob or Eve).  Further, as
mentioned above, since Alice could not have revealed information about $y$,
this information about $y$ must have been revealed strictly before the last
message from Alice, and in particular, strictly before the information about
$x$ was revealed. This is captured in \claimref{frontierorder-inmain} which
implies that (in terms of the simplified presentation above) \FXsimple can
be reached only strictly after passing through a node in \FYsimple.

\paragraph{Some Technical Issues.} Formalizing the above intuitive description
presents several challenges. The most important aspect is the appropriate
definition of the frontier, and the statement regarding the ordering of the
frontiers. For the above curious Bob to have an advantage, the information
revealed about $x$ should have been after the last message from Alice. For
each node $u$ we define $\Aparent(u)$ to correspond to the last message from
Alice; however for a  node $u$ which itself corresponds to a message from
Alice (where the argument relies on the locality property and not the above
curious Bob strategy) $\Aparent(u)$ is defined as its parent node. Another
important issue is that, above we argued in terms of ``the probability of
reaching a segment.'' However, this probability depends on the inputs.  (The
set of nodes in the frontier does not change; only the distribution over
them changes.) Whether these probabilities are similar or different depends
on whether the inputs have already been distinguished or not.  Note that we
use properties of these distributions to reason about the ordering of the
frontiers, and these distributions themselves depend on the ordering of the
frontiers! Much of our technical difficulties arise from circumnavigating
potential circularities.

\subsubsection{Using the Independence Learner}
\label{sec:overview-indeplearner}

As mentioned above, a crucial tool for analyzing protocols using a random oracle is to show that
by making polynomially many queries to the oracle, an eavesdropper Eve can get
sufficient information such that conditioned on this, Alice and Bob's views
in the protocol are almost always close to being independent (up to an inverse polynomially small
error). This is a delicate argument implicitly proved in~\cite{BarakMa09} building on
ideas from~\cite{ImpagliazzoRu89}, and was first explicitly described  in~\cite{DachmanLiMaMa11}.
The view of such an Eve is part of the augmented transcript, with respect to
which the frontiers are defined.

A subtle issue to address when extending this Eve to our case is that Alice
and Bob receive inputs from an arbitrary environment and Eve does not see
the inputs. In particular, Alice and Bob could receive correlated inputs,
and we cannot claim that their views, conditioned on Eve's view, are (almost
always, close to being) independent.  However, we can create an Eve which is
oblivious to the actual inputs, but for {\em every} input pair $(x,y)$ of
inputs,  when the protocol is executed with these  inputs, Alice's and Bob's
views conditioned on Eve's view are (almost always, close to being)
independent.  
For this, we take Eve to be as defined in \cite{BarakMa09} (presented in
\lemmaref{Eve}), but applied to an inputless protocol obtained by considering
our original protocol but with inputs 
$(x,y)$ that are chosen initially at random (say as part of the randomness of the
two parties). Initially this Eve considers the actual input to be of
significant probability (since the inputs come from a polynomially large
domain). In analyzing this Eve, we rely on an argument that with
significant probability, at any round of the protocol, this Eve will
consider the actual input to be a likely input (\lemmaref{inverse}).

In our analysis sketched above, there are two guarantees from this Eve that we rely on,
captured in \claimref{independence} and \claimref{safe}, as described below.

\paragraph{1) Alice's Message Independent of Bob's Input.} Firstly, recall
that the purpose of introducing Eve's view into the transcript was to
restore the ``locality property'' -- i.e., Alice's messages,
conditioned on Eve's view, are independent of Bob's view. More precisely,
we will need the guarantee that at a point where Alice is about to send
a message, if two inputs of Bob, $y$ and $y'$ are both somewhat likely,
then Alice's message is almost independent of which of these two inputs Bob has.
This is stated in \claimref{independence}, and follows from
\lemmaref{IL-Input} proven in \appendixref{EvePlus}.
Note that we need this to hold (and this holds) only at points
where both of Bob's inputs $y$ and $y'$ are
somewhat likely. (In using this claim, the points considered will be above the
frontier \FYsimple so that all inputs for Bob are significantly probable.)

\paragraph{2) Collisions of Private Queries Unlikely.} The second place
where we rely on Eve's properties is in arguing that the curious Bob
strategy outlined above works: i.e., that when curious Bob samples a view
for himself after Alice sends a message, it is unlikely that there will be
an oracle query in either his actual view or in the freshly sampled view
that occurs in Alice's actual view, but is not present in Eve's view.
This is stated in \claimref{safe} and follows from \lemmaref{NoCol}
proven in \appendixref{EvePlus}.
We need this to occur only when the 
``fake'' input $y'$ used for the sampled view is somewhat likely. (Again,
the claim will be applied only to points above the frontier \FYsimple, and
all inputs are somewhat likely there.) We remark that,
just for the actual views, similar statements were already explicitly proven
in \cite{ImpagliazzoRu89,BarakMa09}, bounding the probability of an
``intersection query'' that is not present in Eve's view. The additional
twist in our case is that we need to also consider the view sampled for
a ``fake'' input; further, Bob's views we consider are not at the point Eve finishes
a round of oracle queries, but after a subsequent message from Alice.

It is important to note that Bob's views considered here consist of the
oracle queries he made only up to the point he sent his previous message to
Alice (even though the views include the last message from Alice).
\lemmaref{NoCol} would \emph{not} be true, if instead we consider Bob's
views including oracle queries he makes after receiving Alice's last
message.  The reason is that the last message sent from Alice can simply
tell Bob that Alice has asked a random new query $q$ and Bob might make the
same query immediately afterwards. This way, the information that was
gathered by Eve till the end of the previous round (before Alice sent her
message) is incapable of catching this intersection query.

\section{Preliminaries} \label{sec:Prelim}
 In this section we introduce some basic notation, conventions and definitions.
 (Further conventions needed %
shall be introduced in their respective sections).

\subsection{Secure Evaluation of 2-Party Functions}

\paragraph{2-Party Functions.}
A (deterministic) {\em 2-party function}
$f:\cX\times\cY\mapsto\cZ_A\times\cZ_B$ maps a pairs of inputs $(x,y)$ 
(associated with Alice and Bob respectively) to a pair of outputs $(a,b)$ (for the
two parties, respectively). For most part in our proofs, we shall be dealing
with {\em symmetric} 2-party functions  which
produce two identical outputs (or equivalently, a single output given to
both parties).

For symmetric functions, an {\em Alice-cut} is a partition $(X,\bar X)$ of the input space $\cX$ such
that for any $x\in X$, $\bar x\in \bar X$ and $y\in\cY$ $f(x,y)\neq f(\bar
x,y)$.
The {\em functions associated with an Alice-cut} $(X,\bar X)$ are the two restrictions of
$f$, restricted to domain $X\times\cY$ and to domain $\bar X\times\cY$.
A Bob-cut and functions associated with it are defined similarly.

Now, we define {\em decomposable} functions $f$  in the following recursive manner~\cite{Kushilevitz89, Beaver89}:
\begin{compactenum}
\item A constant function is decomposable.
\item If $f$ has an Alice-cut or a Bob-cut and the two functions associated with that cut
are both decomposable then $f$ is decomposable.
\end{compactenum}
A function is {\em undecomposable} if it is {\em not} decomposable.
Moreover, it is said to be {\em undecomposable at the top-most level}, if
$f:\cX\times\cY\mapsto\cZ$ does not have an Alice-cut or Bob-cut (refer
\appendixref{examples} for some examples).

\paragraph{Secure Function Evaluation.}
A Secure Function Evaluation (SFE) functionality is associated with a
2-party function $f$: the ideal SFE functionality accepts $x$ from Alice,
$y$ from Bob, computes $f(x,y)=(a,b)$ and gives $a$ to Alice and $b$ to Bob.
We shall refer to the SFE functionality and the two-paty function associated
with it, interchangeably.
For most part, we shall consider protocols for SFE
functionalities that are secure against semi-honest adversaries.  Our final
theorems consider the two standard notions of security against active
adversaries as well, namely, standalone security and Universally Composable
(UC) security.  Mostly we work with statistical security, which places no
computational limitations on the parties or environment; but we do state
consequences for our results for security in the computational setting as
well. We omit a detailed description of the standard security definitions.
As it turns out, in our results, there would be no distinction between UC
security and standalone security. (Readers unfamiliar with the details of
the definitions may ignore the few places in our proofs where we discuss
the two notions separately, to establish their similarity.)

\paragraph{Security Definitions.} Security of protocols is defined under the standard simulation paradigm. We
consider semi-honest security in which the adversary and the simulator are
semi-honest (a.k.a.~passive or honest-but-curious), and also active-security. In the latter case
security can be considered in the standalone setting or the universally
composable setting. The statistical difference between the views of the
environment in the real and ideal executions, maximized for each simulator
over all environments, and then minimized over all simulators, will be called the
``security error'' of a protocol.

We can in fact work with a (weaker) game based definition of semi-honest
security which only requires that if $f(x,y)=f(x,y')$ Alice's views in the
two executions with inputs $(x,y)$ and $(x,y')$ should be (statistically)
indistinguishable from each other; similarly Bob's views for executions with
inputs $(x,y)$ and $(x,y)$ should be indistinguishable, if $f(x,y)=f(x',y)$.
This definition is identical to the simulation based definition in the
computationally unbounded setting; but when considering the PPT setting (for
black-box separation results), the weaker security definition makes our
results stronger, and more amenable to being framed in terms of the
definitions in \cite{ReingoldTrVa04}.

\subsection{Random Oracles}
An oracle $\cO$ is specified by a function (from queries to answers) chosen
according to a specified distribution. This choice is made before answering
any query, however for the sake of analysis of the protocol we can choose the randomness of the oracle \emph{along the way} as the parties interact (this is also known as the {\em lazy evaluation} of the oracle).
In this paper, we shall use $\cO$ which are random oracles, i.e. every query is independently mapped to an image chosen uniformly at random.

\paragraph{Security Parameter of $\cO$.} 
We shall associate a security parameter $\spar$ with the queries to  the oracle, and will invariably require that the length of the queries and
their answers is polynomial in $\spar$ (\eg $\cO$ for the security parameter $\spar$ could be a random function from $\bits^\spar$ to $\bits^\spar$). For simplicity, any protocol using the
oracle would make all queries with the same security parameter as the
protocol's own security parameter.

\paragraph{Query Operator.} For any view $V$ of some
oracle algorithm interacting with $\cO$, 
we denote the set of oracle queries made by the algorithm
according to the view $V$ by $\Q(V)$.

\subsection{Frontiers}
\label{sec:frontiers-prelim}

Consider a (possibly infinite) sequence of correlated random variables $(\bfm_1,\bfm_2,\dots)$.
We consider a natural representation of such a sequence as a rooted tree, with each level corresponding to a random
variable $\bfm_i$ and each node $v$ at depth $t$ in the tree is uniquely identified with
an assignment of values $(m_1, m_2, \dotsc, m_t)$ to
$(\bfm_1,\bfm_2,\dotsc,\bfm_t)$, such that $(m_1, m_2, \dotsc, m_{t-1})$ is
equal to the values identified with its parent node.
Then we can identify the sequence of values of these random variables with a
unique path in this tree, starting at the root.

We can identify a set of nodes $S$ in this tree with the event that the
path corresponding to the values taken by the random variables intersects
$S$. A {\em frontier} on this tree corresponds to a set $F$ of nodes
which is ``prefix-free'' (i.e., no two nodes in $F$ are on the
same path starting at the root). We often define a frontier using a
predicate, as the set of nodes which satisfy the predicate but do not have
an ancestor which satisfies the predicate (i.e., the predicate is satisfied
for the ``first time''). Note that the frontier event is deterministic
given a node in the tree (though the event could be in terms of the
probability of other events at that node).

The tree naturally defines an ``ancestor'' partial order of the
nodes in the tree: we say $u \ancest v$ if $u$ occurs somewhere on the path
from the root of the tree to $v$ ($u$ could be identical to $v$). If
$u\ancest v$, but $u\not=v$, then we write $u\strictancest v$.

Invariably, we consider this tree only with sequence of random variables
corresponding to the messages exchanged in a protocol (but possibly augmented
by additional messages added for analysis). Though not necessary, it will be
convenient to consider the underlying process as consisting of picking a
uniformly random input and then executing the protocol.
However, clearly, the tree and frontiers can be used to represent any
sequence of random variables. 

As a simple illustration of the routine arguments we carry out over such a
tree, we state and prove a simple lemma (which gets used later in the
paper).  In Lemma 6.4 of~\cite{ImpagliazzoRu89} it was shown how to obtain
an upper-bound on the conditional probability of an unlikely event under a
sequence of leaking information. The following lemma can be thought of as a
``dual'' statement showing that if the event is noticeable, when it actually
happens, then it remains noticeable conditioned on a sequence of leakages.
More formally we prove the following.

\begin{lemma} \label{lem:inverse}
Consider a sequence of correlated random variables $(\bfm_1,\bfm_2,\dots)$
For any event $X$ jointly distributed with these variables, let $S$ be the
event that
there exists $t$ such
that $\pr[X \mid (m_1,m_2,\dots,m_t)] < \theta$. Then it holds that $\pr[S \mid X] <
\theta/\pr[X]$.
\end{lemma}

\begin{proof}
Consider the tree representing the sequence of random variables
$(\bfm_1,\bfm_2,\dots)$. The event $S$ corresponds to a subset of nodes in
this tree: $S= \set{ v \mid \pr[X\mid v]<\theta }$.
Define $U$ to be the frontier of nodes in $S$ that do not have a strict
ancestor in $S$; namely, $U= \set{ v \mid v \in S \text{ and for all $u$
s.t. } u\strictancest v, u\not\in S }$. Note that $\pr[S \mid X]=\pr[U \mid
X]$. Further,
$$\pr[U \mid X] = \sum_{u\in U} \pr[u \mid X] = \sum_{u\in U} \pr[X \mid u] \pr[u] / \pr[X] < \theta \pr[ U]/\pr[X] \le \theta/\pr[X].\qedhere$$
\end{proof}

A corollary to \lemmaref{inverse} is that in a protocol execution,
the actual inputs of Alice
and Bob will not become ``unlikely'' conditioned on the transcript, except with
small probability.

\section{Transcript Tree and Other Notation}
\label{sec:frontiers-def}

In this section, first we define the tree notation that is used throughout
our analysis. We shall also define the frontiers on this tree that are
central to our analysis.

\paragraph{Augmented Protocol Execution.}
We shall consider two-party protocols \prot where Alice and Bob interact to evaluate a (symmetric)
function $f:\cX\times\cY\rightarrow\cZ$ on their respective local inputs $x\in\cX$ and $y\in\cY$.
We shall assume that $|\cX|$ and $|\cY$ are both polynomial in the security parameter.
Alice and Bob have access to a random oracle \cO.
We ``augment'' the protocol \prot with a 
``public query strategy'' \Eve,
which can
see the publicly generated transcript and can also query the random oracle.
For simplicity, we consider \Eve to be deterministic (as will be the case in
our instantiation of \Eve).
Later, we will instantiate \Eve from \lemmaref{Eve} (applied to an
inputless protocol obtained by using uniformly randomly chosen inputs for
\prot).

When Alice is supposed to generate the next message, she queries the random
oracle at some points.  Based on her local view, she then generates the next
message of the protocol using her next message generation algorithm.
Similarly, Bob also generates the next message of the protocol during his
turns.  Eve, on the other hand, simply performs several queries to the
random oracle and announces all her queries and their corresponding answers
at the end of her turn. For concreteness we shall assume that the protocol
starts with Alice sending a message. Alice and Bob take turns alternately,
with Eve getting a turn after every Alice or Bob message (i.e., the messages
will be sent by Alice, Eve, Bob, Eve, and again Alice, Eve and so on.).

We shall refer to this protocol as the ``augmented protocol'' $(\prot,\Eve)$.

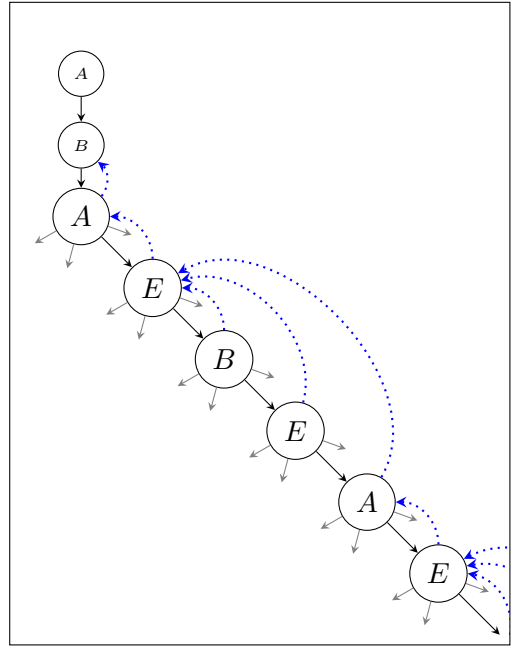
\begin{wrapfigure}[24]{r}{0.40\textwidth}
\vspace{-10pt}
\centering
\begin{tikzpicture}[scale=0.95,>=stealth]
\def\lightshade{gray}
\def\relcolor{blue}
\def\relstyle{dotted}
\newcommand{\tarrow}[2]{\draw[->] (#1) -- (#2);}
\newcommand{\barrow}[2]{
  \draw[->] (#1) -- (#2);
  \coordinate (temp1) at (xyz polar cs:angle=-20, radius=0.75);
  \coordinate (temp2) at (xyz polar cs:angle=210, radius=0.75);
  \coordinate (temp3) at (xyz polar cs:angle=255, radius=0.75);
  \draw[->, \lightshade] (#1) -- ++(temp1);
  \draw[->, \lightshade] (#1) -- ++(temp2);
  \draw[->, \lightshade] (#1) -- ++(temp3);
}
\newcommand{\carrowg}[3]{\draw[->, \relstyle, thick, \relcolor] (#1) to[bend right=#3] (#2);}
\newcommand{\carrow}[2]{\carrowg{#1}{#2}{45};}
\tikzset{n/.style = {circle, draw}}
\tikzset{im_n/.style = {}}
\coordinate (o) at (0,0);
\coordinate (x) at (1,0);
\coordinate (y) at (0,-1);
\coordinate (d) at ($(x) + (y)$);
\draw ++(d) node [n] (v) {\tiny{$A$}}
      ++(y) node [n] (v0) {\tiny{$B$}}
      ++(y) node [n] (v01) {$A$}
      ++(d) node [n] (v011) {$E$}
      ++(d) node [n] (v0111) {$B$}
      ++(d) node [n] (v01111) {$E$}
      ++(d) node [n] (v011111) {$A$}
      ++(d) node [n] (v0111111) {$E$}
      ++(d) node [im_n] (i) {}
      [clip] (o) rectangle (i)
      ++(d) node [im_n] (i1) {}
      ++(d) node [im_n] (i11) {};
\tarrow{v}{v0}; \tarrow{v0}{v01};
\barrow{v01}{v011}; \barrow{v011}{v0111}; 
\barrow{v0111}{v01111}; \barrow{v01111}{v011111}; \barrow{v011111}{v0111111};
\barrow{v0111111}{i}; \barrow{i}{i1}; \barrow{i1}{i11};
\carrow{v01}{v0}; 
\carrow{v011}{v01};
\carrowg{v0111}{v011}{45}; \carrowg{v01111}{v011}{60}; \carrowg{v011111}{v011}{75};
\carrow{v0111111}{v011111};
\carrowg{i}{v0111111}{45}; \carrowg{i1}{v0111111}{60}; \carrowg{i11}{v0111111}{75};
\end{tikzpicture}
\vspace{-80pt}
\caption{
  Schematic representation of the nodes in
	\augtree (including two initial dummy nodes). The nodes are labeled
	$A$, $B$ and $E$, for Alice, Bob and Eve. The  {dotted lines} show the \Aparent relation.
}
\label{fig:nodes}
\vspace{-20pt}
\end{wrapfigure}

\paragraph{Augmented Transcript Tree \augtree.}
Our analysis considers the transcript tree \augtree of an execution of \prot
augmented with a public query strategy \Eve.  The \augtree associated with
an augmented protocol $(\prot,\Eve)$, is the tree as defined in
\sectionref{frontiers-prelim} with the sequence of random variables
$(\bfm_1,\bfm_2,\dotsc)$ being the messages added to the transcript of the
augmented protocol by Alice, Eve and Bob during an execution.  In other
words, the nodes in the transcript tree are all the possible partial
transcripts in the
augmented protocol execution, with a directed edge from a node $u$ to a node
$v$, if the partial transcript associated with $v$ is obtained by adding
exactly one message (from Alice, Bob or Eve) to the partial transcript
associated with $u$.

For convenience we add an initial ``dummy'' round, in which Alice sends a
fixed message followed by Bob sending a fixed message. These correspond to
two dummy nodes at the root of \augtree.
We shall denote 
by \Anode and \Bnode the sets of Alice and Bob nodes,
and by \Achild and \Bchild the sets of (Eve) nodes that
are children of, respectively, Alice nodes and Bob nodes.
The tree \augtree naturally defines an ``ancestor'' partial order of the nodes in the
tree: we say $u \ancest v$ if $u$ occurs somewhere on the path from the root
of the tree to $v$ ($u$ could be identical to $v$). If $u\ancest v$, but
$u\neq v$, then we write $u\strictancest v$. We define
$\Ancest(v)=\{u | w \ancest v \}$.

An important definition we shall use through out is that of \Aparent and
\Bparent nodes.
\begin{definition}[\Aparent]
For every node $v$ in the transcript tree, except the initial dummy Alice node, we define $\Aparent(v)$ as follows:
\begin{itemize}
\item If $v\in\Achild$, then $\Aparent(v)$ is the parent of $v$.
\item If $v\not\in\Achild$, we define $\Aparent(v)$ to correspond to the last message sent by
Alice, before the transcript reached $v$: i.e., $\Aparent(v)=w$ such that
$w\in\Ancest(v)\cap\Achild$, and for all $w'\in \Ancest(v) \cap \Achild$,
$w'\ancest w$.
\end{itemize}
\end{definition}
Note that $\Aparent(v) \in \Achild \cup \Anode$ and $\Aparent(v)\strictancest v$. 
Further, for any node $v$, the sequence $v, \Aparent(v), \Aparent(\Aparent(v)), \dotsc$ ends at the 
initial dummy Alice node.%
\footnote{We added dummy Alice and Bob nodes at the root level to ensure that
\Aparent and \Bparent is well-defined for all the original nodes. Note that no information is exchanged until
after the protocol passes these dummy nodes, and so these nodes will not be part
of any of our frontiers defined later.}
\figureref{nodes} pictorially summarizes the \Aparent relation.

Similarly, for every node $v$ (except the initial dummy Alice and Bob
nodes), we define $\Bparent(v)$ as either the maximal element of
$\Ancest(v)\cap\Bchild$ (if $v\not\in\Bchild$) or the parent of $v$
(if $v\in\Bchild$).
Note that
$\Bparent(v) \in \Bchild \cup \Bnode$ and 
$\Bparent(v)\strictancest v$.

For any partial transcript $w$, we define the views of Alice, Bob and Eve
consistent with the partial transcript $w$.  The Eve view consistent with
$w$ is represented by $\VE{w}$.  We represent the distribution of Alice
views and Bob views conditioned on $w$, when their local inputs are $x$ and
$y$, respectively, by \dVA{x}{w} and \dVB{y}{w} (the bold face
emphasizing that these are distributions). The probability is over the choice
of random tapes for Alice and Bob and the random oracle.
We emphasize that the local views of parties contain only those query-answer
pairs which were generated during next message generation of messages
already present in $w$.  So, if Alice sends the next message in a round and
the resulting transcript was $w$, then Bob's views consistent with $w$ will
contain only query-answer pairs which were generated in previous rounds.
Bob's view gets updated with new query-answer pairs when he sends the next
message in the protocol.

\paragraph{Strictly Above a Set: $u\strictancest F$ and $F_1\strictancest F_2$.}
We shall abuse the $\strictancest$ notation slightly, and use it in the following senses too: if $u$
is a node and $F$ is a set of nodes, we write $u\strictancest F$ (read as
$u$ is strictly above $F$) if $u$ can be reached from the root without
passing through any node in $F$ (i.e., there is no $v\in F$ such that
$v\ancest u$); note that for $u$ to be strictly above $F$, it is not
necessary to have any $v\in F$ such that $u\strictancest v$.  For two sets of
nodes $F_1,F_2$, we define the {\em event} $F_1\strictancest F_2$ to occur
if the transcript path of an execution passes through a node $v\in F_1$
strictly before passing through any node in $F_2$ (it may or may not pass
through a node in $F_2$ afterwards).

\section{Overview of Our Analysis}
\label{sec:frontiers}

Here we sketch the technical details of our frontier analysis (see
\sectionref{overview} for a motivating discussion, and
\sectionref{detailed-frontiers} for the remaining details).

Suppose \prot is a 2-party protocol using a random oracle  \cO that
\simerr-securely realizes a symmetric SFE
functionality $f$ that is not {row or column decomposable at the top
level} (i.e., not even the first step of decomposition is possible; as we
shall see, it is enough to rule out protocols for such functionalities).
Let \Eve be the public query strategy described in \lemmaref{Eve}, with
an adjustable parameter \eps as described there. ($\eps=1/\poly(\spar)$ will be tuned
later in the proof.) Note that in \lemmaref{Eve}, the protocol considered
has no inputs; in order to define \Eve from this, we use an inputless protocol
obtained by running \prot with private inputs chosen uniformly
at random (as part of Alice's and Bob's local randomness).
We shall modify the protocol so that at the end of the protocol, Alice adds
the output of the protocol to the transcript.  (The simulation error \simerr
at most doubles by this modification.)
We consider the transcript tree \augtree 
as described above, for this protocol \prot augmented with \Eve.

Intuitively, we will be
arguing that if some information about $x$ has been revealed by the time the
transcript reaches a node $v$, some information about $x$ or $y$ must have
already been revealed when it reached $\Aparent(v)$.
Similarly, for information about $y$ to be revealed at $v$, some information
about $x$ or $y$ should already have been revealed at $\Bparent(v)$.
Together these requirements yield a contradiction.
To formalize this, we
shall define a frontier \FXsimple (and symmetrically \FYsimple) that
consists of nodes $v$ such that the ``extra information'' revealed about $x$
at $v$ since reaching $\Aparent(v)$ is significant. 

More precisely we define the following two frontiers on this tree, in terms of two
parameters $\delta$ and $\minwt$ (for concreteness, 
consider $\delta=\frac{1}{\numround}$,
where \numround is the depth of the tree \augtree, and $\minwt=\frac{1}{32|\cX||\cY|}$).
\begin{align*}
\bullet\;\; \FX = \{ v | & v \text{ is the first node on the path from root to $v$ s.t. }
\exists y\in \cY, x,x' \in \cX, \pr[y|v]\ge \minwt  \text{ and } \\
& \pr[v|\Aparent(v);x,y] > (1+\delta)\pr[v|\Aparent(v);x',y] \} \\
\bullet\;\; \FY = \{ v | & v \text{ is the first node on the path from root to $v$ s.t. }
\exists x\in \cX, y,y' \in \cY,  \pr[x|v]\ge \minwt \text{ and }  \\
&\pr[v|\Bparent(v);x,y] > (1+\delta)\pr[v|\Bparent(v);x,y'] \},
\end{align*}
Here, $\pr[v|w;x,y]$ denotes the probability (over the random tapes of the parties
and the oracle \cO) of reaching a node $v$ in \augtree, conditioned
on having reached the node $w$, when the parties run
the protocol honestly with inputs $x$ and $y$ respectively.
We shall also write $\pr[x|v]$ and $\pr[y|v]$ to denote the probabilities
of $x$ and $y$ being the inputs for Alice and Bob, respectively, conditioned on a protocol 
execution with a {\em uniformly random input pair} reaching the node $v$.%
\footnote{\label{foot:prconvention}%
In all our equations, we use the convention that the probability of an event conditioned on a
zero-probability event is zero. Alternately, we can avoid this by assuming, adding a negligible
security error, that for any pair of inputs, any node in \augtree is reached
with positive probability.}
Intuitively, the quantity
 $\max_{x,x',y} |\log\pr[v|w;x,y]-\log\pr[v|w;x',y]|$
measures the amount of information about
Alice's input that is revealed at $v$, since passing through $w=\Aparent(v)$.
This quantity is ``significant'' if it is beyond a threshold $\log(1+\delta)$ (where,
for concreteness, $\delta=\nicefrac1{\numround}$, \numround being the depth of
\augtree) and if it is realized by a $y$ which is somewhat likely
(i.e., $\pr[y|v]\ge\minwt$).
In our proofs, it will be useful to consider 
frontiers \FXnomin and \FYnomin which are defined
identically as \FX and \FY, but with $\minwt=0$, i.e. these frontiers are considered without the restriction of
$\pr[y|v]\ge\minwt$ and $\pr[x|v]\ge\minwt$ respectively.

Based on the correctness and the security of the protocol, {\em and using
the fact that $f$ is undecomposable at the top level}, we shall first prove
that these frontiers are almost ``full frontiers'' (when \simerr, the security error
for \prot, is negligible and \minwt is set sufficiently small):
\begin{claim}
\label{clm:frontierfull-inmain}
On an execution over \augtree with a random input pair $(x,y)$,
for any value of \minwt, the
probability that the
transcript does not pass through \FX (or symmetrically, \FY) is at most
$\poly(|\cX||\cY|)\cdot\minwt + O(\simerr)$.
\end{claim}
This is proven as \claimref{frontierfull}.
Given that these frontiers exist, next we prove a restriction on how they can
occur relative to each other, leading to our final contradiction.
Intuitively, the claim states the following: suppose a transcript passes through a
node $u\in\FX$; in a secure protocol not only should $u$ occur only at or
below the frontier \FY, but even $\Aparent(u)$ should occur only at or below
\FY; that is a node in \FY should occur {\em strictly above} $u$.
(Similarly, for $v\in\FY$ and the frontier \FX.)
\begin{claim}
\label{clm:frontierorder-inmain}
Consider running the execution on \augtree with a random input $(x,y)$ where $\eps$ is the parameter of the Independence Learner Eve.
The probability that the transcript passes through a node
$u\in\FX$ such that $\Aparent(u)\strictancest\FY$
is at most
\[\poly(\frac{N|\cX||\cY|}{\minwt})\cdot(\eps^{\Omega(1)}+\simerr) + \poly(|\cX||\cY|)\cdot\minwt.\]
Similarly, the probability that the transcript passes through a node $v\in\FY$
such that $\Bparent(v)\strictancest\FX$
is bounded by the same quantity.
\end{claim}

Once we prove this claim (as \claimref{frontierorder}), the required contradiction follows easily: by
setting $\minwt$ small enough (but $\Omega(1/\poly(|\cX||\cY|))$), and
choosing $\eps$ for the independence learner appropriately (note that this
does not affect \numround), the bounds in above claims can all be driven
below, say, any constant (for sufficiently large values of the security
parameter). Thus with positive probability the transcript must pass through
$u\in\FX$ and $v\in\FY$, with $v\strictancest u$ and $u\strictancest v$,
giving us the desired contradiction.

To prove \claimref{frontierorder-inmain},
technically, it is more convenient to
bound the probability of encountering
$\highFX=\{u|u\in\FX\text{ and } \Aparent(u)\strictancest\FYnomin \}$
(instead of $u\in\FX$ such that
$\Aparent(u)\strictancest\FY$). The difference between these two events can
be bounded relatively easily (see the proof in \sectionref{detailed-frontierorder}
for details). In particular, for this we use the above \claimref{frontierfull-inmain}
(with $\minwt=0$) and a bound on the probability of \FXnomin appearing strictly
above \FX and \FY (proven as \claimref{minvsnomin}):
\begin{align}
\label{eq:minvsnomin-inmain}
\pr[\FYnomin \strictancest (\FX\cup\FY)] \le \minwt \poly(|\cX||\cY|)
\end{align}
Intuitively, the bound above says that if \FYnomin is encountered strictly above \FX, then
it is very likely to occur together with \FY; hence when
a part of \FX occurs at or above \FY (so that its $\Aparent$ is strictly above \FY)
it is very likely to be at or above \FYnomin too. To upper bound the probability
of the former, it is enough to upper bound the probability of the latter.

Bounding $\pr[\highFX]$ (the probability of reaching \highFX with uniformly random
inputs) involves several parts:
\begin{asparaitem}
\item {\bf Part 1:} Firstly, we show that we can concentrate on a $2\times2$
minor of the function $f$: that is,  $\xx0,\xx1\in\cX$ and
$\yy0,\yy1\in\cY$ such that $f(\xx0,\yy0)=f(\xx1,\yy0)$ (but
$f(\xx0,\yy1)\not=f(\xx1,\yy1)$ if \highFX has significant probability).
We show that there exists a segment $\highFXX\subseteq\highFX$ such that
the inputs $(\xx0,\yy1)$ and $(\xx1,\yy1)$ are distinguished at
\highFXX,
and $\pr[\highFX] \le \poly(\nicefrac{|\cX||\cY|}{\minwt}) \pr[\highFXX|\xx0,\yy1]$.%
\footnote{Note that the need for working with \FX and \FY rather than just
\FXnomin and \FYnomin is that in this part we rely on the ``distinguishing
input'' being somewhat likely.  (If $\minwt=0$ the above bound is useless.)}

In the rest of the proof we need to bound $\pr[\highFXX|\xx0,\yy1]$. The
segment \highFXX splits into two parts: nodes $u$ with $\Aparent(u)$ being
an Alice node, denoted by \highSXX, and the ones with $\Aparent(u)$ being a child of an Alice
node, denoted by \highRXX.

\item {\bf Part 2:} Using \lemmaref{Eve} we show that Alice's message cannot
reveal any (significant) information about Bob's input, given the
information already present in the transcript of the augmented execution
(in \lemmaref{IL-Input}).
This is used to bound $\pr[\highSXX|\xx0,\yy1]$.
Note that this part is analogous to the argument when no oracle is present,
though more involved (without oracles,
this property is a trivial consequence of the nature of a protocol).

\item {\bf Part 3:} The most involved part is to bound
$\pr[\highRXX|\xx0,\yy1]$.  Here we want to bound the probability that a
distinction between \xx0 and \xx1 is revealed (when Bob's input is \yy1) at
a node $u$ that is not a child of an Alice node, but at $w=\Aparent(u)$ the
distinction between \yy0 and \yy1 has not been made. Since at $w$, \yy0 and
\yy1 is not distinguished by the transcript, in an execution with his actual
input being \yy0, on hitting the node $w$, Bob can {\em mentally switch his
input to \yy1} --- i.e., sample a view (including answers from the oracle)
consistent with the transcript and input \yy0.  We would like to argue that
then Bob can continue the execution of the protocol (till before Alice
should send the next message) and check if it hits $u$ or not, to
distinguish between \xx0 and \xx1.  However, the execution depends on the
random oracle which in turn is correlated with both parties' inputs. So Bob
cannot sample a correctly distributed random oracle (since he does not know
Alice's input) nor directly use the actual random oracle he has access to
(since it is conditioned on his actual input \yy0 and not \yy1).

The main idea here is that the independence guarantee from \lemmaref{Eve}
can be used to let Bob ``edit'' the actual random oracle (conditioned on
$(x,\yy0)$) to simulate a random oracle conditioned on $(x,\yy1)$ (without
knowing $x$).
The editing involves inserting answers consistent with a
sampled view (with input \yy1), ``deleting'' answers not present in this
sampled view, but is present in the actual view (with input \yy0) and using
the original oracle for queries not answered in the sampled view or the
actual view. (See \figureref{explore} for an illustration.)
The ``safety condition'' in \claimref{safe} assures that  the
queries from the sampled view that are not in Eve's view (for which the
answers from the sampled view are used) and the queries from the original view
that are not in Eve's view (for which random answers are used) are both
unlikely to be in in Alice's view; this lets us show that the oracle resulting from
the editing is correctly conditioned on the input pair $(x,\yy1)$.

The final (passive) attack involves carrying out the above attack at every
node $w$ and checking if the curious exploration hits the segment \highRXX
in any such exploration. We show that if $\pr[\highRXX|\xx0,\yy1]$ has significant
probability then it will be more likely for the exploration to hit \highRXX
in the exploration with input $(\xx0,\yy0)$ than in the exploration with
input $(\xx1,\yy0)$, thereby violating the security condition.
\end{asparaitem}

Throughout the argument, translating intuitive statements about information
and probability is complicated by the fact that the probability of reaching
different nodes depends on the inputs themselves. While intuitively, some of
these distributions must be close to each other until the frontiers
\FXsimple and \FYsimple are crossed, we cannot often leverage this intuition
without being trapped in circular arguments. Nevertheless,
going through several carefully chosen intermediate steps, we can relate the advantage
obtained by Bob in distinguishing \xx0 and \xx1 when using input \yy1, with
that he obtains when using input \yy0 with the above attack.

\section{Detailed Proof of \theoremref{main}}
\label{sec:detailed-frontiers}

In this section we present the remaining details of the proof of 
\theoremref{main}, that were sketched in \sectionref{frontiers}.

Recall the setting introduced in \sectionref{frontiers}: $f$ is a
deterministic symmetric two-party function which is undecomposable at the
top-most level (i.e., not even the first step of decomposition is possible).
Suppose \prot is a semi-honest secure SFE protocol for $f$ using a random
oracle \cO with simulation error \simerr.  We defined an augmented
transcript tree \augtree, and frontiers \FX and \FY in \augtree.
First, we shall state our main technical claim about these frontiers in
\sectionref{detailed-frontierorder}, and show how it follows from several sub-claims that are proven
in subsequent sections. 
Based on \claimref{frontierfull}
and \claimref{frontierorder}, we present
the proof of
\theoremref{main} in \sectionref{detailed-pf-main}. 
The sub-claims used in the proof of \claimref{frontierorder} are proven in 
\sectionref{detailed-frontierfull},
\sectionref{detailed-minvsnomin} and \sectionref{detailed-highFX}.

The technical heart of the
proof appears in  \sectionref{detailed-highFX}, which is
part of the proof of
 \claimref{frontierorder}.

\subsection{Frontier Ordering}
\label{sec:detailed-frontierorder}

\begin{figure}[ht]
\begin{center}
\includegraphics[scale=0.5]{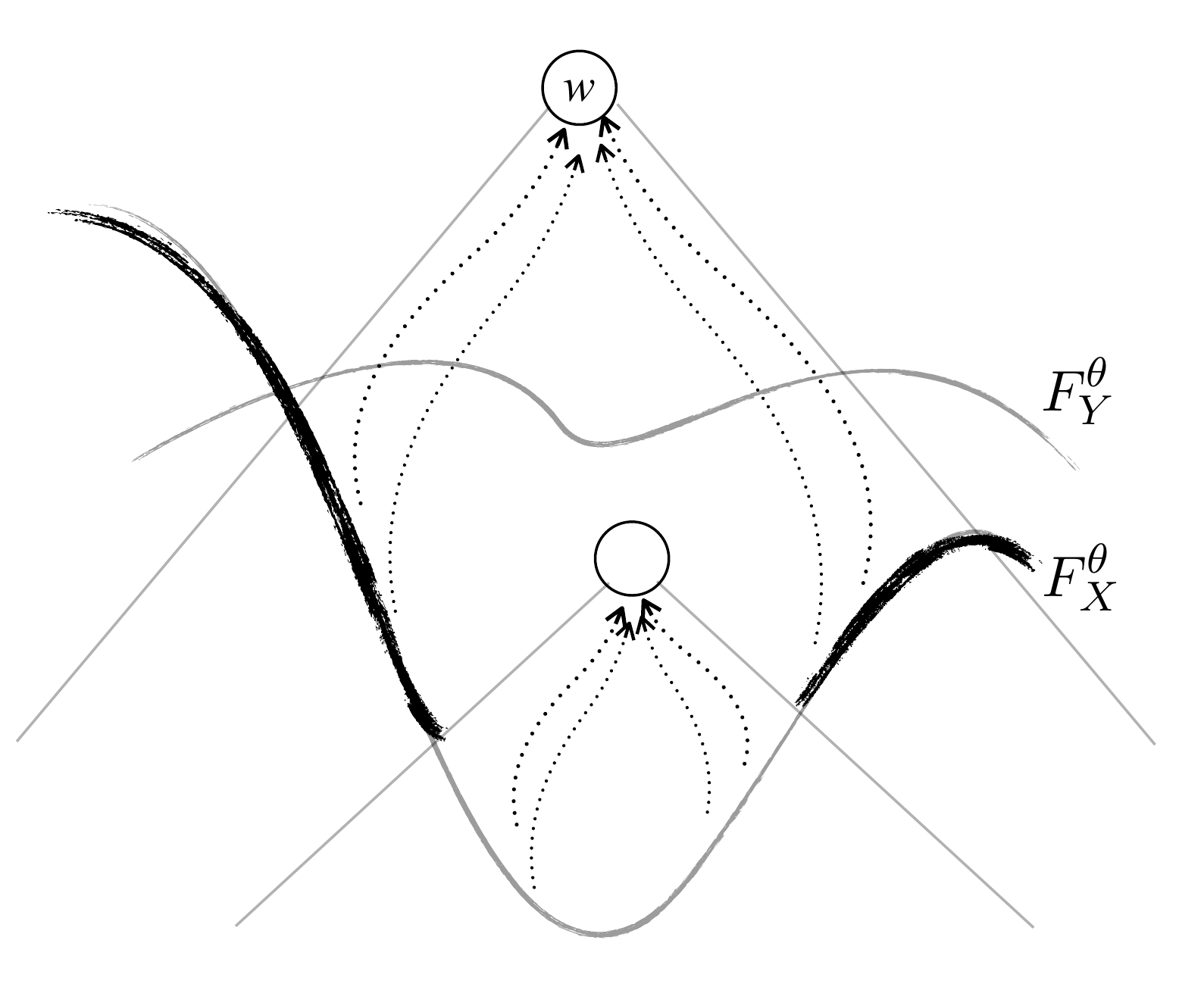}
\end{center}
\caption{A schematic representation of the segment \highFXreal (indicated by
thicker line). The dotted lines connect nodes to their \Aparent nodes (see
\figureref{nodes}).
\highFXreal contains those nodes $u\in\FX$ such that $\Aparent(u)$ occurs strictly above \FY.
We seek to upperbound the probability $\pr[\highFXreal]$ (when the inputs are
uniformly chosen).}
\label{fig:fronts}
\end{figure}

In this section we shall prove the claim regarding the frontier ordering,
\claimref{frontierorder-inmain}.
The claim bounds the probability (with uniformly random inputs) of the
transcript encountering the following part of the frontier \FX:
\[\highFXreal = \{ u | u \in\FX \text{ and } \Aparent(u)\strictancest\FY \}.\]
\figureref{fronts} shows this part schematically.

\begin{claim}
\label{clm:frontierorder}
Let 
$\highFXreal := \{ u | u \in\FX \text{ and } \Aparent(u)\strictancest\FY \}$ and
$\highFYreal := \{ u | u \in\FY \text{ and } \Bparent(u)\strictancest\FX \}.$
Then there exist polynomials $\orderpoly,\orderpoly'$ and $\indeperr,\safeerr = \eps^{\Omega(1)}\poly(|\cX||\cY|\spar)$,
such that for any value of \minwt, 
\begin{align}
\pr[\highFXreal] &\le \orderpoly\left(\numround|\cX||\cY|/\minwt\right)\cdot \big(\indeperr+ \safeerr+ \simerr \big)
+ \orderpoly'(|\cX||\cY|)\cdot \minwt \label{eq:highFXrealbd} \\
\pr[\highFYreal] &\le \orderpoly\left(\numround|\cX||\cY|/\minwt\right)\cdot \big(\indeperr+ \safeerr+ \simerr \big)
+ \orderpoly'(|\cX||\cY|)\cdot \minwt \label{eq:highFYrealbd}
\end{align}
\end{claim}

\begin{proof}
We shall prove \equationref{highFXrealbd} (the second part being
symmetrical).  That is, we are interested in bounding the probability that,
on running the execution on \augtree with uniformly random inputs $(x,y)$, the transcript
reaches a node in 
$\highFXreal = \{ u | u \in\FX \text{ and } \Aparent(u)\strictancest\FY \}$.
We say that the event \highFXreal occurs, if the path from root to the generated transcript passes through a node in \highFXreal.

To obtain an upper bound on $\pr[\highFXreal]$, we first observe that the
event \highFXreal implies the occurrence of one of the following three
events:
\begin{compactenum}
\item Event \missFYnomin: 
  the transcript path does not pass through any node in \FYnomin.
\item Event $\FYnomin \strictancest (\FX\cup\FY)$:
  the transcript path passes through a node $z\in\FYnomin$ and
  $z\strictancest (\FX\cup\FY)$ (i.e., there is no node $v\in\FX\cup\FY$
  such that $v\ancest z$).
\item Event \highFX:
  the path passes through $\highFX$ which is defined similarly to \highFXreal, but replacing \FY by
  \FYnomin.  i.e., 
  \[ \highFX = \{ u | u \in\FX \text{ and }  \Aparent(u)\strictancest\FYnomin \}.\]
\end{compactenum}
  To see this, suppose \highFXreal is encountered, but neither of the first two events occur;
  then transcript path passes through $u\in\highFXreal$, and
  a node $z\in\FYnomin$, and a node $v\ancest z$ such that
  $v\in\FX\cup\FY$. We argue that in this case 
  $\Aparent(u)\strictancest z$; then, since \FYnomin is part of a frontier,
  $\Aparent(u)\strictancest\FYnomin$ and hence $u\in\highFX$. This
  is because:
  \begin{compactitem}
  \item If $v\in\FX$, then $u=v$ (since
  $u\in\highFXreal$ and $v\in\FX$ are on the same path),
  and $\Aparent(u) \strictancest u=v\ancest z$.
  \item If $v\in\FY$, then $v=z$
  (since, $v\in\FY$, $z\in\FYnomin$ and $v\ancest z$ $\implies$ $z=v$)
  and further, since $u\in\highFXreal$ and $v\in\FY$ are
  on the same path, by definition of \highFXreal, $\Aparent(u)\strictancest v = z$.
  \end{compactitem}

Thus, it suffices to upper bound the probabilities of each of these three events.
We will be able to easily bound both $\pr[\missFYnomin]$  and 
$\pr[\FYnomin \strictancest (\FX\cup\FY)]$, (\claimref{frontierfull-inmain}
--- proven as \claimref{frontierfull} --  and
\equationref{minvsnomin-inmain} -- proven as \claimref{minvsnomin} -- respectively). The main technical difficulty is in
bounding $\pr[\highFX]$, which is carried out in \claimref{frontierattack}.
From these three claims we get 
\begin{align*}
\pr[\missFYnomin] &= O(\simerr)  & \text{(By \claimref{frontierfull-inmain}, with $\minwt=0$)}\\
\pr[\FYnomin \strictancest (\FX\cup\FY)] &\leq \poly(|\cX||\cY|)\minwt & \text{(By \equationref{minvsnomin-inmain})} \\
\pr[\highFX] &\leq \poly\left(\frac{N|\cX||\cY|}{\minwt}\right)(\indeperr+\safeerr+\simerr) 
	&\text{(By \claimref{frontierattack})}
\end{align*}

Adding the three, we get the required bound.
\end{proof}

\subsection{Proof of \theoremref{main}}
\label{sec:detailed-pf-main}
The main part of the proof
proves the impossibility of a semi-honest secure SFE protocol,
even using random oracles,
for a symmetric function $f$ that is undecomposable at the top-level.  
We shall shortly see that this is enough.

So, suppose $f$ is a 2-party symmetric function that is
undecomposable at the top-most level, and 
\prot is a semi-honest secure protocol using a random oracle \cO,
for the SFE funcionality evaluating $f$, 
with simulation error \simerr.  This is the setting under which the frontiers in \augtree
are defined, and \claimref{frontierfull} and
\claimref{frontierorder} hold.  The proof follows by a deriving a contradiction
from these two claims (instantiated with suitable parameters).

We shall set $\minwt=\min\{ \frac{1}{8\missfxpoly(|\cX||\cY|)},\frac{1}{8\orderpoly'(|\cX||\cY|)} \}$ 
where the $\missfxpoly$ and $\orderpoly'$ are as in \claimref{frontierfull}
and \claimref{frontierorder}
(in fact, $\minwt=\Theta(\frac{1}{|\cX||\cY|})$, by following the proofs of
the various claims),
and then choose a small enough (but $1/\poly(\spar)$) value of $\eps$ so that
$(\indeperr+\safeerr) \le \frac{1}{8\orderpoly(\numround|\cX||\cY|/\minwt)}$ (which is possible since
$1/\minwt$ is $\poly(\spar)$ and \indeperr and \safeerr are $\eps^{\Omega(1)}\poly(|\cX||\cY|\spar)$), so that (for large enough \spar)
\begin{align*}
\pr[\missFX] + \pr[\missFY] &\leq 2\bigg(\missfxconst\simerr + \missfxpoly(|\cX||\cY|)\minwt\bigg) < \frac13
&\text{By \claimref{frontierfull}} \\
\pr[\highFXreal]  &< \frac13 \qquad \text{and} \qquad \pr[\highFYreal] < \frac13
&\text{By \claimref{frontierorder}}
\end{align*}

So, with non-zero probability, for a random input pair $(x,y)$, the honestly
generated transcript passes through both \FX and \FY, but avoids both events
$\highFXreal$ and $\highFYreal$.
Consider one such transcript $\tau$.
Let $u$ and $v$ be the intersection of this path with the frontiers $\FX$ and $\FY$.
For this transcript $\tau$: $v \ancest \Aparent(u)$ (since
of $u\not\in\highFXreal$) and $\Aparent(u) \strictancest u$ (by definition of
$\Aparent$), i.e. $v \strictancest u$.
Symmetrically, we also get: $u \ancest \Bparent(v)$ and 
$\Bparent(v) \strictancest v$, and hence $u \strictancest v$.
This gives us a contradiction as desired.

\paragraph{\em Extending to all 2-party functions.}
Above we showed that any symmetric 2-party function that is
undecomposable at the top-level does not have an SFE protocol secure against
semi-honest adversaries, in the random oracle model. Now we extend this
to show that the only 2-party functions for which semi-honest secure protocols 
exist in the random oracle model are those for which (perfectly)
semi-honest secure protocols exist in the plain model. We do this in
two steps, first for symmetric 2-party functions and then for general 2-party functions. But first we
state a claim that we will need (in the second step).
\begin{claim}
\label{clm:isomorphic-to-symmetric}
If a (not necessarily symmetric) 2-party function $f_0$ has a  semi-honest secure protocol
in the random oracle model (resp.~plain model), it must be ``isomorphic'' to a symmetric 
2-party function $f_1$ that has a semi-honest secure protocol in the random oracle model
(resp.~plain model).
\end{claim}
This is because, by a result in
\cite{MajiOuPrRo11}, if a 2-party function $f_0$ is not isomorphic to
a certain symmetric 2-party function $f_1$ (namely, the ``common
information function of $f_0$ mentioned in \footnoteref{cifunction}), 
then $f_0$ is complete against semi-honest adversaries. But a
complete functionality cannot have a semi-honest secure protocol in the
plain or random oracle model (as otherwise all functionalities will have
semi-honest secure protocols in the random oracle model, contradicting
the above results.) 

Below are the two steps to complete the proof of \theoremref{main}.
\begin{enumerate}
\item Firstly, we argue that if a {\em symmetric} 2-party function $f_1$ has a semi-honest secure protocol
in the random oracle model, it must be decomposable (and hence has a
perfectly semi-honest secure protocol). This is because, if
$f_1$ is undecomposable, then it has a
minor $f$ which is undecomposable at the top-level. Further, if $f_1$ is semi-honest
securely realizable using a random oracle,
so is every minor of $f_1$,
including $f$, which contradicts our above result.
\item Next, if a general 2-party function $f_0$ has a  semi-honest secure protocol in the random oracle
model, then by \claimref{isomorphic-to-symmetric}, there is a symmetric 2-party
function $f_1$ that is isomorphic to $f_0$ and has a semi-honest secure protocol in the random oracle.
By the previous point, $f_1$ has a perfectly semi-honest secure protocol
in the plain model, and as $f_0$ is isomorphic to $f_1$, so does $f_0$.
\qedhere
\end{enumerate}

\subsection{Bounding probability of events \missFX and \missFY}
\label{sec:detailed-frontierfull}
In this section we prove \claimref{frontierfull-inmain} (restated below).
\begin{claim}
\label{clm:frontierfull}
There exists a constant \missfxconst and a polynomial \missfxpoly such that,
on executing the augmented protocol with a random
input pair $(x,y)$, 
$\pr[\missFX]$ and $\pr[\missFY]$ are both at most
$\missfxconst\simerr+\missfxpoly(|\cX||\cY|)\cdot\minwt$.
\end{claim}

\begin{proof}
We shall just show that $\pr[\missFX]\leq p^* = (5+(1+\delta)^\numround)\simerr+|\cX||\cY|\minwt$
(so that $\missfxconst=(5+(1+\delta)^\numround)$ and $\missfxpoly(\alpha)=\alpha$).
The bound on $\pr[\missFY]$ follows similarly.
We shall, in fact, show the stronger result that $\pr[\missFX|x,y] \leq p^*$, for all $(x,y)\in\cX\times\cY$.

Let $\S$ be the set of all complete transcripts such that none of their
ancestors lie in $\FX$.  First, consider any input pair
$(x,y)\in\cX\times\cY$ such that $f(\cdot,y)$ is not a constant function; we
shall upper-bound the probability $\pr[\S|x,y]$ by $p^*-4\simerr$.

Let the frontier $\U(y)$ be the set of nodes $u$ where, for the first time
on a path from the root, $\pr[y|u]<\minwt$.
Let $\L(y) = \{u\in\S | u\strictancest \U(y) \}$ be the part of \S which
is strictly above $\U(y)$. Then $\pr[\S|x,y] \le \pr[\U(y)|x,y] + \pr[L(y)|x,y]$.
Firstly,
\begin{align*}
\pr[\U(y)|x,y] &= \sum_{u\in\U(y)} \pr[u|x,y] = \sum_{u\in\U(y)} \pr[x,y|u] p[u] / \pr[x,y] 
= |\cX||\cY|  \sum_{u\in\U(y)} \pr[x,y|u] p[u]\\
&\leq |\cX||\cY| \sum_{u\in\U(y)} \pr[y|u] p[u]  < |\cX||\cY| \minwt \sum_{u\in\U(y)} \pr[u] 
\leq |\cX||\cY|\minwt.
\end{align*}

For nodes $v\in\L(y)$, we have $\pr[y|u] \geq \minwt$ for all $u\ancest v$.
Recall that $v$ does not have an ancestor in $\FX$.
So, it must be the case that, for all $x,x'\in\cX$ we have $\pr[v|x,y] \leq (1+\delta)^\numround\pr[v|x',y]$.
Since $f(\cdot,y)$ is not a constant function, there exists $x'\in\cX$ such that $f(x,y)\neq f(x',y)$.
We can partition the set $\L(y)$ into two sets:
\begin{enumerate}
\item $\C(y)$: Those transcripts $v\in\L(y)$ whose associated output is $f(x,y)$, i.e. those transcripts which provide correct output when the input is $(x,y)$, and
\item $\W(y)$: Those transcripts $v\in\L(y)$ whose associated output is $\neq f(x,y)$, i.e. those transcript which provide wrong output when the input is $(x,y)$.
\end{enumerate}

Since, the simulation error is at most \simerr, we can conclude that $\pr[\W(y)|x,y]\leq \simerr$.
Further, observe that the output associated with the transcripts in $\C(y)$ are incorrect for input $(x',y)$.
Therefore, $\pr[\C(y)|x',y]\leq \simerr$.
But, $\pr[\C(y)|x,y]\leq (1+\delta)^\numround\pr[\C(y)|x',y]\leq (1+\delta)^\numround\simerr$.
Now, we can claim that the $\pr[\L(y)|x,y] \leq (1+(1+\delta)^\numround)\simerr$.

Adding these two results, we can conclude that 
$$\pr[\S | x,y] \leq p^*-4\simerr$$

Now, we consider any $(x,y)\in\cX\times\cY$ such that $f(\cdot,y)$ is a constant function.
Since $f$ is undecomposable at the top-most level, there exists $x'\in\cX$ and $y'\in\cY$ such that $f(x',y)=f(x',y')$ and $f(\cdot,y')$ is not a constant function.
Thus, by security condition, we can conclude that the final transcript distributions induced by $(x,y)$ and $(x',y')$ have at most $4\simerr$ statistical distance.
Thus, to complete the proof of the theorem%
\footnote{
  We note that this bound is not restricted only to the uniform distribution over input pairs.
  In fact, for {\em any} input pair distribution such that $\pr[x,y]$ is a function of the output $f(x,y)$, $\pr[\S]\leq p^*$.%
}
:
$$\pr[\S|x,y] \leq \pr[\S|x',y'] + 4\simerr \leq p^*\qedhere$$
\end{proof}

\subsection{Bounding probability of event $\FYnomin \strictancest (\FX\cup\FY)$}
\label{sec:detailed-minvsnomin}

\begin{claim}
\label{clm:minvsnomin}
On executing the augmented protocol with a random
input pair $(x,y)$, $\pr[\FYnomin \strictancest (\FX\cup\FY)]$ is at most
$(1+(1+\delta)^\numround)|\cX||\cY|\minwt$.
The same bound holds for $\pr[\FXnomin \strictancest (\FX\cup\FY)]$.
\end{claim}

\begin{proof}
Let $\S$ be the set of nodes $v\in\FYnomin$ such that for all $u\ancest v$, $u\not\in \FX\cup\FY$, i.e. $v\strictancest (\FX \cup \FY)$.
We shall bound $\pr[\S|x,y]$, for each input pair $(x,y)\in\cX\times\cY$.
Fix an input pair $(x,y)$.
Let $\U(x,y)$ be the frontier of nodes $v$ where for the first time
$\pr[x|v] < \minwt$ or $\pr[y|v]<\minwt$. 
Let $\L(x,y) = \{ u | u\in\S,\text{ and } u\strictancest\U(x,y) \}$
be the part of \S which is strictly above $\U(x,y)$.
We shall bound
$\pr[\S|x,y] \le \pr[\U(x,y)|x,y] + \pr[\L(x,y)|x,y]$, by bounding the two
terms separately.

\begin{align*}
\pr[\U(x,y) | x,y] &= \sum_{v\in \U(x,y)} \pr[v | x,y] 
= \sum_{v\in\U(x,y)} \pr[x,y|v] \cdot \pr[v] / \pr[x,y]\\
&\leq |\cX||\cY| \sum_{v\in\U(x,y)} \min\{\pr[x|v],~\pr[y|v]\}\cdot\pr[v]
< \minwt |\cX||\cY|\sum_{v\in\U(x,y)}\pr[v] \leq \minwt |\cX||\cY|
\end{align*}

To bound $\pr[\L(x,y)|x,y]$,  we partition $\L(x,y)$ into
$\L_{\tx}(x,y)\subseteq\L(x,y)$,
one for each $\tx\in\cX\setminus\{x\}$, such that
for $v\in\L_{\tx}(x,y)$, 
$v$ is included in \FYnomin because 
$\exists y',y''$ such that 
$\pr[v|\Bparent(v);\tx,y'] > (1+\delta)\pr[v|\Bparent(v);\tx,y'']$.
Note that $\tx\neq x$, otherwise $v\in\FY$.
By definition of $\L(x,y)$, we
have $v\strictancest(~\FX\cup\FY\cup\U(x,y)~)$, i.e. $v\strictancest \FY$.
This implies that:
$$\pr[\tx|v]<\minwt$$

Observe that for all $u\ancest v\in\L(x,y)$, we have $\pr[y|u]\geq\minwt$.
But $v\strictancest \FX$.
Which implies:
$$\pr[v|x,y] \le (1+\delta)^\numround \pr[v|\tx,y]$$

Now, $\pr[v|\tx,y] \le \pr[v|\tx]/\pr[y|\tx] = |\cY|\pr[v|\tx]$.
So for $v\in\L_{\tx}(x,y)$,
\begin{align*}
\pr[v|x,y] 
&\le (1+\delta)^\numround |\cY| \pr[v|\tx] = (1+\delta)^\numround |\cX| |\cY| \pr[\tx|v] \pr[v] \\
&\le (1+\delta)^\numround |\cX||\cY| \minwt \pr[v].
\end{align*}
Hence, $\pr[\L(x,y)|x,y] \le (1+\delta)^\numround |\cX||\cY| \minwt \sum_{v\in\L(x,y)} \pr[v] \le
(1+\delta)^\numround |\cX||\cY|\minwt$. 
Putting this together with the above bound on
$\pr[\U(x,y)|x,y]$ we get, for all $(x,y)$,
$\pr[\S|x,y] \le (1+(1+\delta)^\numround) \minwt |\cX||\cY|$. 
Hence,
$\pr[\S] \le (1+(1+\delta)^\numround) \minwt |\cX||\cY|$.
\qedhere
\end{proof}

\subsection{Bounding the probability of event \highFX}
\label{sec:detailed-highFX}
This section carries out the technical heart of the proof.
For convenience we define $\mu=(1+\delta)^\numround$,
$\deltax = (1+\delta)^{1/(|\cX|-1)}-1$ and
$\deltay = (1+\delta)^{1/(|\cY|-1)}-1$.
Note that with $\delta=\frac1{\numround}$,
$\mu=O(1)$ and $\deltax,\deltay = \Omega(\frac1{\numround(|\cX|+|\cY|}))$
(where $|\cX|,|\cY|>1$).

\begin{claim}
\label{clm:frontierattack}
There exist $\indeperr,\safeerr = \eps^{\Omega(1)}\poly(\spar|\cX||\cY|)$,
such that the probability of the augmented protocol with uniformly random
inputs reaching \highFX is

$$\pr[\highFX]  \le \frac{2\mu^2|\cX|^2|\cY|(1+\deltax)\numround}{\minwt\deltax}\left(4\simerr+\indeperr+2\safeerr\right).$$
The same bound, with \deltay instead of \deltax, holds for $\pr[\highFY]$.
\end{claim}

We focus on proving the first part of this claim (the second part being
symmetrical).  That is, we are interested in bounding the probability that,
on executing \prot with uniformly random inputs $(x,y)$, the transcript
reaches a node in  $\highFX = \{ u | u \in\FX \text{ and } \nexists
z\in\FYnomin \text{ s.t. } z\ancest\Aparent(u) \}$.

We break the full proof of the claim into three parts:
\begin{enumerate}
\item {\bf Part 1.}
We shall show that there exist $\highFXX \subseteq \highFX$ such that
$\pr[\highFXX] \ge \pr[\highFX]/(|\cX||\cY|)^2$, and
there are $\xx0,\xx1 \in \cX$ and $\yy0,\yy1 \in \cY$,
such that
$f(\xx0,\yy0)=f(\xx1,\yy0)$, and
$\pr[\highFXX|\xx0,\yy1]$ is comparable to $\pr[\highFXX]$ (with
uniformly random inputs $(x,y)$), and for every $u\in\highFXX$, \yy1
sufficiently distinguishes \xx0 and \xx1. More precisely,
\begin{align}
\pr[\highFXX|\xx0,\yy1] \ge \frac{\minwt |\cY|}{(1+\delta)^\numround} \pr[\highFXX],
\label{eq:xx0yy1heavy}
\end{align}
and
for all $u\in\highFXX$, if $w=\Aparent(u)$, then
$\pr[u|w;\xx0,\yy1] \ge (1+\delta)^{1/(|\cX|-1)} \pr[u|w;\xx1,\yy1]$,
and hence
\begin{align}
\pr[u|w;\xx0,\yy1]-\pr[u|w;\xx1,\yy1] \ge \frac{\deltax}{1+\deltax}\pr[u|w;\xx0,\yy1].
\label{eq:atFX}
\end{align}
where $\deltax=(1+\delta)^{1/(|\cX|-1)}-1$.

\item {\bf Part 2.} We shall also show that $\pr[\highSXX|\xx0,\yy1]$, where
$\highSXX = \highFXX \cap \Achild$, must be ``small'' if the protocol is secure.
(For a node $u\in\highSXX$, $\Aparent(u)\in\Anode$.)

\item {\bf Part 3.} Then we shall show that $\pr[\highRXX|\xx0,\yy1]$, where
$\highRXX = \highFXX \backslash \Achild$, must be small if the protocol is secure.
(For a node $u\in\highRXX$, $\Aparent(u)\in\Achild$.)

Since $\pr[\highFXX|\xx0,\yy1]= \pr[\highRXX|\xx0,\yy1]+ \pr[\highSXX|\xx0,\yy1]$,
Parts~2 and 3 imply that $\pr[\highFXX|\xx0,\yy1]$ is small as well. Further, by
Part~1, $\pr[\highFXX]$ and, thus, $\pr[\highFX]$ is small as well.
\end{enumerate}

The error terms \indeperr and \safeerr appear in Parts 2 and 3 respectively,
from \claimref{independence} and \claimref{safe}. The claims are
consequences of the independence properties obtained by Eve of
\lemmaref{Eve}. Below we state the former claim (and show how it follows
from \lemmaref{IL-Input} proven in \appendixref{EvePlus}),
which states that Alice's message is almost independent of
Bob's input, conditioned on Eve's view thus far.
\begin{claim}
\label{clm:independence}
For all $x\in \cX$, $y,y'\in \cY$, if $W\subseteq \Anode$
is such that for all $w\in W$, $\pr[y|w;x],\pr[y'|w;x] \ge
\sigma$ for $\sigma = \frac1{\poly(|\cX||\cY|)}$, then,
for $\eps \le 1/\poly(\spar|\cX||\cY|)$ (for some polynomial)
and an error parameter $\indeperr=\eps^{\Omega(1)}\poly(\spar|\cX||\cY|)$, 
we have
\begin{align}
\label{eq:independence}
\sum_{w\in W} \pr[w|x,y] \cdot \SD\left(\{\children(w)|w;x,y\},\{\children(w)|w;x,y'\}\right) \le \numround\indeperr,
\end{align}
where $\{\children(w)|w;x,y\}$ and $\{\children(w)|w;x,y'\}$ stand for
the distribution of the next node after $w$ (i.e., Alice's message at $w$)
in \augtree when \prot is executed with inputs $(x,y)$ and $(x,y')$ respectively.
\end{claim}
\begin{proof}
\lemmaref{IL-Input}, stated in terms of a traversal of the tree \augtree,
partitions the nodes at each level in the tree into three sets,
a low-probability set $W^i_0$ such that $\pr[W^i_0|x,y] \le \eps'$,
$W^i_1$ such that for $w\in W^i_1$, $\pr[y|w;x] < \eps'$ or $\pr[y'|w;x] < \eps'$
and $W^i_2$ such that for $w\in W^i_2$,
$\SD\left(\{\children(w)|w;x,y\},\{\children(w)|w;x,y'\}\right) \le \eps'$.
Note that  $W^i_1\cap W=\emptyset$ because (for sufficiently small values of $\eps$),
$\eps'=\eps^{\Omega(1)}\poly(\spar|\cX||\cY|)<\sigma$. So,
\begin{align*}
\sum_{w\in W} \pr[w|x,y] \cdot \SD&\left(\{\children(w)|w;x,y\},\{\children(w)|w;x,y'\}\right) \\
&\le \sum_i \sum_{w\in W^i_0} \pr[w|x,y] + \sum_i \sum_{w\in W^i_2}\pr[w|x,y] \eps' 
\le \numround \eps' + \numround \eps' \le \numround\indeperr,
\end{align*}
where $\indeperr=2\eps'$
\end{proof}

We mention a few other technical inequalities that are useful in the proof.

For $u\in\highFX$, if $w=\Aparent(u)$, then $w$
is strictly above the frontier \FYnomin, and hence
\begin{align}
\pr[w|\xx0,\yy0] \ge \frac1{(1+\delta)^\numround} \pr[w|\xx0,\yy1].
\label{eq:aboveFY}
\end{align}

For any subset $W$ of nodes,
\begin{align}
-2\numround\simerr \le \sum_{w\in W} \pr[w|\xx0,\yy0] - \pr[w|\xx1,\yy0]  \le 2\numround\simerr,
\label{eq:simerrbound}
\end{align}
because $f(\xx0,\yy0)=f(\xx1,\yy0)$ and by the security guarantee of \prot,
restricted to the intersection of $W$ with the frontier corresponding to a
fixed round number, this summation is at most $2\simerr$ (since in the ideal
world, the simulated views are identical, and for each execution, the error
from the simulated distribution is at most \simerr).

It will be useful to relate
$\sum_{w\in W} \left( \pr[w|\xx0,\yy0] \sum_{u\in \S_w} g(u,w) \right)$ to
$\sum_{w\in W} \left( \pr[w|\xx1,\yy0] \sum_{u\in \S_w} g(u,w)\right)$, where
for all $w\in W$, $\sum_{u\in \S_w} g(u,w) \le 1$. This arises for us when
$\S_w$ forms part of a frontier, and $g(u,w)$ is a probability
distribution (possibly conditioned on $w$) or statistical distance
between two probability distributions.
\begin{align}
\sum_{w\in W} & \pr[w|\xx1,\yy0]\sum_{u\in \S_w} g(u,w) \nonumber \\
&= \sum_{w\in W} \pr[w|\xx0,\yy0]\sum_{u\in \S_w} g(u,w)
- \sum_{w\in W} \left( \pr[w|\xx0,\yy0] - \pr[w|\xx1,\yy0] \right) \sum_{u\in \S_w} g(u,w) \nonumber \\
&= \sum_{w\in W} \pr[w|\xx0,\yy0]\sum_{u\in \S_w} g(u,w) \pm  2\numround\simerr
   \qquad \text{ (By \equationref{simerrbound}.)}
\label{eq:simerrbound2}
\end{align}
Here, we applied \equationref{simerrbound} to two subsets of $W$ (where
$\left( \pr[w|\xx0,\yy0] - \pr[w|\xx1,\yy0] \right)$ is positive and
negative, respectively.) and also used the fact that $\sum_{u\in \S_w} g(u,w) \le 1$.

\paragraph{Part 1.}
We define \highFXX and $(\xx0,\xx1,\yy0,\yy1)$.

For any node $u\in\FX$, there exists
$y^*_u\in\cY$ and some $x,x'\in\cX$ such that $\pr[u|w;x,y^*_u] >
(1+\delta)\pr[u|w;x',y^*_u]$, where $w=\Aparent(u)$.  W.l.o.g, we consider $x$ which
maximizes $\pr[u|w;x,y^*_u]$; we call the maximum value $\alpha(u,y^*_u)$.
Since $f$ is not row-decomposable at the top-level,
there exist a sequence of  $t+1\le|\cX|$ values $x_0,\cdots,x_t$ such that
\begin{asparaitem}
\item $x_0=x$, $x_t=x'$ (and hence  $\pr[u|w;x_0,y^*_u] >
(1+\delta)\pr[u|w;x_t,y^*_u]$);
\item for every $i=0,\cdots,t-1$, there exists $y_i\in\cY$ such that
$f(x_i,y_i)=f(x_{i+1},y_i)$.
\end{asparaitem}
Then, there exists an $i$ such that
$\pr[u|w;x_i,y^*_u] > (1+\delta)^{1/t} \pr[u|w;x_{i+1},y^*_u]$
and $\pr[u|w;x_i,y^*_u] > \pr[u|w;x_0,y^*_u]/(1+\delta)$. We will
denote the nodes $(x_i,x_{i+1},y_i)$ by $(x_u,x'_u,y_u)$. Thus, for
every node $u\in\FX$, there are nodes $(x_u,x'_u,y_u,y^*_u)$ such that
\begin{asparaitem}
\item $f(x_u,y_u)=f(x'_u,y_u)$, and
\item $\pr[u|w;x_u,y^*_u] > (1+\delta)^{1/t} \pr[u|w;x'_u,y^*_u]$
and $\pr[u|w;x_u,y^*_u] > \alpha(u,y^*_u)/(1+\delta)$.
\end{asparaitem}

Suppose that  $\pr[\highFX]=p$; i.e., when the protocol is executed with a random input pair $(x,y)$,
with probability $p$, the transcript passes through some $u\in\highFX$.
Since there are at most $|\cX|^2|\cY|^2$ values for the tuples
$(x_u,x'_u,y_u,y^*_u)$, we can find a tuple $(\xx0,\xx1,\yy0,\yy1)$ such
that the transcript passes through $u\in\highFX$ with $(x_u,x'_u,y_u,y^*_u)
= (\xx0,\xx1,\yy0,\yy1)$ with probability at least $p'=p/(|\cX|^2|\cY|^2)$.  We
define $\highFXX \subseteq \highFX$ as containing those $u$ with
$(x_u,x'_u,y_u,y^*_u) = (\xx0,\xx1,\yy0,\yy1)$. Then $\pr[\highFXX] \ge p'$.

For $w=\Aparent(u)$ for $u\in\highFXX$, $w$ is strictly above \FX,
and hence $\pr[w|\xx0,\yy1] \ge \pr[w|\yy1]/(1+\delta)^{\numround-1}$.
(Since $w$ has a child $u$, we upper-bound its depth by $\numround-1$.)

Also, since for $u\in\highFXX$ we have
$\pr[u|w;\xx0,\yy1] \ge \alpha(u,\yy1)/(1+\delta) \ge \pr[u|w;\yy1]/(1+\delta)$, we get that
\begin{align*}
\pr[\highFXX|\xx0,\yy1]
&= \sum_w \pr[w|\xx0,\yy1] \sum_{\substack{u\in\highFXX,\\ \Aparent(u)=w}} \pr[u|w;\xx0,\yy1] \\
&\ge \frac1{(1+\delta)^\numround} \sum_w \pr[w|\yy1] \sum_{\substack{u\in\highFXX,\\ \Aparent(u)=w}} \pr[u|w;\yy1] \\
&\ge \frac1{(1+\delta)^\numround} \pr[\highFXX|\yy1].
\end{align*}
Finally, note that for $u\in\highFXX$, $\pr[\yy1|u]\ge\minwt$ and hence
\[
\pr[\highFXX|\yy1] = \sum_{u\in\highFXX} \pr[u|\yy1] = \sum_{u\in\highFXX} |\cY|\pr[\yy1|u]\pr[u] \ge \minwt|\cY|\pr[\highFXX].
\]
Hence,
\begin{align*}
\pr[\highFXX|\xx0,\yy1] \ge \frac{\minwt |\cY|}{(1+\delta)^\numround} \pr[\highFXX].
\end{align*}

\paragraph{Part 2.}
This part is in fact similar to the argument in~\cite{MajiPrRo09}, except
that we need to rely on the independence guarantee from
\claimref{independence} to say that Alice's message is (almost) independent
of Bob's input, conditioned on the (augmented) transcript so far.
We shall show that $|\pr[\highSXX|\xx0,\yy0] - \pr[\highSXX|\xx1,\yy0]|$
is significant if $\pr[\highSXX|\xx0,\yy1]$ is significant. However, since
$f(\xx0,\yy0)=f(\xx1,\yy0)$, the former must be ``small'', and hence the
latter too must be small.

Since \highSXX is part of a frontier, for all $x,y$,
\[\pr[\highSXX|x,y] = \sum_{u\in\highSXX} \pr[u|x,y]
= \sum_{w\in\Anode} \pr[w|x,y] \sum_{\substack{u\in\highSXX \\ \Aparent(u)=w}} \pr[u|w;x,y].\]
For $u\in\highSXX$, $w=\Aparent(u)$ is $u$'s parent, an Alice node which is
strictly above \FYnomin.

\begin{align*}
\pr[\highSXX|\xx0,\yy0] &= \sum_{w} \pr[w|\xx0,\yy0]\sum_{\substack{u\in\highSXX \\ \Aparent(u)=w}} \pr[u|w;\xx0,\yy0] \\
&= \sum_{w} \pr[w|\xx0,\yy0]\sum_{\substack{u\in\highSXX \\ \Aparent(u)=w}} \pr[u|w;\xx0,\yy1]
    \pm \numround\indeperr & \text{(By \equationref{independence}.)} \\
\end{align*}
Note that \equationref{independence} can be applied above, since the
summation is over $w$ strictly above \FYnomin (since $w=\Aparent(u)$ for
$u\in\highSXX$), and for such $w$, $\pr[y|w;x]>\frac1{(1+\delta)^\numround |\cY|}=\frac1{\poly(|\cX||\cY|)}$.

\begin{align*}
\pr[\highSXX|\xx1,\yy0] &= \sum_{w} \pr[w|\xx1,\yy0]\sum_{\substack{u\in\highSXX \\ \Aparent(u)=w}} \pr[u|w;\xx1,\yy0] \\
&= \sum_{w} \pr[w|\xx1,\yy0]\sum_{\substack{u\in\highSXX \\ \Aparent(u)=w}} \pr[u|w;\xx1,\yy1]
    \pm \numround\indeperr & \text{(By \equationref{independence}.)} \\
&= \sum_{w} \pr[w|\xx0,\yy0]\sum_{\substack{u\in\highSXX \\ \Aparent(u)=w}} \pr[u|w;\xx1,\yy1]
    \pm 2\numround\simerr \pm \numround\indeperr & \text{(By \equationref{simerrbound2}.)} \\
\end{align*}

The above
expressions for $\pr[\highSXX|\xx0,\yy0]$ and $\pr[\highSXX|\xx1,\yy0]$,
combined with \equationref{atFX} and \equationref{aboveFY}
let us relate their difference to $\pr[\highSXX|\xx0,\yy1]$, as follows.
\begin{align*}
\pr[\highSXX|\xx0,\yy0] &- \pr[\highSXX|\xx1,\yy0]  \\
&\ge \sum_{w} \pr[w|\xx0,\yy0]\sum_{\substack{u\in\highSXX \\ \Aparent(u)=w}} \left( \pr[u|w;\xx0,\yy1] - \pr[u|w;\xx1,\yy1] \right) - 2\numround(\simerr+\indeperr) \\
&\ge \left(\frac{\deltax}{1+\deltax}\right) \sum_{w} \pr[w|\xx0,\yy0]\sum_{\substack{u\in\highSXX \\ \Aparent(u)=w}} \pr[u|w;\xx0,\yy1] - 2\numround(\simerr+\indeperr) \\
&\ge \left(\frac{\deltax}{(1+\deltax)(1+\delta)^{\numround}}\right) \sum_{w} \pr[w|\xx0,\yy1]\sum_{\substack{u\in\highSXX \\ \Aparent(u)=w}} \pr[u|w;\xx0,\yy1] - 2\numround(\simerr+\indeperr) \\
&\ge \left(\frac{\deltax}{(1+\deltax)(1+\delta)^{\numround}}\right) \pr[\highSXX|\xx0,\yy1] - 2\numround(\simerr+\indeperr)
\end{align*}

\paragraph{Part 3.}

We shall consider an attack when the protocol is run with inputs
$(\xx0,\yy0)$ or $(\xx1,\yy0)$ (which must be indistinguishable for
security).
We shall show that if $\pr[\highRXX|\xx0,\yy1]$ is significant, then the curious Bob's output is
significantly correlated with Alice's input $x$ (biased more towards 0 when
$x=\xx0$). This will contradict the security of the protocol, since in the
ideal world, Bob's input \yy0 cannot distinguish between Alice's input being
\xx0 or \xx1.

The probability that the execution with input $(\xx0,\yy0)$ reaches a node
$w=\Aparent(u)$ for $u\in\highRXX$ is significant if this
probability is significant in the execution with input $(\xx0,\yy1)$, since
each such $w$ falls above the \FYnomin frontier, and replacing \yy1 with \yy0 causes only
a constant factor change in the probabilities.  In \figureref{curious-bob-algo} we
describe a {\em curious Bob} who can, at such a point, mentally substitute its input
\yy0 with \yy1 and simulate the augmented execution (inluding -- and
this is the non-trivial part -- the answers
from the oracle) till before the next Alice message. The probability that
this simulated execution goes through \highRXX remains significant when
Alice's input is \xx0 (since the simulated execution will have input
$(\xx0,\yy1)$). At the same time, the probability of the execution with
$(\xx1,\yy1)$ hitting each node in \highRXX differs by a significant {\em
factor} from that when Alice's input is \xx0 (\equationref{atFX}).
This will let the curious Bob distinguish between when Alice's input is \xx0
and when it is \xx1, even though Bob's real input is \yy0, leading to a
contradiction.

\setlength{\fboxsep}{10pt}
\begin{figure}
\begin{boxedminipage}{\textwidth}
\vspace{1em}
\begin{center}
{\bf \large{Curious Bob: Learning what Eve learns, with a different input}}
\end{center}

\begin{wrapfigure}{r}{0.45\textwidth}
\centering
\vspace{1em}
\begin{tikzpicture}[scale=0.60]
\def\eve{(0,0) circle (2cm)}
\def\alice{(-2cm,0) ellipse (3cm and 2cm)}
\def\bob{(3cm,0cm) [rotate=30] ellipse (3cm and 2cm)}
\def\bobh{(3cm,0cm) [rotate=-30] ellipse (3cm and 2cm)}
\begin{scope}[even odd rule]
\fill[green!35] \bob;
\end{scope}
\begin{scope}[even odd rule]
\fill[gray!30] \alice;
\end{scope}
\begin{scope}[even odd rule]
\fill[orange!35] \bobh;
\end{scope}
\fill [blue!30] \eve;
\draw\eve;
\draw\alice;
\draw\bob;
\draw\bobh;
\draw (3cm,2.75cm) node {${\VB{\yy0}w}$}
      (3cm,-2.75cm) node {${\VB{\yy1}w}$}
      (-3.5cm,0) node {${\VA{x}{w}}$}
      (-1,0) node (E) {${\VE{w}}$};
\end{tikzpicture}
\caption{Simulating the oracle answers during exploration.
The ovals represent the sets of queries in the views 
\VA{x}{w}, \VE{w}, \VB{\yy0}{w} and \VB{\yy1}{w}.
Queries already answered in $\VE{w}$ (blue)
or in the hypothetical Bob view
$\VB{\yy1}{w}$ (orange) are
answered according to these views.
Answers to the remaining queries in $\cQ(\VB{\yy0}w)$ (green), are freshly sampled, i.e. answered according to $\cO''$.
All other queries are answered using the actual random oracle $\cO$.
When the ``safety'' condition \equationref{safety} holds, 
i.e., the orange and green regions (which have ``edited'' answers) do not intersect
the gray region, this yields
a perfect simulation (see \equationref{compare}).}
\label{fig:explore}
\end{wrapfigure}
Bob is given \yy0 as input, and Alice is given a uniformly random element
from $x\from\{\xx0,\xx1\}$ as input. Alice and Bob execute the protocol
honestly, with access to a random oracle \cO. But at the end Bob carries out the following computation.  

For every Alice node $w$ in the augmented transcript, which is strictly above
\FYnomin, Bob carries out an {\em exploration} as follows.
He samples a view $\VB{\yy1}{w}$ for himself with input \yy1, conditioned on node $w$ 
(and in particular Eve's view $\VE{w}$).
Bob mentally carries out the execution with the
hypothetical view \VB{\yy1}{w}, 
till the
next message from Alice (i.e., Eve queries, followed by Bob's own queries
and his message in the protocol, and then further Eve queries) by simulating
an oracle $\cO'$ defined as follows. 
Below, $\VB{\yy0}{w}$ denotes the actual view of Bob in the protocol at that point,
\cO is the actual oracle and $\cO''$ is a freshly sampled independent random
oracle.
On query $q$,
\begin{itemize}
\item if $q\in \Q(\VB{\yy1}{w}) ~\cup~ \Q(\VE{w})$, answer according to
$\VB{\yy1}{w}$ or $\VE{w}$;%
\footnote{As $\VB{\yy1}w$ is conditioned on $\VE{w}$, if 
$q\in \Q(\VB{\yy1}{w}) \cap \Q(\VE{w})$, both views will have
the same answer for $q$.}
\item else, if $q \in \Q(\VB{\yy0}{w})$, answer according to $\cO''$;
\item else, answer according to $\cO$.
\end{itemize}

Let the set of nodes encountered by Bob during this
exploration (over explorations from every Alice node $w$) be
\expNodes{x}, where $x$ is Alice's input, and Bob substitutes
\yy0 with \yy1 for exploration.  
If $\expNodes{x} \cap \highRXX \not= \emptyset$, then Bob outputs 0; else he outputs 1.
\end{boxedminipage}
\caption{Curious Bob strategy to show that $\pr[\highRXX|\xx0,\yy1]$ is
small.}
\label{fig:curious-bob-algo}
\end{figure}

Before we prove this,
we define a game and state a sub-claim,
which will help us with the analysis (for which we will
derive yet another game based on this).

\paragraph{Game $\game{x}{y}{y'}$.}
An oracle \cO, and random tapes for Alice
and Bob are picked at random.  
Then, Alice and Bob execute the protocol \prot using oracle \cO and the chosen random tapes, with inputs $x$ and $y$ respectively; at each node $w\in\Achild$ in the transcript path, we define $\VA{x}{w}$, $\VB{y}{w}$ and $\VE{w}$ as the views of Alice, Bob and Eve respectively.
Further, at each such node $w$ that is strictly above the frontier \FYnomin
(i.e., $\nexists z \in \FYnomin$, $z\ancest w$.), we pick a random view for
Bob conditioned on $w$ and input $y'$. 
Let $\VB{y'}{w}$ represent that Bob view.

We define the event $\safe{w}$ to occur in this game for a node $w$ where $\VB{y'}{w}$ is sampled (i.e., child of an Alice node that is strictly above \FYnomin), if 
\begin{align}
\Q\left(\VA{x}{w}\right) ~\cap~ \left(\Q(\VB{y}{w}) ~\cup~ \Q(\VB{y'}{w})\right) ~\subseteq~ \Q(\VE{w})
\label{eq:safety}
\end{align}

\begin{claim}
\label{clm:safe}
For any $x\in\cX$, $y,y'\in\cY$, in the game $\game{x}{y}{y'}$,
for an error parameter $\safeerr = \eps^{\Omega(1)}\poly(\spar|\cX||\cY|)$, we have
\begin{align}
\sum_w \pr[w|x,y'] \pcompare{\unsafe{w}|w}{x} \le \numround\safeerr \label{eq:safeerrbound}
\end{align}
(where the summation is over $w$ for which \safe{w} is defined: i.e., $w\in\Achild$ such that
$w\strictancest\FYnomin$).
\end{claim}

\begin{proof}
This follows from \lemmaref{NoCol}.
At every level $L_i$, \lemmaref{NoCol} guarantees
that $\sum_{\substack{w\in L_i \\ \pr[y|w;x]\ge\eps'}} \pr[w \wedge \unsafe{W}] \le \eps'$
for
$\eps' = \eps^{\Omega(1)}\poly(\spar|\cX||\cY|)$.
For $w\strictancest\FYnomin$, we have
$\pr[y|w;x]\ge\frac1{(1+\delta)^\numround|\cY|}> \eps'$.
Hence the sum in the claim is bounded by $\numround \eps'$.
We set $\safeerr=\eps'$.
\end{proof}

\paragraph{Two Experiments: \exptcurious{x} and \exptcompare{x}.}
Now, we define two experiments \exptcurious{x} and \exptcompare{x} as follows:

\exptcurious{x} (which corresponds to the curious attack above) is the same as
\game{x}{\yy0}{\yy1}, but with the following addition. At each node
$w\in\Achild$
above \FYnomin in the transcript path, we carry out an ``exploration'' of Eve's
steps and Bob's step till the next message from Alice (Eve, Bob, Eve), using
the view sampled for \yy1. This exploration is carried out as defined above
for the curious Bob strategy (\figureref{curious-bob-algo}).  For simplifying notation, we make the
following definition. For  a node $u$
and $w=\Aparent(u)$, we define the probability of the {\em exploration} starting
at $w$ visiting $u$ as
\[
\pcurious{u|w}x=
\Pr_{\exptcurious{x}}[u \text{ reached in exploration from } w | w \text{ reached in execution}].
\]
We also define $\pcurious{u}x = \pr[w|x,\yy0] \pcurious{u|w}x$ to be the
probability of the exploration reaching $u\in\highRXX$ (not conditioned on
visiting $w=\Aparent(u)$).

Note that in the right-hand side of the equation, we have $\pr[w|x,\yy0]$, i.e. the node $w$ is generated with Alice interacting with her input $x$ and Bob with his input $\yy0$.
After reaching $w$, Bob samples a new view conditioned on his input being $\yy1$ and proceed to explore till Alice is supposed to send the next message.
This part of the probability, i.e. probability of reaching a node $u$ {\em conditioned} on reaching $w$ is expressed by the term $\pcurious{u|w}x$.
We point out that $\pcurious{u|w}x$ is {\em not necessarily equal to} $\pr[u|w,x,\yy1]$
since the exploration uses a simulated oracle that is simulated without knowing
$x$. (However, as we shall see, it will be closely related to the latter.)

\exptcompare{x} is in fact, the same as \game{x}{\yy1}{\yy0} (note the reversal of
roles for \yy0 and \yy1): i.e., an execution with inputs $(x,\yy1)$, along
with sampling Bob's view for input \yy0 at each node $w\in\Achild$ encountered that is
strictly above \FYnomin. This experiment involves no
exploration. Now, for  a node $u$ 
and $w=\Aparent(u)$, we
define the probability of the execution visiting $u$,
conditioned on it having visited $w$, as
\[
\pcompare{u|w}x=
\Pr_{\exptcompare{x}}[u \text{ reached in the execution } | w \text{ reached in execution}].
\]
We also define $\pcompare{u}x = \pr[w|x,\yy1] \pcompare{u|w}x$ to be the
probability of the execution reaching $u$ 
(not conditioned on visiting $w=\Aparent(u)$).
Note that in this experiment the only significance of $\yy0$ is in defining the event $\safe{w}$.
In particular, there is no exploration phase or switching of inputs, and the
execution considered for defining the probability 
$\pcompare{u|w}{x}$ is simply the same as a faithful execution of the
original augmented protocol. Thus, $\pcompare{u|w}{x} = \pr[u|w,x,\yy1]$.

A priori, there is no direct relation between the probability terms $\pcurious{u|w}x$ and $\pcompare{u|w}{x}$.
This is because the sampling of the Bob view in \exptcurious{x} is not correlated with the view of Alice given Eve's view; while, on the other hand, Bob's view in \exptcompare{x} could possibly be correlated with Alice's view even when Eve view is given.
But, by additionally conditioning on the event $\safe{w}$, these two probabilities are identical.
More formally, we have the following key observation:%
\footnote{\label{foot:switchycomment}We shall use this claim for $w$
strictly above
\FYnomin. It can be seen that if only one of $\pcurious{w,\safe{w}}x$ and
$\pcompare{w,\safe{w}}x$ is positive, then by the convention in
\footnoteref{prconvention}, the node $w$ cannot be
strictly above \FYnomin. Hence the
claim will be applicable.  Alternately, similar to the normal form for
protocols mentioned in \footnoteref{prconvention}, we can assume w.l.o.g
that for all $w\in\Achild$, $\pcurious{w,\safe{w}}x>0$ and
$\pcompare{w,\safe{w}}x>0$, so that the claim holds for all $w$.}
For all $x$, for all $u,w$ such that $w=\Aparent(u)$,
$\pcurious{w,\safe{w}}x>0$ and $\pcompare{w,\safe{w}}x>0$,
\begin{align}
\label{eq:compare}
	\pcurious{u|w,\safe{w}}x = \pcompare{u|w,\safe{w}}x.
\end{align}
This is because, given a node $w$, in either experiment, the set of Alice
views, the set of Bob views with input \yy0 and the set of Bob views with
input \yy1 each compatible with the view in $w$ (individually) are
determined. On conditioning on \safe{w}, the distribution over triplets of
views (one from each of the three sets) is the same in both experiments:
they correspond to pairs of edges in the ``views graph'' at $w$, with both
edges incident on the same Alice view, and the probability of a pair is
(before conditioning) product of the probabilities on the two edges
(according to distributions obtained by conditioning on \yy0 and \yy1), and
the conditioning removes all those pairs of edges that violate the safety
condition; these operations (multiplication and safety condition) are
symmetric in \yy0,\yy1 and hence, both the distributions are the same. Now,
conditioned on \safe{w}, the exploration in \exptcurious{x} for a triplet of
views is identical to the execution in \exptcompare{x} for the same triplet.

Assuming that $\pr[\highRXX|\xx0,\yy1]$ is significant, we are interested in lower-bounding
$\pcurious{\highRXX}{\xx0} - \pcurious{\highRXX}{\xx1}$.

For $x\in\{\xx0,\xx1\}$, we have:
\begin{align*}
\pcurious{\highRXX}{x} = \sum_{u\in\highRXX} \pcurious{u}{x}
 = \sum_w \left( \pr[w|x,\yy0] \sum_{\substack{u\in\highRXX,\\ \Aparent(u)=w}} \pcurious{u|w}x \right)
\end{align*}
Note that the last summation will be over $w\in\Achild$ that are strictly
above \FYnomin, since we consider only those
$w$ for which there exists some $u\in\highRXX$ with $\Aparent(u)=w$.

Fix a node $w$ and consider $u\in\highRXX$ such that $\Aparent(u)=w$. Then,
(using the convention in \footnoteref{prconvention}),
\begin{align*}
\pcurious{u|w}x &= \pcurious{u,\safe{w}|w}x + \pcurious{u,\unsafe{w}|w}x \\
&= \pcurious{u|w,\safe{w}}x\pcurious{\safe{w}|w}x + \pcurious{u,\unsafe{w}|w}x \\
&= \pcurious{u|w,\safe{w}}x - \pcurious{u|w,\safe{w}}x \pcurious{\unsafe{w}|w}x + \pcurious{u,\unsafe{w}|w}x \\
&= \pcurious{u|w,\safe{w}}x + (\pcurious{u|w,\unsafe{w}}x - \pcurious{u|w,\safe{w}}x) \pcurious{\unsafe{w}|w}x
\end{align*}
The sums
$\sum_{u\in\highRXX, \Aparent(u)=w} \pcurious{u|w,\unsafe{w}}x \pcurious{\unsafe{w}|w}x$ and
$\sum_{u\in\highRXX, \Aparent(u)=w} \pcurious{u|w,\safe{w}}x \pcurious{\unsafe{w}|w}x$ are
both bounded by $\pcurious{\unsafe{w}|w}x$. Thus we can write
\begin{align}
\pcurious{\highRXX}{x} & = \sum_w \bigg( \pr[w|x,\yy0] \sum_{\substack{u\in\highRXX,\\ \Aparent(u)=w}} \pcurious{u|w}x \bigg)\nonumber\\
&= \sum_w \bigg( \pr[w|x,\yy0] \sum_{\substack{u\in\highRXX,\\ \Aparent(u)=w}} \pcurious{u|w,\safe{w}}x  \bigg)
   \pm \sum_w \pr[w|x,\yy0]  \pcurious{\unsafe{w}|w}x \nonumber\\
&= \sum_w \bigg( \pr[w|x,\yy0] \sum_{\substack{u\in\highRXX,\\ \Aparent(u)=w}} \pcurious{u|w,\safe{w}}x  \bigg)
   \pm \numround\safeerr \label{eq:FXcuriousbound}
\end{align}
where the last step follows by \claimref{safe}.
Note that $\pcompare{\highRXX}{x}=\pr[\highRXX|x,\yy1]$.

In our derivation below, we shall rely on conditioning the experiments
\exptcurious{x} and \exptcompare{x} on the event \safe{\cdot}. To facilitate our
arguments we relate certain probabilities when conditioned on \safe\cdot and
otherwise.

\begin{claim}
\label{clm:safeuncondition}
The following two inequalities hold:
\begin{align}
\sum_w \pr[w|\xx0,\yy1] \sum_{\substack{u\in\highRXX,\\ \Aparent(u)=w}} \pcompare{u|w,\safe{w}}{\xx0}
&= \bigg( \sum_w \pr[w|\xx0,\yy1] \sum_{\substack{u\in\highRXX,\\ \Aparent(u)=w}} \pcompare{u|w}{\xx0} \bigg)
    \pm \numround\safeerr \label{eq:xx0safeuncondition} \\
\sum_w \pr[w|\xx0,\yy1]  \sum_{\substack{u\in\highRXX,\\ \Aparent(u)=w}} \pcompare{u|w,\safe{w}}{\xx1}
&= \bigg( \sum_w \pr[w|\xx0,\yy1] \sum_{\substack{u\in\highRXX,\\ \Aparent(u)=w}} \pcompare{u|w}{\xx1} \bigg)
        \pm (1+\delta)^\numround(\numround\safeerr + 2\numround\simerr). \label{eq:mixedx-xx1safeuncondition}
\end{align}
\end{claim}

\begin{proof}
Firstly,
\begin{align}
\sum_{\substack{u\in\highRXX,\\ \Aparent(u)=w}} \pcompare{u|w,\safe{w}}{x}
&= \bigg(\sum_{\substack{u\in\highRXX,\\ \Aparent(u)=w}} \pcompare{u|w}{x} \bigg)
    \pm \pcompare{\unsafe{w}|w}x \label{eq:safeunconditionbound}
\end{align}

We get \equationref{xx0safeuncondition} as follows:
\begin{align*}
\sum_w \pr[w|&\xx0,\yy1] \sum_{\substack{u\in\highRXX,\\ \Aparent(u)=w}} \pcompare{u|w,\safe{w}}{\xx0} \\
&= \bigg( \sum_w \pr[w|\xx0,\yy1]
        \sum_{\substack{u\in\highRXX,\\ \Aparent(u)=w}} \pcompare{u|w}{\xx0} \bigg)
        \pm \bigg( \sum_w \pr[w|\xx0,\yy1] \pcompare{\unsafe{w}|w}{\xx0} \bigg) &\text{By \equationref{safeunconditionbound}.}\nonumber\\
&= \bigg( \sum_w \pr[w|\xx0,\yy1]
        \sum_{\substack{u\in\highRXX,\\ \Aparent(u)=w}} \pcompare{u|w}{\xx0} \bigg)
    \pm \numround\safeerr  &\text{By \equationref{safeerrbound}.}
\end{align*}

To prove \equationref{mixedx-xx1safeuncondition}, first we note the
following:
\begin{align}
\sum_w \pr[w|\xx0,\yy0] \pcompare{\unsafe{w}|w}{\xx1}
&\le \sum_w \pr[w|\xx1,\yy0] \pcompare{\unsafe{w}|w}{\xx1} + 2\numround\simerr &\text{By \equationref{simerrbound2}.}
\nonumber \\
&\le \numround\safeerr + 2\numround\simerr &\text{By \equationref{safeerrbound}.}
\label{eq:mixedx-yy0-unsafebound}
\end{align}

\begin{align}
\sum_w \pr[w|\xx0,\yy1] \pcompare{\unsafe{w}|w}{\xx1}
&\le (1+\delta)^\numround \sum_w \pr[w|\xx0,\yy0] \pcompare{\unsafe{w}|w}{\xx1}
\nonumber \\
&\le (1+\delta)^\numround (\numround\safeerr + 2\numround\simerr) &\text{By \equationref{mixedx-yy0-unsafebound}.}
\label{eq:mixedx-yy1-unsafebound}
\end{align}

Hence, we conclude
\begin{align*}
\sum_w \pr[w|\xx0,\yy1] & \sum_{\substack{u\in\highRXX,\\ \Aparent(u)=w}} \pcompare{u|w,\safe{w}}{\xx1} \\
&= \sum_w \bigg( \pr[w|\xx0,\yy1]
        \sum_{\substack{u\in\highRXX,\\ \Aparent(u)=w}} \pcompare{u|w}{\xx1} \bigg)
        \pm \sum_w \pr[w|\xx0,\yy1] \pcompare{\unsafe{w}|w}{\xx1} &\text{By \equationref{safeunconditionbound}.}\nonumber\\
&= \sum_w \pr[w|\xx0,\yy1] \bigg( \sum_{\substack{u\in\highRXX,\\ \Aparent(u)=w}} \pcompare{u|w}{\xx1} \bigg)
        \pm (1+\delta)^\numround (\numround\safeerr + 2\numround\simerr) &\text{By \equationref{mixedx-yy1-unsafebound}.} &\qedhere
\end{align*}
\end{proof} %

To lower bound $\pcurious{\highRXX}{\xx0}-\pcurious{\highRXX}{\xx1}$ we
proceed as follows:

{
\allowdisplaybreaks
\begin{align*}
&\pcurious{\highRXX}{\xx0} - \pcurious{\highRXX}{\xx1} \nonumber \\
&\ge \sum_w \Bigg[  \pr[w|\xx0,\yy0] \Bigg( \sum_{\substack{u\in\highRXX,\\ \Aparent(u)=w}} \pcurious{u|w,\safe{w}}{\xx0}  \Bigg)
\nonumber \\ & \qquad
  -  \pr[w|\xx1,\yy0] \Bigg( \sum_{\substack{u\in\highRXX,\\ \Aparent(u)=w}} \pcurious{u|w,\safe{w}}{\xx1}  \Bigg) \Bigg]
 -2\numround\safeerr  & \text{By \equationref{FXcuriousbound}.}\\
&= \sum_w \Bigg[  \pr[w|\xx0,\yy0] \Bigg( \sum_{\substack{u\in\highRXX,\\ \Aparent(u)=w}} \pcompare{u|w,\safe{w}}{\xx0}  \Bigg)
\nonumber \\ & \qquad
  -  \pr[w|\xx1,\yy0] \Bigg( \sum_{\substack{u\in\highRXX,\\ \Aparent(u)=w}} \pcompare{u|w,\safe{w}}{\xx1}  \Bigg) \Bigg]
 -2\numround\safeerr  & \text{By \equationref{compare} (and \footnoteref{switchycomment}).}\\
&\ge \sum_w \Bigg( \pr[w|\xx0,\yy0]
        \sum_{\substack{u\in\highRXX,\\ \Aparent(u)=w}} \Bigg(
            \pcompare{u|w,\safe{w}}{\xx0} - \pcompare{u|w,\safe{w}}{\xx1}
        \Bigg)
    \Bigg)
\nonumber \\ & \qquad
    -2\numround(\simerr+\safeerr) &\text{By \equationref{simerrbound2}.} \\
&\ge \sum_w \Bigg( \frac{\pr[w|\xx0,\yy1]}{(1+\delta)^\numround}
      \sum_{\substack{u\in\highRXX,\\ \Aparent(u)=w}}
        \Bigg( \pcompare{u|w,\safe{w}}{\xx0} - \pcompare{u|w,\safe{w}}{\xx1} \Bigg)
      \Bigg)
\nonumber \\ & \qquad
    -2\numround(\simerr+\safeerr) &\text{By \equationref{aboveFY}.} \\
&\ge \sum_w \Bigg( \frac{\pr[w|\xx0,\yy1]}{(1+\delta)^\numround}
        \sum_{\substack{u\in\highRXX,\\ \Aparent(u)=w}} \Bigg(
            \pcompare{u|w}{\xx0} - \pcompare{u|w}{\xx1}
        \Bigg)
    \Bigg)
\nonumber \\ & \qquad
    -2\numround(\simerr+\safeerr) -\frac{\numround\safeerr}{(1+\delta)^{\numround}} - (2\numround\simerr+\numround\safeerr) &\text{By \equationref{xx0safeuncondition} and \equationref{mixedx-xx1safeuncondition}.} \\
&= \sum_w \Bigg( \frac{\pr[w|\xx0,\yy1]}{(1+\delta)^\numround}
        \sum_{\substack{u\in\highRXX,\\ \Aparent(u)=w}} \Bigg(
            \pr[u|w;\xx0,\yy1] - \pr[u|w;\xx1,\yy1]
        \Bigg)
    \Bigg)
\nonumber \\ & \qquad
    -2\numround(\simerr+\safeerr) -\frac{\numround\safeerr}{(1+\delta)^{\numround}} - (2\numround\simerr+\numround\safeerr) &\text{By definition of \pcompare{u|w}{x}.} \\
&\ge \sum_w \Bigg( \frac{\pr[w|\xx0,\yy1]}{(1+\delta)^\numround}
        \sum_{\substack{u\in\highRXX,\\ \Aparent(u)=w}}
        \Big(\frac{\deltax}{1+\deltax}\Big) \pr[u|w;\xx0,\yy1]
    \Bigg)
    -4\numround(\simerr+\safeerr) &\text{By \equationref{atFX}.} \\
&\ge \frac{\deltax}{(1+\deltax)(1+\delta)^{\numround}} \pr[\highRXX|\xx0,\yy1]
    -4\numround(\simerr+\safeerr)
\end{align*}
}

\paragraph{Putting things Together}
Let us define $\mu=(1+\delta)^\numround$ and recall that $\deltax = (1+\delta)^{1/(|\cX|-1)}-1$.
From Part 2 and 3, we obtain a lower-bound on the distinguishing advantage obtained in terms of $\pr[\highSXX|\xx0,\yy1]$ and $\pr[\highRXX|\xx0,\yy1]$.
We can assume that this advantages are $\distadv_S$ and $\distadv_R$ respectively.
But we know that simulation error is $\simerr$, so $\distadv_S + \distadv_R \leq 2\simerr$.
Thus, we obtain the following bounds:
\begin{align*}
\pr[\highSXX|\xx0,\yy1] &\leq \frac{(1+\deltax)\mu}{\deltax} \left(\distadv_S + 2\numround(\simerr + \indeperr)\right)\\
\pr[\highRXX|\xx0,\yy1] &\leq \frac{(1+\deltax)\mu}{\deltax} \left(\distadv_R + 4\numround(\simerr + \safeerr)\right)
\end{align*}

Finally, we can obtain a bound on the overall bad event $\pr[\highFX]$:
\begin{align*}
\pr[\highFX] &\leq |\cX|^2|\cY|^2\pr[\highFXX] \\
&\leq \frac{\mu|\cX|^2|\cY|}{\minwt} \pr[\highFXX|\xx0,\yy1]\\
&= \frac{\mu|\cX|^2|\cY|}{\minwt} \left( \pr[\highSXX|\xx0,\yy1] + \pr[\highRXX|\xx0,\yy1] \right)\\
&\leq \frac{2\mu^2|\cX|^2|\cY|(1+\deltax)\numround}{\minwt\deltax} \left( 4\simerr + \indeperr + 2\safeerr \right)
\end{align*}

This completes the proof of \claimref{frontierattack}, and in turn that of
\claimref{frontierorder}. As discussed in \sectionref{detailed-pf-main},
this (combined with \claimref{frontierfull}), is used to prove
\theoremref{main}.

\section{Beyond Semi-Honest Security}
\label{sec:detailed-RO-Fcom}
In this section we prove \theoremref{RO-Fcom}, which tells us that in the
context of building 2-party SFE protocols secure against active adversaries,
a random oracle is only useful as a means for securely realizing the
commitment functionality, denoted by \Fcom. This holds true for both UC and
standalone security.

\begin{restatethm}{\theoremref{RO-Fcom}}
For a deterministic finite 2-party function $f$, the following statements are equivalent:
\begin{compactenum}
\item $f$ has a statistically UC-secure SFE protocol in the random oracle model.
\item $f$ has a statistically standalone-secure SFE protocol in the random oracle model.
\item $f$ has a statistically UC-secure SFE protocol in the \Fcom-hybrid model.
\item $f$ has a statistically standalone-secure SFE protocol in the \Fcom-hybrid model.
\end{compactenum}
\end{restatethm}
\begin{proof}
Clearly, $(1)\implies (2)$ and $(3)\implies(4)$. That $(3)\implies(1)$
and $(4)\implies(2)$ follow from the fact that in the random oracle
model, we can UC-securely implement the \Fcom functionality.  (This
implication holds not only for deterministic SFE, but also for reactive or randomized
functionalities as well.) 

To complete the proof we shall show that $(2)\implies(3)$.  So suppose $f$
has a standalone secure protocol using a random oracle.  Let $f'$ be a
redundancy free function obtained by removing redundant inputs one by one
from $f$ (see \footnoteref{redundant}). Then, it is enough to show
$(2')\implies(3')$ where $(2')$ and $(3')$ are identical to $(2)$ and $(3)$,
but with $f$ replaced by $f'$ (because, $(2)\iff(2')$ and
$(3)\iff(3')$ \cite{MajiPrRo10uc,KraschewskiMu11}).  Now, if $f'$
has a standalone secure protocol in the \Fcom-hybrid model, then the same
protocol is semi-honest secure as well. Further, by replacing \Fcom by a
trivial protocol for commitment with semi-honest security, we obtain a
semi-honest secure protocol for $f'$ in the plain model.  Then, by
\claimref{isomorphic-to-symmetric}, $f'$ must be isomorphic to a symmetric
function $f''$ which has a semi-honest secure protocol in the plain model.
That is $f''$ must be decomposable. Then, by a result in \cite{MajiPrRo09},
$f''$ has a UC secure protocol in the \Fcom-hybrid model.  Since $f''$ is
isomorphic to $f'$, the latter also has UC secure protocol in the
\Fcom-hybrid model, proving $(3')$ as desired.
\end{proof}

\section{Black-Box Separations}
\label{sec:blackbox}

The random oracle model is of interest not only as an abstract theoretical framework,
but also because it models a (strong) one-way function. Thus, informally, the
impossibility results in the random oracle model translate to impossibility
of constructions that rely on a one-way function as its sole computational
primitive. This intuition can be formalized as black-box separation results,
following \cite{ImpagliazzoRu89,ReingoldTrVa04}.

For our black-box separation results, we shall follow the definitions as
introduced by \cite{ReingoldTrVa04} with minor modifications.  Following
\cite{ReingoldTrVa04}, we consider primitives to be specified as pairs of
the form $(F_\cQ, R_\cQ)$.  The set $F_\cQ$ is a set of functions that are
candidate implementations of primitive $\cQ$.  For example, for the one-way
function primitive (represented by $\OWF$) the set $F_\OWF$ consists of all
functions defined over $\zo^*$.  The set $R_\cQ$ is a set of pairs $(Q,M)$,
where $Q$ is a candidate implementation of $\cQ$ and $M$ is an adversary
which breaks the security of $Q$.  (Sometimes we shall abuse the notation
and write $(\Pi,M)\in R_\cQ$ if $\Pi$ implements a function $Q$ such that
$(Q,M)\in R_\cQ$.) Continuing our example of \OWF, $(Q,M)$ would consist of
one-way functions $Q$ where the inverter $M$ inverts non-negligible fraction
of outputs of $Q$.

Next, we recall the definition of fully black-box reductions (or as 
presented below, fully black-box constructions) as introduced in \cite{ReingoldTrVa04}.
Below, we say that a (possibly non-uniform) algorithm is {\em efficient} if it is probabilistic
polynomial time (PPT).
\begin{definition}[Fully Black-box Constructions]
\label{def:fbb}
A {\em fully black-box construction} of a primitive $\cP$ from another
primitive $\cQ$ consists of a pair of efficient oracle algorithms $(\prot,S)$,
such that the following two conditions hold:
\begin{enumerate}
\item {\bf Correct Implementation:} For any $Q\in F_\cQ$, $\prot^Q$
		implements a function $P\in F_\cP$.
\item {\bf Security:} For any $Q\in F_\cQ$ and any (possibly inefficient)
		adversary $A$ that breaks the security of $\prot^Q$, the reduction
		$S^{Q,A}$ breaks the security of $Q$ as an implementation of $\cQ$.
		That is, $\forall A, \forall Q\in F_\cQ$, $(\prot^Q,A)\in R_\cP \implies
		(Q,S^{Q,A})\in R_\cQ$.
\end{enumerate}
\end{definition}
We emphasize that the construction $\prot$ and the reduction $S$ are
efficient. 

\paragraph{Constructions that Preserve the Security Parameter.} 
As is standard in cryptographic constructions, we shall associate a security
parameter with primitives and state security condition in terms of it.
Formally, we shall consider that any primitive $\cP$, the input to any $P\in
F_\cP$ has a security parameter encoded as part of its inputs. 
We prove our separation results with a technical restriction on
blackbox constructions, namely that
the constructions respect the security parameter: that is, in a
black-box construction of $\cP$ from $\cQ$, when the implementation
$\prot^Q$, for $Q\in F_\cQ$, is given an input with security parameter
$\spar$, it always invokes $Q$ with the same security parameter $\spar$.
However, there is no such restriction on the security reduction $S$.

For $Q\in F_\cQ$, we denote by $Q_i$ the restriction of $Q$ to inputs which
have security parameter $i$. We will often identify $Q$ with the infinite tuple
$(Q_1,Q_2,\dotsc)$. For a security parameter respecting construction
$(\prot,S)$, when invoked with security parameter $\spar$, $\prot^Q$ will
access only $Q_\spar$.  There is no such restriction of the security
reduction $S$. When invoked with security parameter $\spar$, $S^{Q,A}$ (for
an adversary $A$ attacking $\prot$) is expected to invert points in the
range of $Q_\spar$.  To perform this inversion,  $S^{Q,A}$ is permitted
access $Q_{\spar'}$, for all values of $\spar'\in\N$ (including $\spar'\neq
\spar$), and in particular can invoke $\prot^Q$ and $A^Q$ with different
security parameters $\spar'$.

This restriction is not as limiting as it may appear at first, since
we can define primitives like one-way function to allow access to
a range of input lengths for a single value of the security
parameter. (See the definition of \OWFz below.)

Below we define the various primitives used to formalize our results.  The
primitives are formally specified by the $F$ and $R$ sets as mentioned
above. We shall specify the functions in $F$ separately for each value of
the security parameter.  We consider the machines $M$ in all the definitions
below as non-uniform machines (with non-uniform advice for each security
parameter); however, one could relax the security definition of any of the
primitives to consider only uniform $M$, and by requiring the fully black-box
construction to also be uniform, our results hold unchanged.

\paragraph{One-Way Function Primitive \OWF.} First, for simplicity, we consider
a one-way function primitive \OWF which considers the security parameter as
the input length itself.%
\footnote{This is the same one-way function primitive as considered
in \cite{ReingoldTrVa04}. However, in the case of security parameter preserving
constructions, this primitive prevents the construction from using the
one-way function with any other input length other than the security
parameter. Later we remove this restriction by considering the primitives
$\OWFz$ defined below.}
\begin{compactitem}
\item $F_\OWF$ consists of all functions from $\zo^*$ to $\zo^*$, and the security
parameter is the length of the input.  
\item $(Q,M)\in R_\OWF$
if there is a non-negligible function $\delta$
such that for infinitely many $\spar\in \N$, $\Pr[Q(M(y)) = y : x \getsr
\zo^\spar, y=Q(x) ] > \delta(\spar)$.
\end{compactitem}

\paragraph{Primitive for Semi-honest Secure SFE Protocol.}
For a 2-party function $f\colon\cX\times\cY\To\cZ_A\times\cZ_B$, we define
the primitive \SFE{f} corresponding to a semi-honest secure protocol for
evaluating $f$. For simplicity, we consider the domain and range of $f$ 
itself to be finite and fixed (independent of the security parameter).%
\footnote{One could consider $f$ to have infinite domains and range, and
define restrictions of $f$,
$f_\spar:\cX_\spar\times\cY_\spar\To\cZ_{A,\spar}\times\cZ_{B,\spar}$, where
$\cX_1 \subseteq \cX_2 \subseteq \dotsi \cX$ etc., with efficient
representations for the subdomains and subrange.  Our results hold as long
as $|\cX_\spar|,|\cY_\spar| \le \poly(\spar)$. We omit such a formalization
for the sake of simplicity.}
A protocol $\prot$ will be identified with the next message
function of the protocol.  One of its inputs is the security parameter
\spar.
\begin{compactitem}
\item $\prot\in F_{\SFE{f}}$ if the protocol defined by $\prot$ is
``correct'', i.e. for all $(x,y)\in\cX\times\cY$, the pair of outputs from
Alice and Bob when they execute $\prot$ with security parameter \spar and
inputs $(x,y)$, is $(a,b)=f(x,y)$ except with probability negligible in
\spar.
\item An adversary \Adv breaks $\prot$, i.e. $(\prot, \Adv)\in
R_{\SFE{f}}$ if there exists $(x,x',y,y')$ such that
\begin{enumerate}
\item $f(x,y)=f(x,y')$ and $|\Pr[\Adv(\VVB\Pi xy)=1]-\Pr[\Adv(\VVB\Pi x{y'})=1]| > \delta(\spar)$, or 
\item $f(x,y)=f(x',y)$ and $|\Pr[\Adv(\VVA\Pi xy)=1]-\Pr[\Adv(\VVA\Pi {x'}y)=1]|>\delta(\spar)$,
\end{enumerate}
where $\VVA\Pi xy$ and $\VVB\Pi xy$ stand for Alice's and Bob's views after
executing $\prot$ with inputs $(x,y)$ and the advantage $\delta(\spar)$ is
non-negligible in \spar.
\end{compactitem}
Note that we used a game-based definition of semi-honest security.
This is in general weaker than the standard simulation based definition of
semi-honest security  (unless simulation with unbounded computational
power is considered, in which case they are identical). Since we are ruling
out blackbox constructions of $\SFE{f}$, using a weaker definition of
security for $\SFE{f}$ makes our result only stronger.

\paragraph{One-Way Function Primitive $\OWFz$.} 
Since we consider only security-parameter preserving constructions, a
construction using the primitive \OWF above can access the one-way function
on inputs of length exactly equal to the security-parameter. This limits the
implications of a separation result, as it leaves open the possibility that
a construction that uses a one-way function on more than one input length
could be secure. To rule out this possibility as well, we consider a
more elaborate primitive and rule out fully black-box construction of
\SFE{f} from this primitive as well. Formally, we define a primitive $\OWFz$
for each polynomial $\zeta$ as follows.

For any function $g:\zo^*\To\zo^*$, let \gz be defined as follows:
$\gz(\spar,x) = g(x)$ if $|x|\le \zeta(\spar)$ and $\gz(\spar,x) = 0$
otherwise. Let $\cW^\zeta_i = \{ \gz(i,\cdot) | g:\zo^*\To\zo^*  \}$.
\begin{compactitem}
\item $F_{\OWFz} = \cW^\zeta_1 \times \cW^\zeta_2 \times \dotsb$.
That is, for $Q\in F_{\OWFz}$, $Q=(Q_1,Q_2,\dotsc)$, the function $Q_\spar$
is of the form $\gz(\spar,\cdot)$ for some function
$g$.
\item $(Q,M)\in R_\OWFz$ if there is a non-negligible function $\delta$
such that for infinitely many $\spar\in \N$, $\Pr[Q(\spar,M(\spar,y)) = y : x \getsr
\zo^1\cup\dotsb\cup\zo^{\zeta(\spar)}, y=Q(\spar,x) ] > \delta(\spar)$.
\end{compactitem}

\begin{theorem}
\label{thm:fully-BBsep-OWF}
For a deterministic two-party function $f$, \SFE{f} (semi-honest secure
protocol for $f$) the following statements are equivalent:
\begin{compactenum}
\item[$(1)$] $f$ has a perfectly semi-honest secure protocol (in the plain model).
\item[$(2)$] \SFE{f} has a security-parameter preserving fully black-box construction from \OWF.
\item[$(3)$] \SFE{f} has a security-parameter preserving fully black-box construction from \OWFz, 
			 for some polynomial $\zeta$.
\end{compactenum}
\end{theorem}
We prove this theorem in \appendixref{blackbox}.

\section{Open Problems and Future Work}
We have shown a black-box separation between one-way functions and
semi-honest SFE protocols for 2-party secure function evaluation for
any function which does not have already have a semi-honest SFE protocol in the plain model.
Intuitively, this introduces new worlds between ``minicrypt''
and ``cryptomania'' \cite{Impagliazzo95}, corresponding to where these
functions have semi-honest SFE protocols. There are several interesting
questions that this gives rise to. We mention a few directions below.

\begin{asparaenum}
\item Our result relies on the combinatorial characterization of
undecomposable function evaluations.  In particular, our strategy is not
able to ``compile out'' the random oracle completely in the context of
2-party deterministic semi-honest function evaluation, i.e., we are not able
to rule out that access to a random oracle could enable secure computation
(of say, a randomized functionality) that cannot be achieved by a protocol
in the plain model.  Understanding the precise power of random oracles in
the context of secure computation in its full generality (especially, for
randomized functions) remains open.

\item The separation of OT from one-way functions (implicit) in
\cite{ImpagliazzoRu89} was strengthened to separate OT from public-key
encryption in \cite{GertnerKaMaReVi00}. In on going work, we give
a similar strengthening of our results, separating every function
which does not have a semi-honest SFE protocol in the plain model
(undecomposable functions, among symmetric functions) from public-key
encryption. This, in particular, would give an
alternate proof for the result in \cite{GertnerKaMaReVi00}.

\item In this work we {\em do not} show that (semi-honest) SFE for the
various functions we separate from one-way functions really correspond to
{\em new} worlds in Impagliazzo's universe. In particular, we do not
separate them from the ``OT protocol'' primitive. Indeed, one could hope to
prove our current results by simply showing that SFE for all the functions
we considered can, in a fully black-box manner, yield an OT protocol. But we
conjecture that such a construction simply does not exist. We leave it open
to fully understand the relationship between the worlds corresponding to
(semi-honest) SFE protocols for the different functions, and in particular,
find out if there is an infinite hierarchy of such distinct worlds.

\end{asparaenum}

\newpage
\bibliographystyle{alpha}
\bibliography{bib,crossrefs} %
\addcontentsline{toc}{section}{References}

\newpage
\appendix
\section{Independence Learners}
\label{app:EvePlus}

\subsection{Some Notations}
Before, we proceed, we introduce some notations used in this section.
\subsubsection{Random Variables.}
We use bold letters to emphasize the nature of a random variable (\eg $\rvx$). 
By $\Supp(\bfx)$ we denote  $\set{x \mid \Pr[\rvx=x]>0}$. 
By $x \getsr \rvx$ we mean that $x$ is sampled according to the distribution of the random variable $\rvx$. 
We usually use the same letter  to denote a sample from a random variable. 
When we say an event occurs with negligible probability denoted by $\negl(\spar)$, we mean it occurs with probability $\spar^{-\omega(1)} $.
We call two random variables $\rvx, \rvy$ (or their corresponding distributions) $\eps$-close if their statistical distance, defined as
$ \SD(\rvx, \rvy) = \frac{1}{2} \cdot \sum_{s\in\Supp(\rvx)\cup\Supp(\rvy)} \adjAbs{ \Pr[\rvx = s] - \Pr[\rvy = s] } $ is at most $\eps$.
By $\rvx \equiv \rvy$ we denote that the random variables $\bfx$ and $\bfy$ are distributed identically.%

By $(\bfx_1,\bfx_2,\dots)$ we denote a (perhaps infinite) sequence of \emph{correlated} random variables where $\bfx_i$ is the random variable of the $i\th$ coordinate. 
For correlated random variables $(\bfx,\bfy)$, by $(\bfx \times \bfy)$ we refer to a new random variable that samples \emph{independent} copies for $\bfx$ and $\bfy$ (\ie sample two pairs $(x_1,y_1) \gets (\bfx,\bfy), (x_2,y_2) \getsr (\bfx,\bfy)$ and output $(x_1,y_2$)). 
For correlated random variables $(\bfx,\bfy)$ and $y \in \Supp(\bfy)$ by $(\bfx \mid \bfy=y)$ we denote the random variable $\bfx$ conditioned on $\bfy=y$. 
When it is clear from the context we simply write $(\bfx \mid y)$ instead of $(\bfx \mid \bfy=y)$.

\subsubsection{Two Party Protocols}
In the proofs in this section, we mostly analyze the 
protocols by rounds rather than frontiers. Hence it will be convenient to
introduce notation involving round numbers (rather than nodes in the
transcript tree). Below we describe notation associated with Alice;
similar notation is associated with Bob as well.
\begin{enumerate}

\item $\mu = (\mu_1,\mu_2,\dots)$ denotes the transcript generated during the interaction where the $i\th$ message is sent by Alice, if $i$ is odd, and it is sent by Bob, if $i$ is even. By $\mu\ii$ we denote $(\mu_1,\dots,\mu_i)$.

\item Let $P_A$ denote the set of oracle query-answer pairs obtained by
Alice from the oracle. By $P_A\ii$ we denote the set of query-answers
obtained by Alice before $\mu\ii$ is sent. $Q_A$ and $Q_A\ii$ are defined
similarly to $P_A$ and $P_A\ii$ while only containing the queries. Namely,
using the query-operator $\Q$ defined earlier it, holds that  $Q_A = \Q(P_A)$
and $Q_A\ii = \Q(P_A\ii)$.
\item $V_A$ denotes the view of Alice which is equal to $(x,r_A,P_A,\mu)$,
where $r_A$ denotes the private random-tape of Alice. By $V_A\ii$ we denote the view of Alice till the message $\mu\ii$ is sent which is equal to $(x,r_A,P_A\ii,\mu\iim)$
\end{enumerate}

\paragraph{A Public Query Strategy.}
For a 2-party protocol $\prot$ in the \cO model, we define a {\em public
query strategy} \Eve as a deterministic algorithm which takes as input a
prefix $\mu\ii$ of the messages of an execution of $\prot$ and a set $P_E\iim$
of query-answer pairs from \cO (standing for the queries that she has asked previously), and then adaptively queries $\cO$ multiple
times.  
The view of \Eve, denoted by $V_E$, is equal to $(\mu,P_E)$. We also define $V\iip = (P_E\iip, \mu  \ii)$ as the view of \Eve before $\mu_{i+1}$ is sent.

We define the query complexity of a public query strategy \Eve for a
protocol $\prot$ to be the maximum number of queries \Eve makes to \cO over
an entire augmented execution of $\prot$ and $\Eve$.

\paragraph{A Round.} For an odd $i$, the $i\th$ round starts right after the $(i-1)\st$ message is received by Alice and starts asking  its oracle queries (which are contained in $Q_A\ii$). When Alice sends $\mu_i$ the $i\th$  round continues when Eve asks its oracle queries (contained in $P_E\ii$). This round ends when Eve is done with asking her oracle queries. For an even $i$, the definition of the $i\th$ round is similar (switching between Alice and Bob).

\subsection{Independence Learner}
The following lemma was implicit in the work of \cite{BarakMa09} and was proved explicitly in \cite{DachmanLiMaMa11} (here, for simplicity, we use this lemma with more relaxed parameters).

\begin{lemma}[Independence Learner for Protocols with No Input \cite{BarakMa09,DachmanLiMaMa11}]
\label{lem:Eve} 

Let  $\prot$ be an $N$-round input-less randomized two-party protocol using a random oracle \cO, with  $m$ query
complexity. Then, for any threshold $0\le \error \le 1$, $\prot$ has a public query strategy \Eve (who only observes the public messages) with query complexity  $\poly(m/\eps)$,  such
that  with probability at least $1-\eps$ over the choice of the view of \Eve: $V_E \getsr \bfV_E$  the following holds. (Recall that $V_E = (P_E,\mu)$ and  $V_E\ii$ is the part of $V_E$ that corresponds to the first $i$ rounds).
\begin{enumerate}
\item {\bf $(1-\eps)$-Independence:} For every $i \in [N]$ the following distributions are $\eps$-close:
$$ \left((\bfV_A\ii  \mid  V_E\ii) \times (\bfV_B\ii  \mid V_E\ii) \right)
\text{\ \ and \ \ }
 \left((\bfV_A\ii , \bfV_B\ii) \mid V_E\ii \right) .
$$
Namely, if we sample the views of Alice and Bob jointly conditioned on  $V_E\ii$, this joint distribution is $\eps$-close to the product distribution in which Alice and Bob's views are sampled independently (each conditioned on the same $V_E\ii$).

\item {\bf $\eps$-Lightness:}
For every $q\not\in Q_E\ii$ (where  $Q_E\ii = \Q(V_E\ii)$) it holds that
$$\pr_{V_A\ii \getsr (\bfV_A\ii \mid V_E\ii)}[q\in \Q(V_A\ii)]\ \le \eps \text{\ \ and \ \ } \pr_{V_B\ii \getsr (\bfV_B\ii \mid V_E\ii)}[q\in \Q(V_B\ii)]\ \le \eps.$$
\end{enumerate}
\end{lemma}

\subsection{Using the Independence Learner}

The Independence Learner of \lemmaref{Eve} is not directly useful in our context.
We need two additional technical properties ensured by the independence learner which are mentioned below.

The first lemma formalizes the intuition that a curious eavesdropper when run with appropriate parameters can ensure that whenever Alice sends a message in the protocol she can only add information about her input and not Bob's input.
\begin{lemma}[Independence Learner for Likely Inputs] \label{lem:IL-Input}
Let $\prot$ be a secure protocol for some secure function evaluation relative to a random oracle $\cO$ and Alice asks $m$ queries to the random oracle.
Suppose $\cX$ and $\cY$ are, respectively, the set of inputs for Alice and Bob.
We run \Eve with input parameter $\eps<1$ over $\Pi$ assuming that $\Pi$ is run with $\tilde x\getsr\cX$ and $\tilde y\getsr\cY$.
Let $x\in\cX$ and $y,{y'}\in\cY$ be fixed inputs.
Then for some $\eps'= \eps^{\Omega(1)}(m\cdot|\cX|)^{O(1)}$, if we run the protocol $\Pi$ with inputs $x$ and $y$ together the curious eavesdropper \Eve, for every
even $i \in [N]$ (\ie Bob sends $\mu_i$), with probability at least
$1-\eps'$ over the choice of the view of \Eve $V_E\ii \getsr
\bfV_E\ii$ at least one of the following holds:
\begin{enumerate}
\item $\pr[y  \mid V_E\ii, x] < \eps'$,
\item $\pr[{y'}  \mid V_E\ii, x] < \eps'$, or
\item $\SD((\bm\mu_{i+1} \mid V_E\ii,x,y) ,  (\bm\mu_{i+1} \mid V_E\ii,x,{y'}) ) \leq \eps'$.
\end{enumerate}
\end{lemma}

The second lemma is slightly more technical.
The curious eavesdropper of \lemmaref{Eve} ensures that all intersection queries are covered with high probability when Alice and Bob execute the protocol with actual inputs $x$ and $y$.
We need a stronger version of this result.
We want to claim that even if Bob pretends to change his input to ${y'}$ and samples a corresponding local view, the intersection queries of this ``hypothetical view'' are also covered by the actual Eve view with high probability.
This ensures that we can sample a consistent random oracle even without the knowledge of actual Alice input $x$ while simulating the hypothetical view.
Looking ahead, this lemma shall be useful when Bob launches a curious attack by changing his private input appropriately.
\begin{lemma}[Bounding Collisions of  Queries for Likely Inputs]
\label{lem:NoCol}
Let $\prot$ be a secure protocol for some secure function evaluation relative to a random oracle $\cO$ in which Alice asks $m$ queries to the random oracle.
Suppose $\cX$ and $\cY$ are, respectively, the set of inputs for Alice and Bob.
We run \Eve with input parameter $\eps<1$ over $\Pi$ assuming that $\Pi$ is run with $\tilde x\getsr\cX$ and $\tilde y\getsr\cY$.
Let $x\in \cX, y,{y'} \in \cY$ be some
fixed inputs.
Suppose we perform the following samplings:
$$(V_E\ii, Q_B\ii,Q_A\iip) \getsr (\bfV_E\ii, \bfQ_B\ii,\bfQ_A\iip \mid x,y) \text{ and } {Q'}_B\ii \getsr (\bfQ_B\ii \mid V_E\ii,\mu_{i+1},{y'}).$$
In the second sampling: the protocol is executed with inputs $x,y$ and Alice's message $\mu_{i+1}$ is generated, and after that we sample a view of Bob for the first $i$ rounds conditioned on $V_E\ii,\mu_{i+1}$ and Bob's input being ${y'}$.
Then for some $\eps' = \eps^{\Omega(1)}(m\cdot|\cX|\cdot |\cY|)^{O(1)}$  with probability at least $1-\eps'$ it holds that either
\begin{enumerate}
\item $\pr[{y'}  \mid V_E\ii, x] < \eps'$, or
\item $ Q_A\iip \cap (Q_B\ii \cup {Q'}_B \ii) \se \cQ(V_E \ii)$.
\end{enumerate}
\end{lemma}

Before proving \lemmaref{IL-Input} and \lemmaref{NoCol} we need to develop some general tools of probability.

\subsection{General Useful Lemmas}

A corollary to \lemmaref{inverse} is that the actual inputs of Alice and Bob will not become ``unlikely'' conditioned on Eve's view, except with small probability.
\begin{corollary}
\label{cor:inputsLikely}
Suppose Alice and Bob run a two party protocol with inputs $x,y$ chosen from an arbitrary distribution and suppose Eve is some public query strategy. Then the probability that  at some point during the protocol it holds that $\pr[(x,y)|u] < \theta$ where $u$ is the view of Eve so far,
 is at most $\theta/\pr[(x,y)]$ (if the inputs are chosen uniformly at random from the sets $\cX,\cY$, this probability is at most $\theta |\cX| |\cX|$).
\end{corollary}

\begin{proof}
\corollaryref{inputsLikely} follows by a direct application of \lemmaref{inverse} by using the event $X$ corresponds to the case that $(x,y)$ are the inputs, and the sequence of random variables $(\bfm_1,\bfm_2,\dots)$ corresponds to the sequence of the bits representing the view of Eve.
\end{proof}

The following lemma states that if two random variables $\rva,\rvb$ are statistically close, they will ``remain close'' even if we condition on a ``likely event'' defined over their supports.
\begin{lemma}
\label{lem:blow}
Let  $\rva, \rvb$ be two random variables such that $\SD(\rva,\rvb) \leq \eps$.
Suppose $E \se \Supp(\rva) \cup \Supp(\rvb)$ be an event such  that $\pr[\rva \in E] \geq \delta>0$ and $\pr[\rvb\in E]>0$.
Define $\rva_E \equiv (\rva \mid E)$ and $\rvb_E \equiv (\rvb \mid E)$.
Then, $\SD(\rva_E,\rvb_E) \leq  \eps / \delta$.
\end{lemma}

\begin{proof}
First, we prove a weaker bound of $3 \eps / 2\delta$ and then will sharpen the analysis to obtain the optimal bound of $\eps / \delta$.

Let $\alpha = \pr[\rva\in E]$ and $\beta=\pr[\rvb\in E]$.
Recall, we are guaranteed that $\alpha\geq\delta>0$ and $\beta>0$.
Moreover, $\sum_{s\in E}\abs{\pr[\rva=s]-\pr[\rvb=s]}\leq 2\eps$ and $\abs{\alpha-\beta}\leq\eps$, because $\SD(\rva,\rvb)\leq\eps$.
Observe that $\pr[\rva_E=s] = \pr[\rva=s]/\alpha$ and $\pr[\rvb_E=s] = \pr[\rvb=s]/\beta$, for $s\in E$.
Therefore, we can perform the following simplification:
\begin{align*}
\SD(\rva_E,\rvb_E) &= \frac12\sum_{s\in E} \abs{\frac{\pr[\rva=s]}{\alpha} - \frac{\pr[\rvb=s]}{\beta}}\\
&\leq \left(\frac12\sum_{s\in E}\abs{\frac{\pr[\rva=s]}{\alpha} - \frac{\pr[\rvb=s]}{\alpha}}\right)
  + \left(\frac12\sum_{s\in E}\abs{\frac{\pr[\rvb=s]}{\alpha} - \frac{\pr[\rvb=s]}{\beta}}\right)\\
&\leq \left(\frac{2\eps}{2\alpha}\right) + \left(\frac{\abs{\alpha-\beta}}{2\alpha}\right)
\quad\leq \frac{3\eps}{2\alpha} \quad\leq \frac{3\eps}{2\delta}
\end{align*}

With a more careful case analysis, the upper bound can be improved to $\eps/\delta$ (which is tight).
Consider these two cases:
\begin{enumerate}
\item Case $\alpha\geq\beta$:
  We shall partition the set $E$ into three sets $E_1$, $E_2$ and $E_3$ as follows:
  \begin{align*}
  E_1 &= \left\{s | s\in E, \pr[\rva=s]/\alpha \geq \pr[\rvb=s]/\beta  \right\}\\
  E_2 &= \left\{s | s\in E, \pr[\rva=s] \geq \pr[\rvb=s]\text{ but } \pr[\rva=s]/\alpha < \pr[\rvb=s]/\beta \right\}\\
  E_3 &= \left\{s | s\in E, \pr[\rva=s] < \pr[\rvb=s] \right\}
  \end{align*}
  Let $u_i=\pr[\rva\in E_i]$ and $v_i=\pr[\rvb\in E_i]$, where $i\in\{1,2,3\}$.
  We shall use the following constraints: $v_2\leq u_2$, $v_3\leq u_3+\eps$ and $v_1\geq u_1-\eps$.
  Now, consider the following manipulation:
  \begin{align*}
  \SD(\rva_E,\rvb_E) &= \frac12\sum_{s\in E} \abs{\frac{\pr[\rva=s]}{\alpha} - \frac{\rvb=s}{\beta}}\\
  &= \frac12\left[ \left(\frac{u_1}\alpha - \frac{v_1}\beta\right) - \left(\frac{u_2}\alpha - \frac{v_2}\beta\right) - \left(\frac{u_3}\alpha - \frac{v_3}\beta\right)\right]\\
  &= \frac{u_1}{\alpha} -\frac{v_1}{v_1+v_2+v_3}\\
  &\leq \frac{u_1}{\alpha} -\frac{v_1}{v_1+u_2+v_3} \qquad \left(\because v_2\leq u_2\right)\\
  &\leq \frac{u_1}{\alpha} -\frac{v_1}{v_1+u_2+u_3+\eps} \qquad \left(\because v_3\leq u_3+\eps\right)\\
  &\leq \frac{u_1}{\alpha} -\frac{u_1-\eps}{u_1+u_2+u_3} \qquad \left(\because v_1\geq u_1-\eps\right)\\
  &= \frac{\eps}{\alpha} \qquad \leq \frac{\eps}{\delta}
  \end{align*}
\item Case $\alpha<\beta$:
  We shall partition the set $E$ into three sets $E_1$, $E_2$ and $E_3$ as follows:
  \begin{align*}
  E_1 &= \left\{s | s\in E, \pr[\rva=s] \geq \pr[\rvb=s]  \right\}\\
  E_2 &= \left\{s | s\in E, \pr[\rva=s] < \pr[\rvb=s]\text{ but } \pr[\rva=s]/\alpha \geq \pr[\rvb=s]/\beta \right\}\\
  E_3 &= \left\{s | s\in E, \pr[\rva=s]/\alpha < \pr[\rvb=s]/\beta \right\}
  \end{align*}
  Let $u_i=\pr[\rva\in E_i]$ and $v_i=\pr[\rvb\in E_i]$, where $i\in\{1,2,3\}$.
  We shall use the following constraints: $v_2> u_2$, $v_3\leq u_3+\eps$ and $v_1\geq u_1-\eps$.
  Now, consider the following manipulation:
  \begin{align*}
  \SD(\rva_E,\rvb_E) &= \frac12\sum_{s\in E} \abs{\frac{\pr[\rva=s]}{\alpha} - \frac{\rvb=s}{\beta}}\\
  &= \frac12\left[ \left(\frac{u_1}\alpha - \frac{v_1}\beta\right) + \left(\frac{u_2}\alpha - \frac{v_2}\beta\right) - \left(\frac{u_3}\alpha - \frac{v_3}\beta\right)\right]\\
  &= \frac{v_3}{v_1+v_2+v_3} -\frac{u_3}{\alpha}\\
  &< \frac{v_3}{v_1+u_2+v_3} -\frac{u_3}{\alpha} \qquad \left(\because v_2> u_2\right)\\
  &\leq \frac{u_3+\eps}{v_1+u_2+u_3+\eps} -\frac{u_3}{\alpha} \qquad \left(\because v_3\leq u_3+\eps\right)\\
  &\leq \frac{u_3+\eps}{u_1+u_2+u_3} -\frac{u_3}{\alpha} \qquad \left(\because v_1\geq u_1-\eps\right)\\
  &= \frac{\eps}{\alpha} \qquad \leq \frac{\eps}{\delta}
  \end{align*}
\end{enumerate}
This completes the proof that $\SD(\rva_E,\rvb_E)\leq \eps/\delta$.
Equality holds if and only if, $\{s|\pr[\rva=s]\neq\pr[\rvb=s]\}\subseteq E$ and $\pr[\rva\in E]=\delta$.
\end{proof}

The following lemma states that if two random are close to being independent iff they are close to the product of their marginal distribution.

\begin{lemma} \label{lem:closeToMargin}
Let $(\bfa,\bfb)$ be jointly distributed random variables such that 
$\SD((\bfa,\bfb), (\bfu \times \bfv))\leq \eps$ for some random variables $\bfu$ and $\bfv$. Then it holds that
$\SD((\bfa,\bfb), (\bfa \times \bfb))\leq 3 \eps$.
\end{lemma}
\begin{proof}
$\SD((\bfa,\bfb), (\bfu \times \bfv))\leq \eps$ implies that $\SD(\bfa,\bfu) \leq \eps$ and $\SD(\bfb,\bfv) \leq \eps$. Therefore, by two applications of triangle inequality it holds that:
$\SD((\bfa,\bfb), (\bfa \times \bfb)) \leq \SD((\bfa,\bfb), (\bfu \times \bfv)) +
\SD((\bfu \times \bfv), (\bfa \times \bfv)) + 
\SD((\bfa \times \bfv), (\bfa \times \bfb)) \leq 3\eps.$
\end{proof}

The following lemma states that whenever two random variables $(\bfa,\bfb)$ are close to being independent, then they will remain so, even if we sample $\bfa$ conditioned on some partial leakage $c$ as a function  of $\bfb$.
\begin{lemma}\label{lem:stillProd}
Let $(\bfa,\bfb)$ be jointly distributed random variables such that $\SD((\bfa,\bfb), (\bfa \times \bfb))\leq \eps$.
Suppose $c = f(b)$ is a possibly randomized function of $b$, where the random tape for $f(\cdot)$ is chosen uniformly and independently at random.
Given a  sample for $(b,f(b)=c)$, let $\bfa'$ be another random variable sampled from the distribution $(\bfa \mid c)$. Then it holds that $\SD((\bfb,\bfa'), (\bfb \times \bfa)) \leq \eps$.
\end{lemma}

\begin{proof}
Suppose $f(b;r)$ is the deterministic function where $r$ is the random tape used to evaluate the randomized function $f$.
This case reduces to the deterministic case as follows:
\begin{align*}
\SD((\rvb, \rva'), (\rvb\times\rva)) &\leq \SD((\rvb, \rva',\rvr), (\rvb\times\rva\times\rvr))
  = \SD( ((\rvb\times\rvr), \rva'), ((\rvb\times\rvr)\times\rva) )
\end{align*}
Henceforth, we can assume, without loss of generality, that $f$ is a deterministic function. In this case:
{\allowdisplaybreaks
\begin{align*}
2\SD((\rvb,\rva') , (\rvb\times\rva))
  &= \sum_{a}\sum_{b} \pr[\rvb=b] \cdot \abs{\pr[\rva=a|c=f(b)] - \pr[\rva=a]}\\
  &= \sum_{a}\sum_{c} \sum_{b\in f^{-1}(c)}\pr[\rvb=b] \cdot \abs{\pr[\rva=a|c]-\pr[\rva=a]}\\
  &= \sum_{a}\sum_{c} \pr[\rvc=c]\cdot \abs{\pr[\rva=a|c]-\pr[\rva=a]}\\
  &= \sum_{a}\sum_{c} \abs{\pr[\rva=a,\rvc=c] - \pr[\rva=a]\pr[\rvc=c]}\\
  &= \sum_{a}\sum_{c} \abs{\sum_{b\in f^{-1}(c)}\left(\pr[\rva=a,\rvb=b]-\pr[\rva=a]\pr[\rvb=b]\right)}\\
  &\leq \sum_{a}\sum_{c}\sum_{b\in f^{-1}(c)} \abs{\pr[\rva=a,\rvb=b]-\pr[\rva=a]\pr[\rvb=b]}\\
  &= 2\SD((\rvb,\rva) , (\rvb\times\rva)).&\qedhere
\end{align*}
}
\end{proof}

\subsection{Proving \lemmaref{IL-Input} and \lemmaref{NoCol}}

We shall prove both \lemmaref{IL-Input} and \lemmaref{NoCol} both using the following intermediate lemma.
\begin{lemma} \label{lem:Inter}
Suppose $V_E\ii$ is the view of \Eve by the end of the $i\th$ round with respect to the two party protocol in which the inputs are chosen at random and is such that the $(1-\eps)$-Independence and $\eps$-Lightness properties hold conditioned on $V_E\ii$. 
Suppose $x \in \cX, y \in \cY$  are such that $\pr[x,y \mid V_E\ii] \geq \gamma$ and $m$ is the total number  Alice's queries. Then both of the following hold:
\begin{enumerate}
\item  $\pr[Q_A\iip \cap Q_B\ii \not\se Q_E\ii \mid V_E\ii, x,y] \leq O(m \eps / \gamma)$. %
\item The following two are $O(m\eps/\gamma)$-close:
$$ (\bfV_A\iip, \bfV_B\ii \mid V_E\ii,x,y) \text{  and  } 
((\bfV_A\iip \mid V_E\ii,x) \times (\bfV_B\ii \mid V_E\ii,y)).$$
\end{enumerate}
\end{lemma}

Before proving \lemmaref{Inter} we shall see how it can be used to prove \lemmaref{IL-Input} and \lemmaref{NoCol}.

\subsubsection{Proof of \lemmaref{IL-Input}}
For simplicity, we shall use another parameter $0<\sigma<1$ and prove the following result:
With probability at least
$1-\eps-\sigma |\cX|$ over the choice of the \Eve view  $V_E\ii \getsr
\bfV_E\ii$ at least one of the following holds:
\begin{enumerate}
\item $\pr[y  \mid V_E\ii, x] < \sigma$,
\item $\pr[{y'}  \mid V_E\ii, x] < \sigma$, or
\item $\SD((\bm\mu_{i+1} \mid V_E\ii,x,y) ,  (\bm\mu_{i+1} \mid V_E\ii,x,{y'}) ) \leq\  O(m \eps/\sigma^2)$ where $m$ is the number of oracle queries asked by Alice during the protocol.
\end{enumerate}
Then \lemmaref{IL-Input} follows by setting $\eps = \sigma^3$ and taking $\eps' = \max(\sigma, \eps+\sigma |\cX|,m \eps/\sigma^2)$ in the above mentioned statement.

By \lemmaref{Eve}, with probability at least $1-\eps$ over the choice of $V_E\ii$, the $(1-\eps)$-Independence and $\eps$-Lightness properties both hold.
\corollaryref{inputsLikely} implies that with probability at least $1-\sigma|\cX|$, we shall have $\pr[x|V_E\ii]\geq \sigma$.
By union bound, both these events hold with probability at least $1-\eps-\sigma|\cX|$.
Henceforth, we shall assume that both these conditions hold for our choice of $V_E\ii$.

For our choice of $V_E\ii$, if one of the first two cases of \lemmaref{IL-Input} holds then we are done.
Suppose this is not the case.
Then, we have $\pr[y  \mid V_E\ii, x] \geq \sigma$ and $\pr[{y'}  \mid V_E\ii, x] \geq \sigma$.
Therefore we can conclude that  \emph{both pairs} of inputs $(x,y)$ and $(x,{y'})$ are ``likely'' conditioned on $V_E \ii$.
More formally:
$$\pr[x,y \mid V_E\ii] \geq \pr[x \mid V_E\ii] \cdot \pr[y \mid V_E\ii , x] \geq \sigma^2
\text{ and similarly } \pr[x,{y'} \mid V_E\ii] \geq \sigma^2$$

So, currently we are considering $V_E\ii$ such that $\pr[x,y|V_E\ii]\geq \sigma^2$, $\pr[x,{y'}|V_E\ii]\geq \sigma^2$; and $(1-\eps)$-Independence and $\eps$-Lightness guarantees hold. Therefore \lemmaref{IL-Input} follows by the second part of \lemmaref{Inter} because $(\bfV_A\iip \mid V_E\ii,x)$ is independent of $y$ and ${y'}$ and $\mu_{i+1}$ is a function of $V_A\iip$.

\subsubsection{Proof of \lemmaref{NoCol}}

Similarly to the proof of \lemmaref{IL-Input}, we use another parameter $0 < \sigma < 1$ and prove the following statement:
With probability $1-O(\eps+\sigma|\cX| \cdot |\cY| + m \eps/\sigma^2)$
over the  samples at least one of the following is true:
\begin{enumerate}
\item $\pr[{y'}  \mid V_E\ii, x] < \sigma$, or
\item $ Q_A\iip \cap (Q_B\ii \cup {Q'}_B \ii) \se \cQ(V_E \ii)$.
\end{enumerate}
\lemmaref{NoCol} follows by setting $\sigma^3 = \eps$ and $\eps' = \eps+\sigma
|\cX| \cdot |\cY| + m \eps/\sigma^2$ in the above mentioned statement.

Recall that with probability at least $1-\eps$, the sampled Eve view $V_E\ii$ has the $(1-\eps)$-Independence and the $\eps$-Lightness properties.
Henceforth, we shall restrict ourselves to such $V_E\ii$.
By \corollaryref{inputsLikely} we conclude that with probability at least $1-\sigma|\cX|$ it holds that $\pr[x \mid V_E\ii] \geq \sigma$.
If $\pr[{y'}|V_E\ii, x]<\sigma$ for this Eve view $V_E\ii$, then we are done.
So, assume on the contrary that $\pr[{y'}|V_E\ii, x]\geq\sigma$, which implies that:
$$\pr[x, {y'} \mid V_E\ii] \geq \pr[x \mid V_E\ii] \cdot \pr[{y'} \mid V_E\ii,x] \geq \sigma^2$$

Since $(x,y)\getsr\cX\times\cY$, we can apply \corollaryref{inputsLikely} directly to conclude that with probability at least $1-\sigma^2 |\cX| |\cY|$, $\pr[x,y \mid V_E\ii] \geq \sigma^2$.
By union bound, we can assume that all these properties hold with probability $1- O(\eps+\sigma |\cX| \cdot |\cY|)$.
Henceforth, we shall assume that $V_E\ii$ satisfies these conditions.

First, using $\pr[x,y \mid V_E\ii] \geq \sigma^2$ and by a direct application of \lemmaref{Inter} we can conclude that with probability $1-O(m \eps /\sigma^2)$, it holds that $Q_A\iip \cap Q_B\ii \se \cQ(V_E \ii)$.
Thus, it suffices to show that with probability $1-O(m \eps /\sigma^2)$, it holds that $Q_A\iip \cap {Q'}_B\ii \se \cQ(V_E \ii)$, in which case \lemmaref{NoCol} would trivially follow by a union bound from these two results.

Lets define $\wt{V}_B\ii$ and $\wt{Q}_B\ii$ similar to $V_B\ii$ and $Q_B\ii$ with the only difference that we sample them conditioned on the input ${y'}$. 
Then the same exact proof as for the case of likely input $y$, can be applied to the case of likely input ${y'}$ and conclude that with probability $1-O(m \eps /\sigma^2)$, it holds that $Q_A\iip \cap \wt{Q}_B\ii \se \cQ(V_E \ii)$. 
We emphasize that the distributions $\wt{V}_B\ii$ and ${V'}_B\ii$ are not identical.
Although, both are sampled based on Bob input being ${y'}$ and Eve view being $V_E\ii$, the latter is additionally conditioned on the next message $\mu_{i+1}$ of Alice.
Here, we shall be leveraging \lemmaref{stillProd}.

So, consider an Eve view $V_E\ii$ with the following properties:
\begin{enumerate}
\item $(1-\eps)$-Independence and $\eps$-Lightness properties hold, and
\item $\pr[x, {y'}|V_E\ii, x,  {y'}]\geq\sigma^2$.
\end{enumerate}

Let $(\bfV_A\iip,\wt\bfV_B\ii)$ represent the joint Alice-Bob views when Alice has input $x$ and Bob has input ${y'}$.
By \lemmaref{Inter}, we know that this distribution is $O(m\eps/\sigma^2)$ close to the distribution $(\bfV_A\iip|V_E\ii,x)\times(\wt\bfV_B\ii|V_E\ii,{y'})$.
Let $(\bfV_A\iip,{\bfV'}_B\ii)$ represent the joint Alice-Bob views when Alice has input $x$, Bob has input ${y'}$ as picked in our experiment, i.e. Bob's view is additionally conditioned on the next message $\mu_{i+1}$.
Considering $\mu_{i+1}$ as a leakage on $V_A\iip$, we can conclude that $(\bfV_A\iip,{\bfV'}_B\ii)$ is also $O(m\eps/\sigma^2)$ close to $(\bfV_A\iip|V_E\ii,x)\times(\wt\bfV_B\ii|V_E\ii,{y'})$, by \lemmaref{stillProd}.
Consequently, the distributions $(\bfV_A\iip,\wt\bfV_B\ii)$ and $(\bfV_A\iip,{\bfV'}_B\ii)$ are $O(m\eps/\sigma^2)$ close.

Recall that the probability of the event $Q_A\iip \cap \wt{Q}_B\ii \se \cQ(V_E \ii)$ when Alice-Bob views are sampled according to $(\bfV_A\iip,\wt\bfV_B\ii)$ is $1-O(m\eps/\sigma^2)$.
So, the probability of the same event when Alice-Bob joint views are sampled according to $(\bfV_A\iip,{\bfV'}_B\ii)$ is also $1-O(m\eps/\sigma^2)$.
This concludes the proof of \lemmaref{NoCol}.

\subsubsection{Proof of \lemmaref{Inter}}
Finally we prove \lemmaref{Inter}.
Recall that with respect to the Eve view $V_E\ii$, $(1-\eps)$-Independence and $\eps$-Lightness hold, when the protocol is run with uniformly chosen $x\in\cX$ and $y\in\cY$.
Consider the space of all Alice and Bob private views and random oracles such that $V_E\ii$ is produced as the view of \Eve.
We know by \lemmaref{Eve} that the distribution of $(\bfV_A\ii, \bfV_B\ii | V_E\ii)$ is $\eps$ close to a distribution $(\bfU_A\times\bfU_B)\equiv(\bfV_A\ii|V_E\ii)\times(\bfV_B\ii|V_E\ii)$.

Additionally, we are also given that $\pr[x,y|V_E\ii]\geq\gamma$.
Now, consider the event $E$ such that $x$ and $y$ are actually the local inputs in sampled Alice and Bob views $V_A\ii$ and $V_B\ii$.
By \lemmaref{blow}, we can conclude that $(\bfV_A\ii, \bfV_B\ii | V_E\ii, x, y)$ is $\eps/\gamma$ close to the distribution $(\bfU_A\times\bfU_B |  x, y)\equiv (\bfU_A|x) \times (\bfU_B|y)$.
Now, observe that when Alice and Bob views are sampled according to $(\bfV_A\ii, \bfV_B\ii | V_E\ii, x, y)$, then they also satisfy $\eps/\gamma$-Lightness property.
Otherwise we can use the fact that $\pr[x,y|V_E\ii]\geq\gamma$ to show that $(\bfV_A\ii, \bfV_B\ii | V_E\ii)$ does not satisfy the $\eps$-Lightness property.
Now, since the distribution $(\bfV_A\ii, \bfV_B\ii | V_E\ii, x, y)$ has $\eps/\gamma$-Lightness property and is $\eps/\gamma$ close to the product distribution $(\bfU_A|x) \times (\bfU_B|y)$,
this implies that the distribution $(\bfV_A\ii, \bfV_B\ii | V_E\ii, x, y)$ satisfies $(1-\eps')$-Independence and $\eps'$-Lightness properties, where $\eps'=\eps/\gamma$.
Next, based on these properties, we shall first prove the first part of \lemmaref{Inter} by showing that $\pr[\cQ(V_A\iip) \cap \cQ(V_B\ii) \not\subseteq \cQ(V_E\ii) | V_E\ii, x, y] \leq O(m\eps')$.

We define several hybrid experiments where the distribution of $V_A\iip$ and $V_B\ii$ is defined differently in each of them.
We are interested in comparing the probability $p_i$ of the bad event $B$ defined as $\cQ(V_A\iip) \cap \cQ(V_B\ii) \not\subseteq \cQ(V_E\ii)$  in  the game $\G_i$.

\paragraph{$\G_0$:} In this game the views $V_A\iip$ and $V_B\ii$ are jointly sampled consistent with $V_E\ii$ and local inputs $x$ and $y$.

\paragraph{$\G_1$:} This game is indeed a perfect lazy simulation of $\G_0$:

\begin{enumerate}
\item Sample $(V_A\ii, V_B\ii)$ according to the distribution $(\bfV_A\ii,\bfV_B\ii|V_E\ii,x,y)$.
\item Start the next message generation algorithm for Alice.
  If any query $q$ asked by Alice is already contained in $\cQ(V_A\ii) \cup \cQ(V_E\ii) \cup \cQ(V_B\ii)$, then it is consistently answered.
  Otherwise, a uniformly random answer is provided.
\end{enumerate}
So, the probability $p_1$ of the bad event $\cQ(V_A\iip) \cap \cQ(V_B\ii) \not\subseteq \cQ(V_E\ii)$ in this game is still equal to $p_0$.

\paragraph{$\G_2$:} In this game
\begin{enumerate}
\item Alice and Bob views are drawn according to $(\bfV_A\ii|V_E\ii,x) \times (\bfV_B\ii|V_E\ii,y)$.
\item Start the next message generation algorithm for Alice with respecting the answers to Bob's private queries. Namely,
  if any query $q$ asked by Alice is already contained in $\cQ(V_A\ii) \cup \cQ(V_E\ii) \cup \cQ(V_B\ii)$, then it is consistently answered.
  Otherwise, an uniformly random answer is provided.
\end{enumerate}

By $(1-\eps')$-Independence we know that $\G_1$ and $\G_2$ are $\eps'$ close, so $p_1\leq p_2+\eps'$.

Now, we shall bound $p_2$.
Recall that $\eps'$-Lightness of $(\bfV_B\ii|V_E\ii,y)$ implies that any query not already answered in $V_E\ii$ occurs with probability at most $\eps'$ in a Bob view $V_B\ii\getsr(\bfV_B\ii|V_E\ii,y)$.
So, the probability of $m$ new queries of Alice hitting any query of Bob view $V_B\ii\getsr(\bfV_B\ii|V_E\ii,y)$ is at most $m\eps'$, by union bound.
So, $p_2\leq m\eps'$.
This implies that $p_0=p_1\leq p_2+\eps' \leq (m+1)\eps'$.
This completes the proof of the first part of \lemmaref{Inter}.

\paragraph{Proving the Second part of \lemmaref{Inter}.} 
In our previous hybrids, we showed that the joint distribution of views $(\bfV_A\iip,\bfV_B\ii)$ in $\G_0$ and $\G_2$ are $\eps'$ far.

Consider the following $\G_3$ as the next hybrid following $\G_2$:
In this game
\begin{enumerate}
\item Alice and Bob views are drawn according to $(\bfV_A\ii|V_E\ii,x) \times (\bfV_B\ii|V_E\ii,y)$.
\item Start the next message generation algorithm for Alice without respecting the answers to Bob's private queries. Namely,
  if any query $q$ asked by Alice is already contained in $\cQ(V_A\ii) \cup \cQ(V_E\ii)$, then it is consistently answered.
  Otherwise, an uniformly random answer is provided.
\end{enumerate}

If the bad event $\cQ(V_A\iip) \cap \cQ(V_B\ii) \not\subseteq \cQ(V_E\ii)$ does not occur, then the distribution of Alice-Bob joint views sampled in $\G_2$ and $\G_3$ are identical.
Further, the distribution of Alice views in $\G_3$ is identical to $(\bfV_A\iip|V_E\ii,x)$.
Note, that by the same argument at in $\G_3$, the probability $p_3$ of the bad event is at most $m\eps'$, because the argument was independent of how Alice queries were answered.
So, the joint distribution of views $(\bfV_A\iip,\bfV_B\ii)$ in $\G_2$ and $\G_3$ are at most $\max\{p_2,p_3\}\leq m\eps'$ far.

Therefore, the statistical distance between $(\bfV_A\iip,\bfV_B\ii|V_E\ii,x,y)$ and $(\bfV_A\iip|V_E\ii,x)\times(\bfV_B\ii|V_E,y)$ is at most $(m+1)\eps'$.
Thus, the second part of \lemmaref{Inter} follows.

\section{Some Examples for Intuition}
\label{app:examples}

\subsection{Undecomposable Functions}
We give examples of some representative undecomposable functions in \figureref{OR}, \figureref{spiral} and \figureref{weave}

\begin{figure}[H]
\centering
\begin{tikzpicture}
\tikzset{sim/.style={rounded corners, rectangle, draw, thick, dotted, inner sep = 0}}
\pgfsetmatrixcolumnsep{8pt}
\pgfsetmatrixrowsep{8pt}
\matrix at (0,0) [left delimiter = (, right delimiter = )]
{
  \node (n00) {0}; & \node (n01) {1}; \\
  \node (n10) {1}; & \node (n11) {1}; \\
};
\node [sim, fit = (n01) (n11)] {};
\node [sim, fit = (n10) (n11)] {};
\end{tikzpicture}
\caption{A Complete (and Undecomposable) Function.}
\label{fig:OR}
\end{figure}

\begin{figure}[H]
\centering
\begin{tikzpicture}
\tikzset{sim/.style={rounded corners, rectangle, draw, thick, dotted, inner sep = 0}}
\pgfsetmatrixcolumnsep{8pt}
\pgfsetmatrixrowsep{8pt}
\matrix at (0,0) [left delimiter = (, right delimiter = )]
{
  \node (n11) {1}; & \node (n12) {1}; & \node (n13) {2}; \\
  \node (n21) {4}; & \node (n22) {0}; & \node (n23) {2}; \\
  \node (n31) {4}; & \node (n32) {3}; & \node (n33) {3}; \\
};
\node [sim, fit = (n11) (n12)] {};
\node [sim, fit = (n13) (n23)] {};
\node [sim, fit = (n32) (n33)] {};
\node [sim, fit = (n21) (n31)] {};
\end{tikzpicture}
\caption{An Incomplete but Undecomposable Function (Minimum $|\cX|+|\cY|$).}
\label{fig:spiral}
\end{figure}

\begin{figure}[H]
\centering
\begin{tikzpicture}
\tikzset{sim/.style={rounded corners, rectangle, draw, thick, dotted, inner sep = 0}}
\pgfsetmatrixcolumnsep{8pt}
\pgfsetmatrixrowsep{8pt}
\matrix at (0,0) [left delimiter = (, right delimiter = )]
{
  \node (n11) {1}; & \node (n12) {1}; & \node (n13) {3}; & \node (n14) {4}; \\
  \node (n21) {3}; & \node (n22) {2}; & \node (n23) {2}; & \node (n24) {4}; \\
  \node (n31) {3}; & \node (n32) {4}; & \node (n33) {1}; & \node (n34) {1}; \\
  \node (n41) {2}; & \node (n42) {4}; & \node (n43) {3}; & \node (n44) {2}; \\
};
\node [sim, fit = (n11) (n12)] {};
\node [sim, fit = (n22) (n23)] {};
\node [sim, fit = (n33) (n34)] {};
\node [sim, fit = (n13) ] (open13) {}; \draw [color=white, line width=5pt] (open13.north west) -- (open13.north east);
\node [sim, fit = (n43) ] (open43) {}; \draw [color=white, line width=5pt] (open43.south west) -- (open43.south east);
\node [sim, fit = (n21) (n31)] {};
\node [sim, fit = (n32) (n42)] {};
\node [sim, fit = (n14) (n24)] {};
\node [sim, fit = (n41) ] (open41) {}; \draw [color=white, line width=5pt] (open41.north west) -- (open41.south west);
\node [sim, fit = (n44) ] (open44) {}; \draw [color=white, line width=5pt] (open44.south east) -- (open44.north east);
\end{tikzpicture}
\caption{An Incomplete but Undecomposable Function (Minimum $|\cZ|$).}
\label{fig:weave}
\end{figure}

\subsection{Decomposable Example}

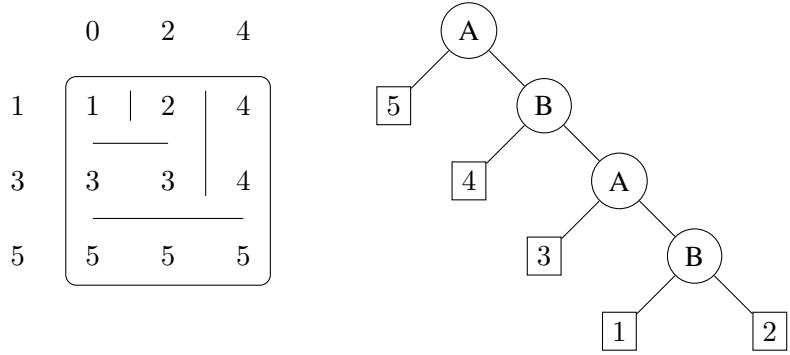
\begin{figure}[H]
\centering
\begin{tikzpicture}
\coordinate (o) at (0,0);
\coordinate (x) at (1,0);
\coordinate (y) at (0,-1);
\coordinate (dy) at ($0.2*(y)$);
\draw (o) node {}
    ++(x) node (y1) {$0$}
    ++(x) node (y2) {$2$}
    ++(x) node (y3) {$4$}
      (o) 
    ++(y) node (x1) {$1$}
    ++(x) node (m11) {$1$}
    ++(x) node (m12) {$2$}
    ++(x) node (m13) {$4$}
      (o) 
    ++($2*(y)$) node (x2) {$3$}
    ++(x) node (m21) {$3$}
    ++(x) node (m22) {$3$}
    ++(x) node (m23) {$4$}
      (o) 
    ++($3*(y)$) node (x3) {$5$}
    ++(x) node (m31) {$5$}
    ++(x) node (m32) {$5$}
    ++(x) node (m33) {$5$};
\node [rectangle, rounded corners, draw, fit = (m11) (m33)] {};
\draw ($(m21)!.5!(m31)$) -- ($(m23)!.5!(m33)$);
\draw ($(m13)!.5!(m12) - (dy)$) -- ($(m23)!.5!(m22) + (dy)$);
\draw ($(m11)!.5!(m21)$) -- ($(m12)!.5!(m22)$);
\draw ($(m11)!.5!(m12) - (dy)$) -- ++($0.4*(y)$);
\coordinate (p) at ($(o) + 6*(x)$);
\coordinate (s) at ($(x) + (y)$);
\coordinate (t) at ($(y) - (x)$);
\draw (p) node (r) [circle, draw] {A}
    ++(s) node (r1) [circle, draw] {B}
    ++(s) node (r11) [circle, draw] {A}
    ++(s) node (r111) [circle, draw] {B}
    ++(s) node (r1111) [rectangle, draw] {$2$}
      (r111)
    ++(t) node (r1110) [rectangle, draw] {$1$}
      (r11)
    ++(t) node (r110) [rectangle, draw] {$3$}
      (r1)
    ++(t) node (r10) [rectangle, draw] {$4$}
      (r)
    ++(t) node (r0) [rectangle, draw] {$5$};
\draw (r) -- (r0)
      (r) -- (r1)
      (r1) -- (r10)
      (r1) -- (r11)
      (r11) -- (r110)
      (r11) -- (r111)
      (r111) -- (r1110)
      (r111) -- (r1111);
\end{tikzpicture}
\caption{Decomposition of a Decomposable Function}
\label{fig:big-max}
\end{figure}

Let us consider the example of computing maximum of Alice and Bob inputs, where Alice's input set is $\{1, 3, 5\}$ and Bob's input set is $\{0, 2, 4\}$.
This function is decomposable and its decomposition provides a perfectly semi-honest secure protocol, see \figureref{big-max}.
The semi-honest protocol is as follows:
\begin{figure}[H]
\centering
\begin{minipage}{.9\textwidth}
Protocol to compute maximum of Alice and Bob inputs:
\begin{enumerate}
\item If Alice's input is $5$, then she announces the outcome to be $5$; Otherwise she asks Bob to proceed.
\item If Bob's input is $4$, then he announced the outcome to be $4$; Otherwise he asks Alice to proceed.
\item If Alice's input is $3$, then she announces the outcome to be $3$; Otherwise she asks Bob to proceed.
\item Now, Alice's input is $1$ for certain. If Bob's input is $2$, then he announces the outcome to be $2$; Otherwise the outcome is $1$.
\end{enumerate}
\end{minipage}
\end{figure}

\section{Black-box Separation Proof}
\label{app:blackbox}

\paragraph{Proof of \theoremref{fully-BBsep-OWF}.}

It is immediate that $(1)\implies(2)\implies(3)$. We shall show that
$(3)\implies(1)$. In fact, for clarity, first we shall show $(2)\implies(1)$
before extending the argument to show $(3)\implies(1)$.

We rely on the following claim.
\begin{claim}
\label{clm:bb-sep-claim}
Let $f$ be a a deterministic two-party function which does not have a
perfectly semi-honest secure protocol.  For any security-parameter
preserving fully black-box construction $(\prot,S)$ of \SFE{f} from \OWF,
there exist $Q\in F_\OWF$ and an oracle algorithm \Adv such that
$(\prot^{Q}, \Adv^{Q}) \in R_\SFE{f}$ and $( Q, S^{Q, \Adv^Q}) \nin R_\OWF$.
\end{claim}

Before proving this claim, we note that it indeed shows $(2)\implies(1)$, as
follows. Suppose, for the sake of contradiction, $(\prot,S)$ is a
security-parameter preserving PPT-secure fully black-box construction of \SFE{f} from
\OWF, for some deterministic two-party function $f$ which does not have a
perfectly semi-honest secure protocol.
For $(\prot,S)$, let $Q\in F_\OWF$ and \Adv be as guaranteed in \claimref{bb-sep-claim}.
Let $A$ stand for $\Adv^{Q}$.
The claim guarantees that $(\prot^Q,A) \in
R_\SFE{f}$.  Consequently, by the security guarantee of 
fully black-box construction,
we have $(Q,S^{Q,A}) \in R_\OWF$. 
But this contradicts the guarantee from \claimref{bb-sep-claim}.

\begin{proof}[Proof of \claimref{bb-sep-claim}]
Let $\cU_\spar = \{g \colon \zo^\spar \To \zo^\spar\}$ denote the set of all
length preserving functions over $\zo^\spar$.  Let $\cU = \cU_1 \times \cU_2
\times \dotsi$.

Note that since \prot is
security-parameter preserving, $\prot^{Q}$ accesses only $Q_\spar$. Implicit
in the proof of \theoremref{main} is an adversary $\Adv$ such that, for
$Q_\spar \getsr \cU_\spar$ the adversary $\Adv^{Q_\spar}$ breaks the
security of $\prot^{Q_\spar}$ (as an implementation of $\SFE{f}$) with
advantage $\delta(\spar) > 1/\poly(\spar)$, by asking $\poly(\spar)$ queries
to $Q_\spar$. This will be the adversary \Adv in the statement of the claim.

Next, we need to find a deterministic function $Q$ such that 
$\Adv^Q$ breaks $\prot^Q$, but there does not exist any efficient
reduction $S$ such that $S^{Q,\Adv^Q}$ breaks $Q$ as a \OWF implementation.

We show the existence of such a $Q$ by the probabilistic method.  For this,
first we define $\cV_\spar\subseteq\cU_\spar$ for each $\spar\in\N$ as
follows.  
As mentioned above, $\Adv^{Q_\spar}$ has an advantage of $\delta(\spar) >
1/\poly(\spar)$ in breaking $\prot^{Q_\spar}$, where
$Q_\spar\getsr\cU_\spar$.  Then, by an
averaging argument, for a subset $\cV_\spar \subseteq \cU_\spar$ with
$\frac{|\cV_\spar|}{|\cU_\spar|} \geq \delta(\spar)$, it holds that for all
$Q_\spar \in \cV_\spar$,  $\Adv^{Q_\spar}$
has an advantage at least $\delta(\spar)/2$ in the \SFE{f} security game
for $\prot^{Q_\spar}$.
Now, we pick $Q_\spar \getsr \cV_\spar$ independently for each security
parameter $\spar$.  $Q$ will be the composite oracle $(Q_1,Q_2,\dotsc)$.

By construction, $(\prot^Q,\Adv^Q)\in R_{\SFE{f}}$ with probability 1, since
for all $Q_\spar\in\cV_\spar$, $\Adv^{Q_\spar}$ has a significant advantage
(as a function of $\spar$) in the security game.  To complete the proof, we
need to show that with positive probability $Q$ is such that
$(Q,S^{Q,\Adv^Q})\nin R_\OWF$. 

Consider again $Q_\spar\getsr \cU_\spar$ (rather than
$Q_\spar\getsr\cV_\spar$, which we shall return to shortly).  For each
\spar, for each choice of $Q_{\ol\spar} = (Q_1,\dotsc, Q_{\spar-1},
Q_{\spar+1},\dotsc)$, define the (inefficient) machine $T_{Q_{\ol\spar}}$
such that $T_{Q_{\ol\spar}}^{Q_\spar}$ simulates $S^{Q,\Adv^Q}$: for this,
$T_{Q_{\ol\spar}}$ internally simulates all of $Q$
except $Q_\spar$, which it accesses through oracle calls. Even though
$T_{Q_{\ol\spar}}$ is inefficient, since $S$ is efficient, the number of
oracle queries it makes is bounded by $\poly(\spar)$. W.l.o.g, we can assume
that a machine $T^{Q_\spar}$ can invert an input $y$ with respect to its
oracle, only if one of its oracle queries is answered by $y$ (by adding a
final query, in which it queries the oracle at its output). But when
$Q_\spar\getsr\cU_\spar$ this happens with only negligible probability for a
machine making polynomially many queries, because each distinct query is
answered by a \spar-bit string chosen uniformly at random which has a
probability of $\frac1{2^\spar}$ of being equal to $y$.

Thus, if $Q_\spar\getsr
\cU_\spar$, then for each choice of $Q_{\ol\spar}$, the probability that
$S^{Q,\Adv^Q}$ has a non-negligible advantage in breaking $Q$ at $\spar$ is
$\nu(\spar)$ for some negligible function $\nu$. Then if if $Q_\spar\getsr
\cV_\spar$, this probability is at most $\nu(\spar)
\frac{|\cU_\spar|}{|\cV_\spar|}$ which is also negligible (since
$\frac{|\cV_\spar|}{|\cU_\spar|} \ge \delta(\spar)/2$).

Then, by a union bound over all $\spar\ge\spar_0$ for a sufficiently large
value of $\spar_0$, the probability that $S^{Q,\Adv^Q}$ has a non-negligible
advantage in breaking $Q$ at some $\spar\ge\spar_0$ is 
$\sum_{\spar=\spar_0}^\infty \nu(\spar) < 1$
(and can in fact be made arbitrarily close to 0, by choosing $\spar_0$ large
enough).  In particular, there exists $Q$ such that $S^{Q,\Adv^Q}$ does not
have a non-negligible advantage in breaking $O$ at infinitely many values of
\spar.  That is, $(Q,S^{Q,\Adv^Q})\nin R_\OWF$.
\end{proof}

\paragraph{Extending to \OWFz.} The above argument can be easily extended
to show $(3)\implies(1)$, to complete the proof. Fix a polynomial $\zeta$.
Then, in the above argument consider the
set $\cW_\spar := \{ \gz(\spar,\cdot) | g:\zo^*\To\zo^\spar  \}$ (i.e., set
of functions that map $x$, $|x|\le\zeta(\spar)$ to $y\in\zo^\spar$), instead
of $\cU_\spar$. We remark that for the adversary from the proof of
\theoremref{main} it was not crucial that the random oracle has input domain
$\zo^\spar$, or that the oracle is length-preserving,
as long as the queries are answered independent of each
other. The rest of the argument, including the fact that an inverter making
polynomial queries to an oracle $W_\spar\getsr\cW_\spar$ can have only a
negligible success probability, remains unchanged.

\end{document}